\documentclass{elsart}
\usepackage{amssymb,natbib,epsfig,graphicx,subfigure,longtable}

\newcommand{\ii}{\mathrm{i}}
\newcommand{\jcd}{Christensen-Dalsgaard}
\newcommand{\be}{\begin{equation}}
\newcommand{\ee}{\end{equation}}
\newcommand{\bea}{\begin{eqnarray}}
\newcommand{\eea}{\end{eqnarray}}
\begin{document}
\begin{frontmatter}
\title{Helioseismology and Solar Abundances}
\author[ad1]{Sarbani Basu\thanksref{corr}},\and
\author[ad2]{H. M. Antia}
\thanks[corr]{Corresponding author. E-mail: sarbani.basu@yale.edu.}
\address[ad1]{Department of Astronomy, Yale University, PO Box 208121, \cty New Haven, CT 06520, \cny USA}
\address[ad2]{Tata Institute of Fundamental Research, Homi Bhabha Road, \cty Mumbai 400 005, \cny India}

\begin{abstract}
Helioseismology has allowed us to study the  structure of the
Sun in unprecedented detail. 
One of the triumphs of the theory of stellar evolution was that helioseismic 
studies had shown that the structure of solar models is very similar to that of 
the Sun. However, this agreement has been spoiled by recent revisions of the 
solar heavy-element abundances. Heavy element abundances  determine the opacity of the 
stellar material and hence, are an important input 
to stellar model calculations.
The models with the new, low abundances do not satisfy helioseismic constraints. 
We review here how heavy-element abundances affect solar models, 
how these models are tested with helioseismology, and the impact of the new 
abundances on standard solar models. We also discuss the attempts made to improve 
the agreement of the low-abundance models with the Sun and discuss how 
helioseismology is being used to determine the solar heavy-element abundance.
A review of current literature shows that
attempts to improve agreement between solar models with low heavy-element
abundances and seismic inference have been unsuccessful so far.  The low-metallicity
models that have the least disagreement with seismic data require changing all input physics to
stellar models beyond their acceptable ranges.
Seismic determinations of the solar heavy-element abundance yield results that are  consistent with
the older, higher values of the solar abundance, and hence,  no major changes
to the inputs to solar models are required to make higher-metallicity solar models
consistent with helioseismic data.
\end{abstract}

\end{frontmatter}
\newpage

\tableofcontents

\newpage

\section{Introduction}
\label{sec:intro}

Arthur Eddington began his book ``The Internal Constitution of the
Stars'' saying that {\it ``At first sight it would seem that the deep interior of the sun and stars
is less accessible to scientific investigation than any other region of the universe. Our
telescopes may probe farther and farther into the depths of space; but how can we
ever obtain certain knowledge of that which is hidden behind substantial barriers? What appliance
can pierce through the outer layers of a star and test the conditions within?''} (Eddington 1926).
 Eddington went
on to say that perhaps we should not aspire to directly ``probe'' the interiors of the Sun and stars,
but instead use our knowledge of basic physics to determine what the structure of a star should
be. This is still the predominant approach in studying stars today. However,
we now also have the ``appliance'' that can  pierce through the outer layers of the Sun and
give us detailed knowledge of what the internal structure of the Sun is. This `appliance' is
helioseismology, the study of the interior of the Sun  using solar oscillations.
While solar neutrinos can probe the solar core, helioseismology provides us a much more
detailed and nuanced picture of the entire Sun.

The first definite
observations of solar oscillations were made by Leighton et al.\ (1962), who detected 
roughly periodic oscillations in Doppler velocity with periods of about 5 minutes. 
Evans \& Michard (1962) confirmed the initial observations. The early observations
were of limited duration, and the oscillations were generally interpreted as phenomena
in the solar atmosphere. Later observations that resulted in power spectra as a function
of wave number (e.g., Frazier 1968) indicated that the oscillations may not be mere
surface phenomena. The first major theoretical advance in the field came when 
Ulrich (1970) and Leibacher \& Stein (1971) proposed that the oscillations were
standing acoustic waves in the Sun, and 
predicted that power should be concentrated along ridges in a wave-number v/s frequency
diagram. 
Wolff (1972) and Ando \& Osaki (1975) strengthened the hypothesis of standing waves
by showing that oscillations in the observed frequency and wave-number range
may be linearly unstable and hence, can be excited.
Acceptance of  this interpretation of the observations as normal modes of solar
oscillations 
was the result of the observations of Deubner (1975), which first showed ridges in
the wave-number v/s frequency diagram. Rhodes et al.~(1977) reported similar observations.
These observations did not, however, resolve the individual modes of solar oscillations, despite
that, these data were used to draw initial inferences about solar structure and
dynamics.
Claverie et al.~(1979) using Doppler-velocity observations, integrated over
the solar disk were able to resolve the individual modes of oscillations
corresponding to the largest horizontal wavelength. They found a series of
almost equidistant peaks in the power spectrum just as was expected from theoretical
models.
However, helioseismology as we know it 
today did not begin till Duvall \& Harvey (1983) 
determined frequencies of a reasonably large  number of solar oscillation
modes covering a wide range of horizontal wavelengths. Since then many
sets of solar oscillations frequencies have been published.
A lot of early helioseismic analysis was based on frequencies determined
by Libbrecht et al.~(1990) from
observations made at the Big Bear Solar Observatory in the period 1986--1990.
Accurate determination of solar oscillations
frequencies requires long, uninterrupted observations of the Sun, that
are possible only with a network of ground based instruments or from an instrument
in space.
The Birmingham Solar Oscillation Network
(BiSON;  Elsworth et al.~1991; Chaplin et al.~2007a) was one of the first such networks.
BiSON, however, observes the Sun in integrated light and hence is capable of observing only very large horizontal-wavelength modes.
The Global Oscillation Network Group (GONG), a ground based network
of telescopes,  and the Michelson Doppler Imager (MDI) on board the Solar and Heliospheric
Observatory (SOHO) have now collected data for more than a decade and have given us
an unprecedented opportunity to determine the structure and dynamics of the Sun in great
detail. Data from these instruments have also allowed us to probe whether or not the Sun changes on the time
scale of a solar activity cycle.

Helioseismology has proved to be an extremely important tool in studying the Sun.
Thanks to helioseismology, we know the most important features of the
structure of the Sun extremely well. We know what the sound-speed and density
profiles are (see e.g., Christensen-Dalsgaard et al.~1985, 1989; Dziembowski et al.~1990;
D\"appen et al.~1991; Antia \& Basu~1994a; Gough et al.~1996; Kosovichev et al.~1997; 
Basu et al.~1997, 2000; etc.),
which in turn means that we can determine the radial distribution 
of pressure. We can also determine the profile of the adiabatic index (e.g., Antia \& Basu 1994a; 
Elliott 1996; Elliott \& Kosovichev 1998). 
Inversions of solar oscillations frequencies have allowed us to determine a number of other fundamental facts about the
Sun. We know, for instance, that the position of  base of the solar convection
zone can be determined
precisely (Christensen-Dalsgaard et al.~1991; Basu \& Antia 1997; Basu 1998).
Similarly, we can determine the helium abundance in the solar  convection zone
(D\"appen \& Gough 1986; Christensen-Dalsgaard \& P\'erez Hern\'andez 1991; Kosovichev et al.~1992;
Antia \& Basu 1994b).
In addition to these structural parameters,
helioseismology has also revealed what the rotational profile of the Sun is
like. It had been known for a long time that the rotation rate at the solar surface depends
strongly on latitude, with rotation being fastest at the equator and slowest at the poles.
Only with helioseismic data however, we have been able to probe the rotation of 
the Sun as a function of depth (Duvall et al.~1986; Thompson et al.~1996; Schou et al.~1998b;
etc.).

The ability of helioseismology to probe the solar interior in such detail has allowed
us to use the Sun as a laboratory to test different inputs that are used to
construct solar models. For instance, helioseismic inversions have 
allowed us to study the equation of state of stellar material
(Lubow et al.~1980; Ulrich 1982; \jcd\ \& D\"appen 1992;
Basu \& Christensen-Dalsgaard 1997; Elliott \& Kosovichev 1998; 
Basu et al.~1999) and to test opacity calculations
(Korzennik \& Ulrich 1989; Basu \& Antia 1997; Tripathy \& \jcd~1998).
Assuming that opacities, equation of state, and nuclear energy generation rates are
known, one can also infer the temperature and hydrogen-abundance profiles
of the Sun (Gough \& Kosovichev 1988; Shibahashi 1993; Antia \& Chitre 1995, 1998;
Shibahashi \& Takata 1996; Kosovichev 1996 etc.).
These studies also provide a test for nuclear reaction rates (e.g., Antia \&
Chitre 1998; Brun et al.~2002) and the  heavy-element abundances in the
convection zone (e.g., Basu \& Antia 1997; Basu 1998) and the core (e.g., Antia \& Chitre 2002).

One of the major inputs into solar models is the abundance of heavy elements. The
heavy-element abundance, $Z$, affects solar structure by affecting radiative opacities.
The abundance of some specific elements, such as oxygen, carbon, and nitrogen can also
affect the energy generation rates through the CNO cycle. The effect of $Z$ on opacities changes the
boundary between the radiative and convective zones, as well as the structure
of radiative region;  the effect of $Z$ on 
energy generation rates can change the structure of the core. 
The heavy element abundance of the Sun is believed to be known to a much better accuracy than
that of other stars, however, there is still a lot of uncertainty and that
results in uncertainties in solar models. It is not only the total $Z$ that affects structure,
the relative abundance of different elements has an effect as well. Elements that affect
core opacity  are, in the order of importance, iron, sulfur, silicon and
oxygen. The elements that contribute to opacity in the region near the base
of the convection zone and
thereby affect the position of the base of the solar convection zone are,
 again in the order of importance, oxygen, iron and neon.
Although, the main effect of heavy element abundances is through opacity,
these abundances also affect the equation of state. In particular, the adiabatic index
$\Gamma_1$ is affected in regions where these elements  undergo
ionization. This effect is generally small, but in the convection zone where the
stratification is adiabatic and hence the 
structure is determined by equation of state rather than opacity, 
 this effect can be significant.

The importance of the solar heavy-element abundance does not merely lie in being able
to model the Sun correctly, it is  often used
as the standard against which heavy-element abundances of other stars are
measured. Thus the predicted structure of those stars too become uncertain if 
the solar heavy-element abundance is uncertain. Given
that for most stars other than the Sun, we usually only know the position on the
HR diagram, an error in the solar  abundance could lead to errors in the
predicted mass and age of the stars. 
Stellar evolution calculations are used throughout astronomy to 
classify, date, and interpret the spectra of individual stars and of galaxies, 
and hence errors in metallicity affect age determinations, and other 
derived parameters of stars and star clusters.
The exact value of the solar heavy-element abundance determines
the amount of heavy elements that had been present in the solar neighborhood
when the Sun was formed. This,
therefore, determines the chemical evolution history of galaxies.

Solar models in the 1990's were generally constructed with the solar heavy element
mixture of Grevesse \& Noels (1993). The ratio of the mass fraction of heavy elements 
to hydrogen in the Sun was determined to be $Z/X= 0.0245$. Grevesse \& Sauval (1998; henceforth GS98)
revised the abundances
of oxygen, nitrogen, carbon and some other elements, and that resulted in
$Z/X=0.023$.
In a series of papers Allende-Prieto et al.~(2001, 2002) and
Asplund et al.~(2004, 2005a) have revised the spectroscopic
determinations of the solar photospheric composition. In particular,
their results indicate that  carbon, nitrogen and oxygen abundances
are lower by about 35\% to 45\% than those listed by
GS98.  The revision of the oxygen
abundance leads to a comparable change in the abundances of neon and
argon since these abundances are generally measured through the abundances
ratio for Ne/O and Ar/O.
Additionally, Asplund (2000) also determined a somewhat
lower value (by about 10\%) for the photospheric abundance of silicon compared
with the GS98 value. 
As a result, all the elements for which abundances
are obtained from meteoritic measurements have seen their abundances
reduced by a similar amount. These measurements have been summarized
by Asplund, Grevesse \& Sauval (2005b; henceforth AGS05). The net result of these changes
is that $Z/X$ for the Sun is reduced to $0.0165$ (or $Z=0.0122$), about 28\% lower than
the previous value of GS98 and almost 40\% lower than the old value of
Anders \& Grevesse (1989).
The change in solar abundances implies large changes in solar structure as well as
changes in quantities derived using solar and stellar models, and therefore, warrants
a detailed discussion of the consequence of the changes, and how one can test
the new abundances.
In this paper we review  the effects of 
solar abundances on  solar models and how the models with lower abundances
stand up against helioseismic tests. 

The review is written in a pedagogical style with  detailed
explanations of how the analysis is done. However, the review is organized in
such  a manner that not all readers need to read all sections unless they
want to. We start with a description of how solar models
are constructed (\S~\ref{sec:models}), this section also describes the sources
of uncertainties in solar models (\S~\ref{subsec:ssmerr}). In \S~\ref{sec:helioseis} we
describe how helioseismology is used to test solar models as well as how
helioseismology can be used to determine solar parameters like the convection-zone
helium abundance, the depth of the convection zone and how input physics, like
the equation of state, can be tested. In \S~\ref{sec:res} we describe what helioseismology
has taught us about the Sun  and inputs to solar model thus far. Thus readers who are
more interested in helioseismic results rather than techniques can go directly to
this section. In \S~\ref{sec:solarz} we give a short review of how solar abundances
are determined and the results obtained so far.
The consequences of the new abundances are described in \S~\ref{sec:abund}. This section
also includes a brief summary of the changes in the solar neutrino outputs 
(\S~\ref{subsec:radint}) and some consequences of the new abundances on models
of stars other than the Sun (\S~\ref{subsec:stellar}), though the latter discussion
is by no means complete and comprehensive. Readers who are only interested
in how the lower solar abundances affect the models would perhaps wish to
go straight to this section. The next section, \S~\ref{sec:recon}, is devoted to 
reviewing the numerous attempts that have been made to reconcile the low-metallicity
solar models with the Sun by changing different physical inputs. In \S~\ref{sec:seisz} we
describe attempts that have been made to determine solar metallicity using helioseismic
techniques. 
In \S~\ref{sec:problems} we discuss some possible reasons for the discrepancy between
 helioseismically determined abundances  and
the new spectroscopic abundances  and we present some final thoughts in
\S~\ref{sec:concl}.

\section{Making solar models}
\label{sec:models}

The Sun is  essentially similar to other stars. The internal structure
of the Sun and other stars obey the same principles, and hence
we use the theory of stellar structure and evolution to make
models of the Sun. 
However, since we have more observational constraints
on the Sun, these constraints have to be met before we can call a
model a solar model. Otherwise, the result is simply a model of a star that has the
same mass as the Sun. As in the case of other stars, we know the effective temperature and
luminosity of the Sun, but unlike most stars, we also have independent estimates
of the age and radius of the  Sun. 
Thus to be called a solar model, a $1M_\odot$ model
 must have the correct radius and
luminosity at  its current age. This makes modelling the Sun somewhat
different from modelling other stars. The commonly adopted values of
solar mass, radius, luminosity and age of the Sun are listed in
Table~1.

There are many excellent text-books on stellar structure and
evolution. 
The equations governing stellar structure and evolution  have been  discussed in detail
by Kippenhahn \& Weigert (1990), Huang \& Wu (1998), Hansen et al.~(2004),
Weiss et al.~(2004), etc. We, therefore, only give a quick overview.

\subsection{Equations of stellar structure and evolution}
\label{subsec:struceq}

The most common assumption involved in making solar and stellar models is that 
stars are not merely spherical, but that they are also
spherically symmetric, i.e., their internal structure is only a
function of radius
and not of latitude or longitude. This assumption implies that rotation
and magnetic fields do not unduly change stellar structure.
This is a good approximation for most stars and can be applied to the
Sun too.
The measured oblateness of the
Sun is about a part in $10^5$ (Kuhn et al.~1998). The latitudinal
dependence of solar  structure is also small (Antia et al.~2001).
This assumption allows us to express the properties of a star with a
set of one-dimensional (1D) equations, rather than a full
set of three-dimensional (3D) equations. These equations can be
derived from very basic physical principles.

%%%%%%-------------- TABLE -------------
\begin{table}[t]
\caption{Global parameters of the Sun\label{tab:glob}}
\begin{center}
\begin{tabular}{lcl}
\noalign{\smallskip}
\hline
{\hbox{Quantity}}&{\hbox{Estimate}}&{\hbox{Reference}}\\
\hline
Mass ($M_\odot$)$^*$ & $1.98892(1\pm 0.00013)\times 10^{33}$ g & Cohen \& Taylor (1987)\\
Radius ($R_\odot$)$^+$ & $6.9599(1\pm0.0001)\times 10^{10}$ cm & Allen (1973)\\
Luminosity ($L_\odot$) & $3.8418(1\pm 0.004)\times 10^{33}$ ergs s$^{-1}$ & Fr\"ohlich \& Lean (1998),\\
 & & Bahcall et al.~(1995)\\
Age & $4.57(1\pm 0.0044) \times 10^{9}$ yr & Bahcall et al.~(1995)\\
\hline
\noalign{\smallskip}
\end{tabular}

{$^*$} {\small Derived from the values of $G$ and $GM_\odot$}\hfill\\

{$^+$} {\small See  Schou et al.~(1997), Antia (1998) and 
Brown \& Christensen-Dalsgaard
(1998) 
 for a more recent discussion about the exact value of the solar radius.}\\
\end{center}
\end{table}
%-----------------------------

The first equation is a result of conservation of mass
and can be written as
\be
{\d m\over \d r}=4\pi r^2 \rho,
\label{eq:mass}
\ee
where $\rho$ is  the density, and $m$ is the mass enclosed in radius $r$.
Since stars expand or contract  over their lifetime, it is generally
easier to use equations with mass $m$ as the independent variable since
for most stars the total mass does not change much during their
lifetimes. The Sun for instance, loses about $10^{-14}$ of its
mass per year. Thus in its expected main-sequence lifetime of about 10 Gyr,
the Sun will lose only about 0.01\% of its mass. The radius on the other hand is
expected to change significantly. Thus Eq.~(\ref{eq:mass})
is usually  re-written as
\be
{\d r\over \d m}= {1\over 4\pi r^2\rho}\>.
\label{eq:drdm}
\ee
The next equation is a result of conservation of momentum.  In
the stellar context it implies that
 any acceleration of a mass shell is caused by a mismatch between
outwardly acting pressure and inwardly acting gravity. However,
pressure and gravity  balance each other throughout most
of a star's life, and under these conditions  we can write the so-called equation of
{\em hydrostatic equilibrium}, i.e.,
\be
{\d P\over \d m}=-{Gm\over 4\pi r^4}\>.
\label{eq:hydro}
\ee
The third equation is conservation of energy.
Since stars are not just passive spheres of gas, but produce energy through
nuclear reactions in the core, the energy equation needs to be considered.
In a stationary state, energy $l$ flows through a shell
of radius $r$ per unit time  as  a result of
nuclear reactions in the interior. If $\epsilon$ be the energy released
 per unit mass per second by nuclear reactions, and $\epsilon_\nu$ the energy
lost by the star because of neutrinos streaming out of the star without
depositing their energy, then,
\be
{\d l\over \d m}=\epsilon -\epsilon_\nu.
\label{eq:nuc}
\ee
Since stars expand (or contract) at certain phases of their lives, the equation needs
to be re-written to include the energy used (or released) due to expansion (or contraction).
Thus:
\be
{\d l\over \d m}=\epsilon-\epsilon_\nu -C_P{\d T\over \d t}+
{\delta\over\rho}{\d P\over \d t}\>,
\label{eq:nucf}
\ee
where $C_P$ is the specific heat at constant pressure, 
$t$ is time,    and $\delta$, given by the equation of state, is defined as
\be 
\delta = -\left({\partial\ln\rho\over\partial\ln T}\right)_{P,X_i},
\ee
where $X_i$ denotes composition.
The last two terms on the
right-hand side of Eq.~(\ref{eq:nucf}) are often referred to  together as 
$\epsilon_g$, $g$ for gravity, because they denote the gravitational release of energy.
The next  is the equation of
energy transport which determines the temperature at any point.
In general terms, and with the help of Eq.~(\ref{eq:hydro}), this equation
can be
 written quite trivially as
\be
{\d T\over \d m}=-{Gm T \over 4\pi r^4 P}\nabla,
\label{eq:dtdm}
\ee
where
$\nabla$ is the dimensionless ``temperature gradient'' $\d\ln T/\d\ln P$.
The difficulty lies in determining what $\nabla$ is.
In the radiative zones, under the approximation of diffusive radiative
transfer, $\nabla$ is given by
\be
\nabla=\nabla_{\rm rad}={3\over 64\pi \sigma  G}{\kappa l P\over m T^4}\>,
\label{eq:gradrad}
\ee
where, $\sigma$ is the Stefan-Boltzmann constant and 
 $\kappa$ is the opacity.

The situation is more complicated if energy is transported by convection. Deep inside the
star, $\nabla$  is usually the
adiabatic temperature gradient $\nabla_{\rm ad}\equiv (\partial\ln T/\partial\ln P)_s$
($s$ being the specific entropy),  which is determined
by the equation of state. In the outer layers, one has to use some
approximate formalism, since there is no ``theory'' of stellar convection as such.
Convection can be described by solving the Navier-Stokes equations. However the
mismatch between time scales involved with convective transport
of energy  (minutes to hours to days) and time scales of stellar evolution
(millions to billions of years) makes solving the Navier-Stokes equations along
with the equations of stellar evolution computationally impossible today. As a result,
several approximations are used to describe convection in stellar models.
One of the
most common formulations used in the calculation of convective flux in stellar models
 is the so-called ``mixing length theory'' (MLT).
The mixing length theory was first proposed by Prandtl (1925).
His model of convection was analogous to heat transfer by particles; the
transporting particles are the macroscopic eddies and their mean free path is
the ``mixing length''. This was applied to stars by Biermann (1951),
 Vitense (1953), and B\"ohm-Vitense (1958).
Different mixing length formalisms have slightly different assumptions about
what the mixing length is. The main assumption in the usual mixing length formalism is that the size of the
convective eddies at any radius is the mixing length $l_m$, where
$l_m=\alpha H_P,$ and
$\alpha$, a constant,  is the so-called `mixing length parameter',
and $H_P$ is the pressure scale height given by $-\d r/\d\ln P$. Details of how under these
assumptions $\nabla$ is calculated can be found in any standard textbook, such as
Kippenhahn \& Weigert (1990). There is no {\it a priori\/} way to determine
$\alpha$, and it is one of the free parameters in stellar models. 
A different prescription for calculating convective flux based on a treatment of
turbulence was given by Canuto \& Mazzitelli (1991).
These are the so-called `local' prescriptions, as the convective flux at any depth is
determined by the local values of $T$, $P$, $\rho$, etc. There have been attempts
to formulate nonlocal treatments for calculating convective flux
(e.g., Xiong \& Chen 1992; Balmforth 1992), but these treatments introduce  free parameters
to quantify the nonlocal effects and there are no standard ways of determining those
parameters.

Whether energy is transported by radiation or by convection
depends on the value of $\nabla_{\rm rad}$. For any given
material there is a maximum value of $\nabla$ above which the material
is convectively unstable. This  maximum is  $\nabla_{\rm ad}$. If $\nabla_{\rm rad}$
obtained from Eq.~(\ref{eq:gradrad}) exceeds $\nabla_{\rm ad}$, convection sets in.
This is usually referred to as the `Schwarzschild Criterion'.
Regions
where energy is transported by radiation are usually referred to as
radiative zones, and regions where a part of the energy is transported by convection
are referred to as convection zones.

The last important equation concerns the change of chemical composition with
time. There are three main reasons for the change in chemical composition at
any point in the star. These are:
(1) nuclear reactions; (2) the changing boundaries of convection zones;
and (2) Diffusion and gravitational settling (usually simply referred to as
diffusion) of helium and heavy elements.

If $X_i$ is the mass fraction of any isotope $i$, then the change in $X_i$ with time
because of nuclear reactions can be written as
\be
{\partial X_i\over \partial t}=
{m_i\over\rho}\left[\sum_j r_{ji}-\sum_k r_{ik}\right],
\label{eq:nucx}
\ee
where $m_i$ is the mass of the nucleus of each isotope $i$, $r_{ji}$ is the rate at which
isotope $i$ is formed from isotope $j$, and $r_{ik}$ is the rate at which isotope $i$
is lost because it turns into a different isotope $k$. The rates $r_{ik}$ are
inputs to models.

Convection zones are chemically homogeneous since eddies of matter move carrying their
composition with them and when they break-up, the material gets mixed
with the surrounding. They become chemically homogeneous over very short time
scales compared to the time scale of a star's evolution. If a convection zone
exists in region between two spherical shells of masses $m_1$ and $m_2$, the average abundance
of any species $i$ in the convection zone is:
\be
{\bar X_i}={1\over m_2-m_1}\int_{m_1}^{m_2} X_i\;\d m\>.
\label{eq:barx}
\ee
Thus the rate at which ${\bar X_i}$ changes will depend on nuclear reactions in the convection
zone, as well as the rate at which the mass limits  $m_1$ and $m_2$ change. One can therefore
write:
\bea
\!\!\!\!{\partial{\bar X_i}\over \partial t} & = &
{\partial\over\partial t}\left({1\over m_2-m_1}\int_{m_1}^{m_2} X_i\; \d m\right)\nonumber\\
&=& {1\over m_2-m_1}\left[ \int_{m_1}^{m_2} {\partial X_i\over\partial t}\;\d m+
{\partial m_2\over \partial t}(X_{i,2}-{\bar X_i}) -
{\partial m_1\over \partial t}(X_{i,1}-{\bar X_i})\right],\phantom{as}
\label{eq:convx}
\eea
where $X_{i,1}$ and $X_{i,2}$ is the mass fraction of element $i$ at $m_1$ and
$m_2$ respectively.

The gravitational settling of helium and heavy elements can be described by the
process of diffusion and the change in abundance can be found with the help of the
diffusion equation:
\be
{\partial X_i\over \partial t}=D\nabla^2 X_i\>,
\label{eq:diff}
\ee
where $D$ is the diffusion coefficient, and $\nabla^2$ is the Laplacian operator.
The diffusion coefficient hides the complexity of the process and includes, in addition
to gravitational settling, diffusion due to composition and temperature gradients.
All three processes are generally simply called `diffusion'.
The coefficient $D$ depends on the isotope under consideration.
Typically, $D$ is an input to stellar model calculations and is not
calculated from first principles in the stellar evolution code. Among the more
commonly used prescriptions for calculating diffusion coefficients are those
of Thoul et al.~(1994) and Proffitt \& Michaud (1991).

Equations~(\ref{eq:drdm}), (\ref{eq:hydro}), (\ref{eq:nucf}), (\ref{eq:dtdm})
 together with the
equations relating to change in abundances, form the full set of equations
that govern stellar structure and evolution. In the most general case,  Eqs.~(\ref{eq:drdm}), (\ref{eq:hydro}),
 (\ref{eq:nucf}) and (\ref{eq:dtdm}) are solved for a given $X_i$ at a given time
$t$. Time is then
advanced, Eqs.~(\ref{eq:nucx}), (\ref{eq:convx}) and (\ref{eq:diff}) are
solved to give new $X_i$, and equations
(\ref{eq:drdm}), (\ref{eq:hydro}),
 (\ref{eq:nucf}) and (\ref{eq:dtdm}) are solved  again. Thus we consider two independent
variables, mass $m$ and time $t$, and we look for solutions in the interval
$0\le m \le M$ (stellar structure) and $ t \ge t_0$ (stellar evolution).

Four boundary conditions are required to solve the stellar
structure equations. Two (on radius and luminosity) can be applied quite 
trivially at the center. The remaining conditions (on temperature and pressure) need to be
applied at the surface. The boundary conditions at the surface are much
more complex than the central
boundary conditions  and are  usually determined  with the aid of simple
stellar-atmosphere models.
It is very common to define the surface using the Eddington approximation. 
The initial conditions needed to start evolving a star depend on where we start the
evolution. If the evolution begins at the pre-main sequence phase, i.e., while the
star is still collapsing, the initial structure is quite
simple. Temperatures  are low enough to make the
star fully convective and hence chemically
homogeneous. If evolution is begun at the ZAMS, i.e., the Zero Age Main Sequence, which is the point
at which  hydrogen fusion
begins, a ZAMS model must be used. 

\subsection{Input microphysics}

Equations~(\ref{eq:drdm}), (\ref{eq:hydro}), (\ref{eq:nucf}), (\ref{eq:dtdm}) and the
equation for abundance changes are simple, but hide their complexity in the form
of the external inputs (usually referred to as `microphysics') needed to solve the
equations. We discuss these below:

\subsubsection{The Equation of State}

There are five equations in six unknowns, 
$r$, $P$, $l$, $T$, $X_i$, and
$\rho$. None of the equations tells us how density,  $\rho$, changes with time or mass. Thus
we need a relation connecting density to the other quantities. This is given by
the equation of state which  specifies the relation between density, pressure,
temperature and composition. Although the ideal gas equation of state is good enough for
making simple models, it  does not apply in all regions of
stars. In particular, the ideal gas law does not include effects of ionization, radiation
pressure, pressure ionization, degeneracy, etc. Modern equations of state are usually
given in tabular form as functions
of $T$, $P$ (or $\rho$) and composition,  and interpolations are done
 to obtain the thermodynamic quantities needed to solve the equations.
Among the popular equations of state used to construct solar models are
the OPAL equation of state (Rogers et al.~1996; Rogers \& Nayfonov 2002), the
so-called MHD (i.e., Mihalas, Hummer \& D\"appen) equation of state
(D\"appen et al.~1987, 1988a; Hummer \& Mihalas 1988;
Mihalas et al.~1988) and the CEFF equation of state (Guenther et al.~1992; 
Christensen-Dalsgaard \& D\"appen 1992). CEFF stands for
Coulomb corrected Eggleton, Faulker and Flannery equation of state, and is basically the
equation of state described by Eggleton et al.~(1973), along with
correction for Coulomb screening.

\subsubsection{Opacity}

In order to calculate $\nabla_{\rm rad}$ (Eq.~\ref{eq:gradrad}), we need to know the opacity. The opacity,
$\kappa$, is a measure of how opaque a material is to photons. Like the modern equations of state,  Rosseland
mean opacities are given in tabular form as a function of density, temperature and
composition. Among the tables used to model the Sun are the OPAL tables (Iglesias \& Rogers 1996),
and the OP tables (Badnell et al.~2005; Mendoza et al.~2007). 
The OPAL opacity tables include contributions from 19 heavy elements whose
relative abundances  (by numbers) with respected to hydrogen  are larger than about $10^{-7}$. 
The OP opacity calculations include only 15 of these heavy elements, as
P, Cl, K and Ti are not included.
These tables do not properly include all contributions from molecules, and hence these
tables  are usually 
supplemented with more accurate low-temperature opacity tables.
For solar models, opacity tables by Kurucz (1991) or Ferguson et al.~(2005) are often used
at low temperatures.

\subsubsection{Nuclear Reaction Rates}

Nuclear reaction rates are required to compute
energy generation, neutrino fluxes and composition
changes. Two major sources of reaction rates
for solar model construction are
the compilations of Adelberger et al.~(1998) and Angulo et al.~(1999).
The rates of most relevant  nuclear reactions  are obtained by extrapolation from
laboratory measurements, but in some cases the reaction rates are based
on theoretical calculations.
These reaction rates need to be corrected for electron screening
in the solar plasma. Solar models generally use the weak (Salpeter 1954)
or intermediate (Mitler 1977) screening approximations to treat electron
screening.

\subsection{Constructing standard solar models}
\label{subsec:ssm}
The mass of a star is the most fundamental quantity needed to model a star.
Once the mass is known, one can, in principle, begin making models of the
star and determine how it will evolve. 
In reality, several other quantities
are required. These include $Z_0$, the initial heavy-element abundance of the star,
as well as the initial helium abundance $Y_0$.  Both these quantities
affect evolution since they affect the equation of state and
opacities. Also required is an estimate of the mixing length
parameter $\alpha$. Once these quantities are known, or chosen by some means, the
model is evolved by solving the equations. For most stars, the models are evolved
till they reach a given temperature and luminosity. The age of the star is assumed to be
the age of the model, and the radius of the star is assumed to be the radius of the model.

The Sun is modeled in a slightly different manner since the age, luminosity and radius
are known independently.
Thus to be called a solar model,
a $1M_\odot$ model must have a luminosity of $1L_\odot$ and a radius of
$1R_\odot$ at the age of 4.57 Gyr. The way we ensure that we get a solar model is
to vary the mixing length parameter
$\alpha$ and the initial helium abundance $Y_0$ till we get a model with the
required characteristics. Mathematically speaking, we have two unknown parameters
($\alpha$ and $Y_0$) and two constraints (radius and luminosity) at 4.57 Gyr, and hence
this is a well defined problem. However, since the equations are non-linear, we
need an iterative method to determine $\alpha$ and $Y_0$. The value of
$\alpha$ obtained in this manner for the Sun is often used to model
other stars. In addition to $\alpha$ and $Y_0$, very often
initial $Z$ is adjusted to get the observed $Z/X$ in the solar envelope.
The solar model so obtained does not have any free parameters,
since the two unknowns,  $\alpha$ and $Y_0$, are determined to match solar constraints.

\begin{figure} 
\begin{center}
\includegraphics[width=.99\textwidth]{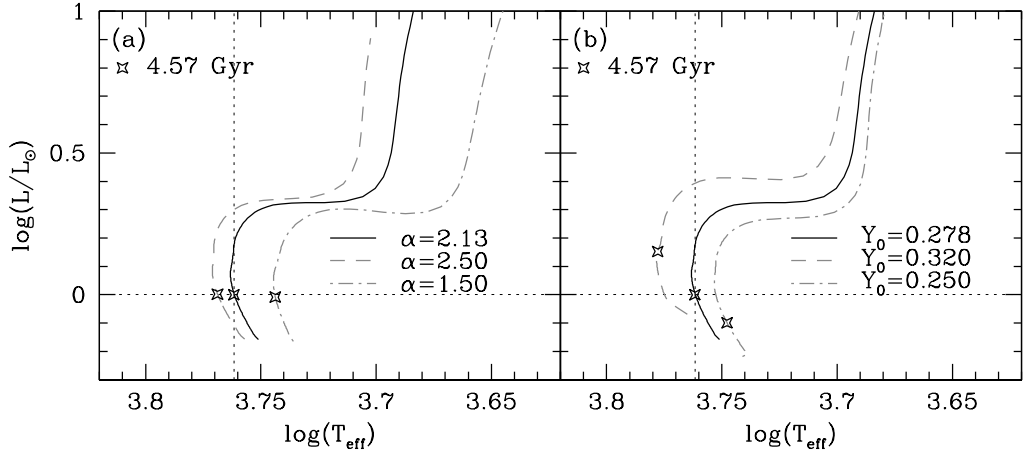}
\caption{Panel (a): The effect of the mixing length parameter on the evolution
of a $1M_\odot$ star. All models have $Y_0=0.278$. Panel (b): The effect of the initial helium abundance $Y_0$ on
the  evolution of a $1M_\odot$ star. All models have $\alpha=2.13$. In both panels the intersection of the dotted lines
mark the position of the Sun. The star on each curve marks 4.57 Gyr, the current age of the
Sun.}
\label{fig:ssmalhe}
\end{center}
\smallskip
\end{figure}

Fig.~\ref{fig:ssmalhe}(a) shows a series of evolutionary tracks for a $1M_\odot$
model constructed using different values of $\alpha$. All models have the same $Y_0$ (0.278).
All models have been constructed with YREC, the Yale Rotating Evolutionary
Code in its non-rotating configuration (Guenther et al.~1992). They were constructed using
the  OPAL equation
of state (Rogers \& Nayfonov 2002) and OPAL opacities (Iglesias \& Rogers 1996).
The models were constructed to have $Z/X=0.023$ (i.e., the GS98 value)
at the age of the current Sun.
As can be seen from the figure,
only one of the models (with $\alpha=2.13$) satisfies solar constraints at 4.57 Gyr.
Fig.~\ref{fig:ssmalhe}(b) shows models with different $Y_0$ but the same
$\alpha$ (2.13). Again, only one model (with $Y_0=0.278$)
satisfies solar constraints, and thus is the only solar
model of the three models shown.

The concept of standard solar models (SSM) is very important in solar physics.
Standard solar models are models where only standard input physics
such as equations of state, opacity, nuclear reaction rates, diffusion
coefficients etc., are used. The parameters $\alpha$ and
$Y_0$ (and sometime $Z_0$) are adjusted to match the current solar radius
and luminosity (and surface $Z/X$). No other input
is adjusted to get a better agreement with the Sun. Thus a standard solar
model does not have any free parameters. By comparing standard
solar models constructed with different input physics with the Sun we can
put constraints on the input physics. One can use helioseismology to
test whether or not the structure of the model agrees with that of the Sun.
The model in Fig.~\ref{fig:ssmalhe} that satisfies current solar constraints
on luminosity, radius and age is a standard solar model.

A solar model turns out to be quite simple. Like all
stars of similar (and lower) masses, it has an outer convection zone and
an inner radiative zone. In the case of the Sun, the convection zone
occupies the outer 30\% by radius. The outer convection zone is a result of
large opacities caused by relatively low temperatures.
The temperature gradient required to transport energy by radiation in this region
exceeds the adiabatic temperature gradient
resulting in a convectively unstable layer.
Convective eddies ensure that the convection zone is 
chemically
homogeneous. In models that
incorporate the diffusion and gravitational settling of helium and
heavy-elements, the  abundances of these elements  build up below the 
convection-zone base. 

\subsection{Sources of uncertainty in  standard solar models}
\label{subsec:ssmerr}

Standard solar models constructed by different groups are usually not identical,
and are only as good as the input physics.
The models depend on  nuclear reaction rates, 
radiative opacities, equation of state, diffusion coefficients, surface boundary
conditions, 
uncertainties in any of these inputs result in uncertainties in solar models.
Even the numerical scheme used to construct a model can introduce some
uncertainties.
There have been many investigations of the effect of uncertainties in
inputs on standard solar models.

Boothroyd \& Sackmann (2003) did a systematic investigation of the 
effect of some of the uncertainties in input physics on standard solar models.
They also investigated the effect of the
number of mass zones and time steps used  to calculate the models.
They found that models with 2000 spatial zones, about what is usually
used to calculate solar models, did only marginally worse than
a model with 10,000 zones. 
Their coarse-zone models agreed with the
adopted solar radius to a part in $10^5$ and with $Z/X$ to a part in $10^4$. Their
fine-zoned models were better than a part in $10^5$ in radius and a few parts in
$10^5$ in $Z/X$. They found that the rms relative difference
in sound speed and density of the coarse-zoned models relative to
the fine-zoned models was 0.0001 and 0.0008 respectively, but that the difference in the
number of zones  had no 
effect on the adiabatic index $\Gamma_1$.

Bahcall et al.~(2001) tested the effects of $2\sigma$ changes
in $L_\odot$ and found negligible effects on most quantities, though there
were minor effects on neutrino fluxes. Boothroyd \& Sackmann (2003) 
found that a shift of 0.8\% in $L_\odot$ produces a fractional change in the sound speed of 
less than three parts in $10^4$ that drops  to one part in $10^4$ for $r > 0.3 R_\odot$.
Uncertainties in solar radius have a much larger effect. Basu (1998) showed that 
using a solar radius different from the standard value of $R_\odot = 695.99$ Mm 
causes a small, but significant change in the sound-speed and density profiles of the
models. If the radius is reduced to 695.78 Mm (Antia 1998), then in the regions that can be 
successfully probed with helioseismology,
the rms relative sound-speed difference between this and a model with the standard radius
 is 0.00014 with a maximum difference of
0.0002. The rms relative density difference is 0.0015 and the maximum difference
is 0.0019.
The effect of uncertainties in solar age are minor (Morel et al.~1997; Boothroyd \& Sackmann 2003).
A change  of 0.02 Gyr in the  solar age of 4.57 Gyr yield small effects according
to Boothroyd \& Sackmann (2003), with rms relative sound-speed differences
of about 0.0001 and rms relative density differences of 0.001.
The effect of changing the value of solar mass is more subtle since the product $GM_\odot$
is known very accurately.  Any change in the value of solar mass has
to be compensated by an opposite change in the value of $G$ in order to keep $GM_\odot$ the same. The effect of a change
in the value of the gravitational constant $G$ will be similar --- compensated 
by the need to change the value of $M_\odot$.  It turns out
that solar models are not very sensitive to these changes.
\jcd\ et al.~(2005) found that changing $G$ by 0.1\%, changes the position
of the base of the convection zone by $0.00005R_\odot$ and the helium abundance at the
surface by 0.0003. Similarly, the relative change in sound speed is less
than $10^{-4}$, while that in the density is less than 0.002.

Uncertainties and changes in nuclear reaction rates predominantly affect the
core. However, the effect in not completely limited to the core and can be felt in the solar
envelope too, and results  in the  change in the position of the convection-zone base. Boothroyd \& Sackmann (2003)
found that 
 a change of 5\% in the p-p reaction rate results in rms relative sound-speed
changes of 0.0009 and rms relative density changes of $0.018$. The relative sound-speed change in the
core can be as much as 0.003, but only 0.0014 in the regions that can be successfully 
probed by helioseismology.
Brun et al.~(2002) found that changing the the
${}^3$He - ${}^3$He and ${}^3$He - ${}^4$He reaction rates by 10\% results in
less than 0.1\% change in sound speed and less tha 0.6\% in density.
Recently, the cross section of the $^{14}$N($p,\gamma)^{15}$O reaction rate 
was reduced (Formicola et al.~2004). This changes the core structure, and even changes the
position of the base of the convection zone by $0.0007R_\odot$ (Bahcall et al. 2005c).

It is difficult to quantify the uncertainties in input physics like the
equation of state and opacities since  these depend on a number of quantities such as  temperature, density, composition,
etc. A measure of the uncertainties can be obtained by using two independent
estimates.
Basu et al.~(1996) found that changing the input equation of state from OPAL to MHD changes
the position of the base of convection zone in the model by $0.0009R_\odot$.  Guzik \& Swenson (1997) tested
a number of different equations of state and also found that OPAL and MHD 
models differ significantly and that the use of the MHD equation of state gives
rise to higher pressure in parts of the solar envelope. This was confirmed by
Boothroyd \& Sackmann (2003).

One of the largest effects on a solar model is that of radiative opacity. Opacities determine
the structure of the radiative interior, and in particular, the position of the base of
the convection zone. 
Neuforge-Verheecke et al.~(2001a) compared models using the 1995 OPAL opacities
with LEDCOP opacities from Los Alamos  to find fractional sound speed difference as
large as 0.003. Boothroyd \& Sackmann (2003) found that the differences can be 
much larger, depending on how much the opacities are changed. Opacities are often given in 
tabular form and require interpolation when used in stellar models, and interpolation
errors can also play a role.  Neuforge-Verheecke et al.~(2001a) suggest that $T$ and $\rho$ 
interpolation errors in the opacity can be as much as a few percent. 
Bahcall et al.~(2004) found that different state-of-the-art interpolation schemes used to
interpolate between  existing OPAL tables yield opacity values that differ by up to  4\%. 
Interpolation
errors also play a role in defining the position of the base of the convection zone in solar models.
Bahcall et al.~(2004) did a detailed study of how accurately one could
calculate the depth of the convection zone in a model, and concluded
that  radiative
opacities need to be known to an accuracy of 1\% in order to get an accuracy
of 0.14\% in the position of the convection-zone base. 

Another input whose uncertainties affect solar structure is the diffusion coefficient.
Boothroyd \& Sackmann (2003) showed that a 20\% change in the helium diffusion
rate leads to an rms relative sound-speed difference of 0.0008 and 
rms relative density difference of 0.007. The effect on the rms differences,
of changing the heavy-element diffusion coefficients was found to be smaller, with a
40\% change causing an rms relative sound-speed difference of 0.0004 and density
of 0.004. Increasing  diffusion coefficients also  change
the position of the base of the convection
zone and the convection-zone helium abundance. Montalb\'an et al. (2004)
found that increasing the heavy element diffusion coefficients by 50\%
changes the position of the convection-zone base by 0.006$R_\odot$ and the
convection-zone helium abundance by 0.009.

The formulation used to calculate  convective flux in the convection zone
also affects solar models. However, since the temperature gradient in most of the convection zone
is close to the adiabatic value the
formulation used  for calculating convective flux does not play much
of a role in those parts of the convection zone. 
Differences arise only in regions close to the solar surface where convection is
inefficient. If the
prescription of Canuto \& Mazzitelli (1991) is used instead of Mixing
length theory, the difference is confined to outer 5\% of solar radius
(Basu \& Antia 1994),
with maximum relative difference in sound speed of 6\% close to the solar
surface, though the difference is less than 1\% below 
$0.99R_\odot$ and less than 0.15\% below $r=0.95R_\odot$.
Similarly, the maximum relative difference in density can
exceed 10\% near the surface, but is less than 2\% below 
 $0.99R_\odot$ and less than 0.2\% below $r=0.95R_\odot$.
Both these differences fall off rapidly with depth.

Since helioseismology allows 
us to determine the position of the base of the convection zone as well
as the convection-zone helium abundance, Delahaye \& Pinsonneault (2006) have
presented a table listing the theoretical errors on the position of the
convection-zone base and helium abundance, and we refer
the reader to that as a convenient reference.
The effect of these uncertainties on neutrino fluxes has been discussed by
Bahcall \& Pe\~na-Garay (2004), and  Bahcall et al.~(2006).

In addition to all the sources of uncertainty discussed above, the solar heavy-element 
abundance $Z$ plays an important role in the structure of solar models, and that
is the subject of this review. This is probably the largest source of
uncertainty in current solar models.

There are many physical processes not included in `standard' solar models
since there are no standard formulations derived from first principles that
can be used to model the processes.
Models that include these effects rely on  simple formulations with free parameters  and hence
 are not regarded as standard solar models, which by definition
have no free parameters (since the mixing length parameter and the initial helium
abundance are fixed from known solar constraints).
Among the missing processes are effects
of rotation on structure and of  mixing induced by rotation. There are other 
proposed mechanisms for mixing in the radiative layers of the Sun,
such as mixing caused by waves generated at the convection-zone base (e.g., Kumar et al. 1999). These
processes also affect the structure of the model. For example, Turck-Chi\`eze et al.~(2004)
found that mixing below the convection-zone base can change both the 
position of the convection-zone base (gets shallower) and the
helium abundance (abundance increases). Accretion and mass-loss at some
stage of solar evolution can also affect the solar models.
Castro et al.~(2007) have investigated the effects of accretion.

Another non-standard input is the effect of magnetic fields. 
Magnetic fields are extremely important in the Sun, but standard solar models
do not include magnetic fields. Part of the reason, of course, is that we do
not know the configuration of magnetic fields inside the Sun. However, given that
we know that the solar cycle exists means that there are changes in the
Sun that take place on much shorter time scales than the evolutionary
time scale of the Sun. There are attempts to include convection and magnetic
fields in solar models (e.g., Li et al. 2006), however, the effort is extremely computationally 
intensive, and it is not likely that such models will become the norm anytime soon.

\section{Helioseismology}
\label{sec:helioseis}

As mentioned in the previous section one can put constraints on the inputs that go into
constructing the model by comparing the structure of standard solar models
with that of the Sun. Helioseismology gives us the means to do such a 
detailed comparison. In order to do so, oscillation frequencies of solar
models need to be calculated first. In this section, we describe the basic
equations used to describe solar oscillations and indicate how frequencies of solar
models can be calculated. We then
outline  how helioseismic techniques are used
to determine the structure of the Sun and compare solar models with the Sun.

\subsection{The Basic Equations}
\label{subsec:baseeq}

To a good approximation, solar oscillations can be described as linear
and adiabatic. Each solar-oscillation mode
has a velocity amplitude of the order of 10 cm/s at the surface, which is
very small compared to the sound speed at the surface as determined from solar models.
Except in the regions close to the solar surface, the oscillations are
very nearly adiabatic since the thermal time-scale is much larger
than the oscillation period. Although the adiabatic approximation breaks
down near the surface, non-adiabatic effects are generally ignored
because there are many other uncertainties associated with
treatment of these near-surface layers.
As explained later, the effect of these uncertainties can be
filtered out by other means. 
It is reasonably straightforward to calculate the frequencies of a
solar model.  The equations are written in spherical polar coordinates
and the variables are separated by writing the solution
in terms of spherical harmonics. The resulting set of equations
form an eigenvalue problem, with the frequencies being the eigenvalues.
Details of the equations,  how they are solved, and the properties of
the oscillation have been described by Cox (1980), Unno et al.~(1989),
Christensen-Dalsgaard \& Berthomieu (1991), Gough (1993), 
Christensen-Dalsgaard (2002) etc.  Here we give a short overview of the
basic equations.

The basic equations of fluid dynamics, i.e., the continuity equation,
the momentum equation and the energy equation (in the adiabatic approximation)
and the Poisson's equation to describe the gravitational field, can be
applied to the solar interior. These equations are:
\bea
{\partial\rho\over\partial t}+\nabla\cdot(\rho{\bf v})&=&0,\label{eq:cont}\\
\rho\left({\partial v\over\partial t}+{\bf v}\cdot\nabla {\bf v}\right)&=&
-\nabla P -\rho\nabla \Phi,\label{eq:mom}\\
{\partial P\over \partial t}+{\bf v}\cdot\nabla P&=&
c^2\left({\partial\rho\over \partial t}+{\bf v}\cdot\nabla\rho
\right), \label{eq:adia}\\
\nabla^2\Phi&=&4\pi G\rho\label{eq:pois},
\eea
where  ${\bf v}$ is the velocity of the fluid element, $c=\sqrt{\Gamma_1 P/\rho}$ is
the sound speed, $\Phi$ is the gravitational potential, and $G$ the
gravitational constant. 
The equations describing solar oscillations are obtained by a linear 
perturbation analysis of Eqs.~(\ref{eq:cont}--\ref{eq:pois}).
Since time does not appear explicitly in the equations, the time dependence
of the different perturbed quantities can be assumed to have an oscillatory form and
separated out.
 Thus we can write the perturbation to pressure as:
\be
P(r,\theta,\phi,t)=P_0(r)+P_1(r,\theta,\phi)\e^{-\ii\omega t},
\label{eq:pres}
\ee
where the subscript 0 denotes the equilibrium, spherically symmetric, quantity which by definition
does not depend on time,
and the subscript 1 denotes the perturbation. As is customary, we have used 
spherical polar coordinates centered at the solar
center with $r$ being the radial distance, $\theta$ the colatitude, and $\phi$
the longitude. Here, $\omega$ is the frequency of the oscillation.
Perturbations to other quantities, such as density, can be expressed in the
same form.
These are the Eulerian perturbations, which are evaluated at a specified
point. 
Velocity is given by as ${\bf v}={\partial{\vec\xi}/ \partial t}$, 
where ${\vec\xi}$ is the
displacement from equilibrium position.
Substituting the perturbed quantities in the basic equations
(\ref{eq:cont}--\ref{eq:pois}), and keeping only  linear terms
in the perturbations,
we get:
\bea
\rho_1+\nabla\cdot(\rho_0{\vec \xi})&=&0,\label{eq:cont1}\\
 -\omega^2\rho{\vec\xi}&=&-\nabla P_1-\rho_0 \nabla\Phi_1-\rho_1\nabla\Phi_0\>,
\label{eq:mom1}\\
P_1+{\vec\xi}\cdot\nabla P_0&=&c^2_0\left(\rho_1+
{\vec\xi}\cdot\nabla\rho_0\right),\label{eq:adia1}\\
\nabla^2\Phi_1&=&4\pi G\rho_1.\label{eq:pois1}
\eea
Eliminating $P_1$ and $\rho_1$, and expressing the gravitational potential as
an integral, we can combine Eqs.~(\ref{eq:cont1} -- \ref{eq:pois1}) to  get one
equation to describe linear, adiabatic oscillations:
\bea
-\omega^2\rho\vec{\xi}=\rho{\mathcal L}\vec{\xi}
&=&\nabla(c^2\rho\nabla\cdot\vec{\xi}+\nabla P\cdot\vec{\xi})
-\vec g\nabla\cdot(\rho\vec{\xi}) \nonumber \\
& & \qquad - G\rho\nabla\left(
{\int_V{\nabla\cdot(\rho\vec{\xi})\;\d V\over |\vec r -\vec{r'}|}}
\right).
\label{eq:herm}
\eea
Here for convenience we have dropped the subscript $0$ from the equilibrium quantities
since the perturbations, $\rho_1$, $P_1$ etc., do not occur in this equation.
This equation, supplemented by appropriate boundary conditions at the center
and the solar surface, defines an eigenvalue problem for the operator
$\mathcal L$, with frequency $\omega$ as the eigenvalue.
The different
modes are uncoupled in the linear approximation,  and hence, the equations can be 
solved for each mode separately.
Furthermore, it can be shown that the radial and angular dependences can be separated by expressing
the perturbations in terms of spherical harmonics. Thus the perturbation
to any 
 scalar quantity can be written as:
\be
P(r,\theta,\phi,t)=P_0(r)+P_1(r)Y_\ell^m(\theta,\phi)\e^{-\ii\omega t},
\label{eq:p1}
\ee
The displacement vector  can be expressed as:
\be
{\vec\xi}=\left(\xi_r(r)Y_\ell^m(\theta,\phi),
\xi_h(r){\partial Y_\ell^m\over\partial\theta},
{\xi_h(r)\over \sin\theta}{\partial Y_\ell^m\over\partial\phi}\right)
\e^{-\ii\omega t},
\label{eq:xi}
\ee
where $\xi_r$ and $\xi_h$ are respectively the radial and horizontal components of
the displacement. A detailed discussion of properties of the different
oscillation modes is given by Unno et al.~(1989), Gough (1993) and \jcd~(2002).

The different modes of solar oscillations are described by three numbers that
characterize the perturbations that  define the effect
of the mode. These are (1) the radial order $n$ which is related to the number
of nodes in the radial direction, (2) the degree $\ell$ which is related to
the horizontal wavelength of the mode and is approximately
the number of nodes on the solar surface, and (3) the azimuthal order $m$
which defines the number of nodes along the equator.
The radial order, $n$, can have any integral value.
Positive values of $n$ are used to denote acoustic modes, i.e.,
the so-called p-modes (p for pressure, since the dominant restoring
force for these modes is provided by the pressure gradient). Negative values of
$n$ are used to denote modes for which buoyancy provides the main
restoring force. These are usually referred to as g-modes (g for gravity).
Modes with $n=0$ are the so-called fundamental or f-modes. For large $\ell$, f-modes are  essentially
surface gravity modes whose frequencies are largely independent of the stratification
of the solar interior. As a result, f-mode frequencies are normally not used for
determining the structure of the Sun, however, these  have been used to
draw inferences about the solar radius (e.g., Schou et al.~1997; Antia 1998;
Lefebvre et al.~2007).
Only p- and f-modes have been reliably detected in the Sun and we
shall confine our discussion to these modes.
The degree $\ell$ and the azimuthal order $m$ describe the angular dependence
of the mode as determined by $Y_\ell^m(\theta,\phi)$.
The degree $\ell$ is either 0 (the radial mode)
or positive (non-radial modes). The azimuthal 
order   $m$ can have $2\ell+1$ values with  $-\ell\le m \le \ell$.
While $\ell$ and $m$ can be determined by spherical harmonic transform
of Doppler or intensity images of the solar surface, $n$ can only be determined
from the power spectrum of the spherical harmonic transforms by counting
the ridges in the power spectra. The
positions of the peaks in the power spectrum, when compared with asymptotic expressions
or frequencies of solar models, can also give an estimate of $n$.
The frequencies of solar oscillations are usually expressed as the cyclic
frequency $\nu=\omega/(2\pi)$.

If the Sun were spherically symmetric, the frequencies
would be independent of $m$. Rotation and magnetic
fields lift the $(2\ell+1)$-fold degeneracy of modes with the same $n$ and $\ell$,
giving rise to the so-called frequency splittings. The frequencies $\nu_{n\ell m}$ of the modes within a
multiplet are usually  expressed in terms of the splitting coefficients
\be
\nu_{n\ell m}=\nu_{n\ell}+\sum_{j=1}^{J_{\rm max}} a_j^{n\ell}
{\mathcal P}_j^{\ell}(m).
\label{eq:aj}
\ee
Here, $\nu_{n\ell}$ is the mean frequency of a given $(n,\ell)$ multiplet and
$a_j^{n\ell}$ are the  `splitting coefficients', often referred
to as `$a$-coefficients'.
In this expression,
${\mathcal P}_j^{\ell}(m)$ are orthogonal polynomials in $m$ of degree $j$.
The orthogonality of these polynomials  is defined over the discrete values of $m$.
In the expansion, $J_{\rm max}$ is
generally much less than $2\ell$. Although this expansion
reduces  the number of data points that are available for use, the splitting coefficients
can be determined to much higher precision than the individual frequencies
$\nu_{n\ell m}$.
Unfortunately,
 different workers have used
different normalizations of ${\mathcal P}_j^{\ell}(m)$ (e.g.,
Ritzwoller \& Lavely 1991; Schou et al.~1994) and 
there is no unique definition of splitting coefficients either. 
Early investigators (e.g., Duvall et al.~1986) commonly used Legendre
polynomials, whereas now it is common to use the
Ritzwoller-Lavely formulation (Ritzwoller \& Lavely 1991)
where the basis functions are orthogonal polynomials related
to Clebsch-Gordan coefficients.
The readers are referred to Pijpers (1997) for details on how the different
polynomials and splitting coefficients are related.

The mean frequency, $\nu_{n\ell}$, in Eq.~(\ref{eq:aj})  is determined
by the spherically symmetric structure of the Sun, and hence can be used
to determine solar structure.
The odd-order  coefficients $a_1, a_3, \ldots$ depend principally on the
rotation rate (Durney et al.~1988; Ritzwoller \& Lavely 1991)
and reflect the advective, latitudinally symmetric part
of the perturbations caused by rotation.  Hence, these
are used to determine the rotation rate inside the Sun.
The even order $a$ coefficients on the
other hand, result from a number of different causes, such as
magnetic fields (Gough \& Thompson 1990; Dziembowski \& Goode 1991), 
asphericities in solar structure (Gough \& Thompson 1990),
and the second order effects of rotation 
(Gough \& Thompson 1990; Dziembowski \& Goode 1992).
In this review we shall only concentrate on
the spherically symmetric part of the solar structure and hence, only the mean frequencies
$\nu_{n\ell}$ are relevant.

As mentioned earlier, most of the observed oscillations are acoustic or p-modes, 
which are
essentially sound waves. These modes are stochastically excited by 
convection in the very shallow layers of the Sun (Goldreich \& Keeley 1977;
Goldreich et al.~1994).
As these waves travel 
inwards from the solar
surface, they pass through regions of increasing sound speed, a result of 
increasing temperatures. This causes  the waves to be
refracted away from the vertical direction until they
undergo a total internal reflection. The depth at which this
happens is called the `lower turning point' of the mode, and is approximately
the position at which $\omega^2=
{\ell(\ell+1)c^2(r) / r^2}$, where  
$c(r)$ is the sound speed at radial
distance $r$ from the center. Thus, low-degree modes
penetrate to the deep interior, but high degree modes are trapped in
the near-surface regions.
Each solar-oscillation mode is trapped in a different region inside the
Sun and the frequency of the mode is determined by the structure variables, like
sound speed and density, in the trapping region.
By considering different modes it is possible to determine the solar
structure in the region that is covered by the observed set of modes,
which includes almost the entire Sun. 
This is done by solving the inverse problem (Gough \& Thompson 1991; \jcd\ 2002).

\begin{figure}[t]
\begin{center}
\includegraphics[width=.8\textwidth]{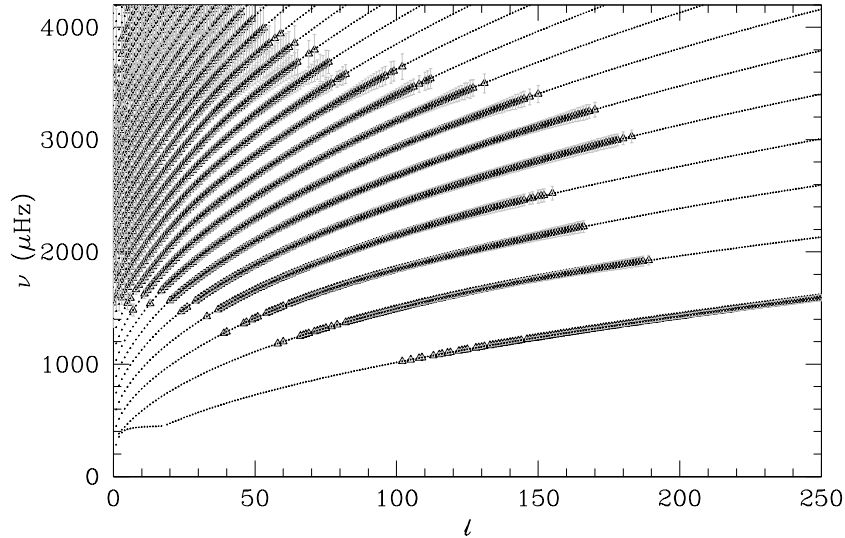}
\end{center}
\caption[]{The frequencies of a solar model plotted  as a function of degree $\ell$ 
are shown by dots which have merged into a lines which can be identified
with ridges in power spectrum.
The triangles with error bars are the observed frequencies obtained from the
first 360 days of observations by the MDI instrument. The error bars
represent $5000\sigma$ errors. The lowermost ridge
corresponds to f-modes.}
\label{fig:lnu}
\smallskip
\end{figure}

There are a number of projects that provide readily
available solar oscillation frequencies, the 
chief among these are the ground-based Global Oscillation Network Group (GONG) 
project (Hill et al.~1996)\footnote{GONG frequencies can be downloaded from
http://gong.nso.edu/data/}, and the Michelson Doppler Imager (MDI) instrument on board the Solar
and Heliospheric Observatory (SOHO) (Scherrer et al.~1995)\footnote{MDI frequencies
are available at http://quake.stanford.edu/$\tilde{\; }$schou/anavw72z/}. These projects
determine frequencies of both low- and intermediate-degree modes,
and typical sets include 3000 modes for different values of $n$ and $\ell$, with
$\ell$ up to $\approx 200$ and $\nu_{n\ell}$ in the range 1 to 4.5 mHz. 
Most of the frequencies are determined to a precision of about 1 part in $10^5$
which allows one to make stringent tests of solar models.
The most precise sets of low degree data ($\ell=0$, 1, 2 and 3), i.e.,
data that probe the solar core, are
however, obtained from unresolved observations of the solar disc such as those made by
 the Birmingham Solar Oscillation Network
(BiSON;  Elsworth et al.~1991; Chaplin et al.~2007a) or the  Global Oscillations at Low Frequencies (GOLF) instrument (Gabriel et
al.~1995a)  on board SOHO.
Figure~\ref{fig:lnu} shows the frequencies of one set of solar oscillation data
plotted  as a function
of the degree $\ell$ for different  radial orders $n$. 
This figure 
also plots the frequencies of a solar model. On the scale of the figure
 there is very good agreement between the two, which essentially
confirms the mode identification and also gives us confidence in the 
model.

\subsection{Determining solar structure from seismic data}
\label{subsec:inv}

To determine solar structure from solar oscillation frequencies one
begins with the  equation for solar oscillations as given by  Eq.~(\ref{eq:herm}).
In Eq.~(\ref{eq:herm}), $\omega$ is the observed quantity and we would like to
determine sound speed $c$ and density  $\rho$ (and hence pressure
 $P$ assuming hydrostatic equilibrium).
However, the displacement eigenfunction  $\vec{\xi}_{n,\ell}$ can only be
observed at the solar surface and
therefore the equations cannot
be inverted directly. The way out of this is to recognize  that
Eq.~(\ref{eq:herm}) defines an eigenvalue problem of the form
\be
{\mathcal L}\vec{\xi}_{n,\ell}=-\omega^2_{n,\ell}\vec{\xi}_{n,\ell}\>,
\label{eq:herl}
\ee
${\mathcal L}$ being a differential operator in Eq.~(\ref{eq:herm}).
Under specific boundary conditions, namely
$\rho=P=0$ at the outer boundary, 
the eigenvalue problem defined by Eq.~(\ref{eq:herm}) is Hermitian
(Chandrasekhar 1964) and hence,
the variational principle can be used to linearize Eq.~(\ref{eq:herm})
around a
known solar model (called the ``reference model'') to obtain
\be
{\delta\omega^2_{n,\ell}}=-{\int_V\rho\vec{\xi}_{n,\ell}^\star\cdot\delta
{\mathcal L}
\vec{\xi}_{n,\ell}\;\d V
\over \int_V \rho\vec{\xi}_{n,\ell}^\star\cdot\vec{\xi}_{n,\ell} \;\d V},
\label{eq:moddif}
\ee
where $\delta\omega_{n,\ell}^2$ is the difference in the squared frequency of an oscillation  mode 
of  the reference model 
 and the Sun, $\delta {\mathcal L}$ 
is the perturbation to the operator $\mathcal L$ (defined by Eq.~\ref{eq:herm}) as a result of 
the differences between the reference model and the Sun,
and $\vec{\xi}_{n,\ell}$ is the displacement eigenfunction for the
known solar model (and thus can be calculated). One such equation
can be written for each mode of oscillation characterized by $(n, \ell)$,  
and the set of equations
can be used to calculate the difference in structure between
the solar model and the Sun, and thus determine the structure of the Sun.
The denominator on the right hand side of Eq.~(\ref{eq:moddif}) is
usually denoted as $I_{n,\ell}$, and 
is often called the mode inertia  since it can be shown that the
time-averaged kinetic energy of a mode
is proportional to $\omega^2_{n,\ell} I_{n,\ell}$. 

Equation~(\ref{eq:moddif}) implies that for a given difference in structure,
the resulting differences in
frequencies of modes with a high inertia are  less than those
of modes with lower inertia. For modes of a given frequency, lower
degree (i.e., deeply penetrating) modes have higher mode inertias
than higher degree (i.e., shallow) modes. 
The mode inertia therefore,  is a convenient weighing factor to
quantify the effect of any perturbation on the frequency of a 
mode. 
Since computed  eigenfunctions of a solar model are arbitrary to a
constant multiple, it is customary to use  $I_{n,\ell}$ 
normalized by the value at the surface, sometimes
referred to as $E_{n,\ell}$.
When  modes are represented as 
spherical harmonics, $E_{n,\ell}$ can be shown to be
\be
E_{n\ell}={4\pi\int_0^R\left[|\xi_r(r)|^2+\ell(\ell+1)|\xi_h(r)|^2\right]
\rho_0r^2\;\d r \over
M\left[|\xi_r(R_\odot)|^2+\ell(\ell+1)|\xi_h(R_\odot)|^2\right]}\>,
\label{eq:enl}
\ee
where $\xi_r$ and $\xi_h$ are the radial and horizontal components
of the displacement eigenfunction, $M$ the total  mass and
$\rho_0(r)$ the density profile of the model. It is convenient to
define another measure of inertia, denoted by $Q_{n\ell}$, which is defined
as
\be
Q_{n\ell}={E_{n\ell}\over E_0(\nu_{n\ell})}\>,
\label{eq:qnl}
\ee
where $E_0(\nu)$ is  $E_{n\ell}$ for $\ell=0$ modes interpolated
to the frequency $\nu$. Thus  $Q_{n\ell}=1$ for $\ell=0$ modes and
less than one for modes with higher $\ell$.
 Multiplying frequency differences
with $Q_{n\ell}$ is equivalent to inversely scaling the frequency differences
with mode inertia.

Linearizing Eq.~(\ref{eq:herm}) around a known solar model
by applying the variational principle results in an
equation that relate the frequency differences between the Sun and the
solar model to the differences in structure between the Sun and the model:
\be
{\delta \nu_{n\ell}\over \nu_{n\ell}}=\int_0^R \mathcal{K}_{c^2,\rho}^{n\ell}(r)
{\delta c^2\over c^2}(r)\;\d r+
\int_0^R \mathcal{K}_{\rho, c^2}^{n\ell}(r) {\delta \rho\over \rho}(r)\;\d r\;,
\label{eq:inva}
\ee
where, $\delta c^2/c^2$ and $\delta\rho/\rho$ are the relative
differences in the squared sound speed and density between the Sun and the
model. The functions  
$\mathcal{K}_{c^2,\rho}^{n\ell}(r)$ and  $\mathcal{K}_{\rho, c^2}^{n\ell}(r)$
are the kernels of the inversion 
 that relate the changes in frequency to the changes in $c^2$ and $\rho$
respectively. These are known functions of the reference solar model.

\begin{figure}[t]
\begin{center}
\includegraphics[width=.85\textwidth]{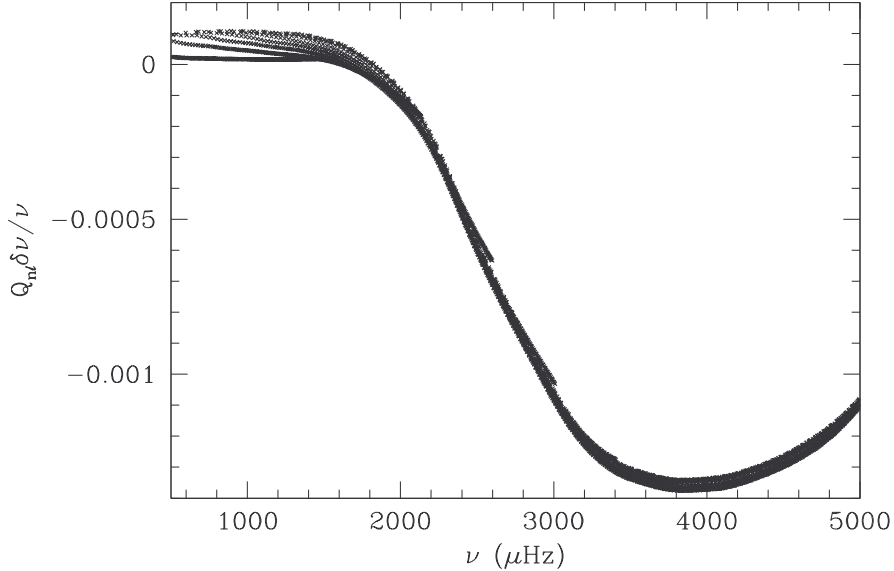}
\end{center}
\caption[]{The scaled, relative differences between the  frequencies of two solar models
constructed using different formulations of  convective
flux. Other input physics is the same for both models.
One of the  models was constructed using the
mixing length theory and the other using  the Canuto \& Mazzitelli (1991) formulation
 for calculating convective flux.}
\label{fig:mltcm}
\smallskip
\end{figure}

Eq.~(\ref{eq:inva}), unfortunately, is not enough to represent the differences between the
Sun and the models. There is an additional complication that arises due to uncertainties
in modelling layers just below the solar surface.
Eq.~(\ref{eq:moddif}) implies that we can invert the
solar frequencies provided we know how to model the Sun properly, and if the frequencies
can be described by the  equations for adiabatic oscillations.
Neither of these two assumptions is completely correct.
For example, our treatment of convection in surface layers is known to be
approximate.
Simulations of convection seem to indicate that there are significant
departures between the temperature gradients obtained from the mixing length approximation
and that from a full treatment of convection. The deviations mainly occur
 close to the solar surface (see e.g., Robinson et al.~2003 and
references therein).
This results in differences in the density and pressure profiles. Not using
a full treatment of convection also means that several physical
effects, such as turbulent pressure and turbulent kinetic energy, are missing
from the models. Another source of error is the fact that the adiabatic
approximation that is used to calculate the frequencies of the models
breaks down near the surface where the thermal time-scale is comparable
to the period of oscillations.
This implies that the RHS of Eq.~(\ref{eq:inva})
does not fully account for the frequency difference $\delta\nu/\nu$ between
the Sun and the model.
Fortunately, all these uncertainties are localized in a thin layer
near the solar surface.
For modes which are not of very high degree ($\ell \simeq 200$ and lower),
the structure of the wavefront near the surface is almost independent of the
degree, the wave-vector being almost completely radial. This implies
that any additional difference  in frequency due to errors in the surface structure
has to be a function of frequency alone once  mode inertia has been
taken into account.  It can also be shown (e.g., Gough 1990)
that surface perturbations cause the difference in frequency to be a slowly varying
function of frequency which can be modeled as a sum of low
degree polynomials. This effect is shown in Fig.~\ref{fig:mltcm}
which shows the  frequency differences, scaled inversely by their
mode inertia,  between two models which differ
only near the surface due to differences in their convection formalisms.
It can be seen that all points tend to fall on a curve  which is a 
a function of frequency.
Thus Eq.~(\ref{eq:inva}) is modified  to represent
the difference between the model and the Sun and is re-written as
\be
{\delta \nu_{n\ell}\over \nu_{n\ell}}=\int_0^R \mathcal{K}_{c^2,\rho}^{n\ell}(r)
{\delta c^2\over c^2}(r)\;\d r+
\int_0^R \mathcal{K}_{\rho, c^2}^{n\ell}(r) {\delta \rho\over \rho}(r)\;\d r+
{F(\nu_{n\ell})\over E_{n\ell}}\>,
\label{eq:inv}
\ee
where $F(\nu_{n\ell})$ is a slowly varying function of frequency that
arises due to the errors in modelling the near-surface regions 
(Dziembowski et al.~1990; Antia \& Basu 1994a). In addition to satisfying Eq.~(\ref{eq:inv}),
the differences in density should integrate to zero since otherwise
the total solar mass will be modified. This is ensured by putting an additional
constraint on $\delta\rho$.

Instead of sound speed and density we can write Eq.~(\ref{eq:inv}) in terms of other
pairs of independent structure variables, such as the adiabatic index $\Gamma_1$ and density,
or $\Gamma_1$ and $P/\rho$.
It can be shown
that once sound speed and density are known, other structure variables
that are required for the adiabatic oscillations equations can be
calculated. For example, pressure can be calculated from the equation
of hydrostatic equilibrium. The equation of state is not required
for this purpose, since the only thermodynamic index that occurs in the oscillation 
equations is the adiabatic index, 
$\Gamma_1=(\partial \ln P/\partial \ln\rho)_S=c^2\rho/P$, which can be
directly calculated once $\rho$, $P$ and $c$ are known.
However, other thermodynamic quantities like temperature, composition etc., do not occur
in the equations of adiabatic oscillations, and hence, these  cannot
be obtained directly through inversions.
In order to estimate
temperature and composition one has to assume that 
input physics such as the equation of
state, opacities and nuclear energy generation rate are known exactly. Additionally,
the equations of thermal equilibrium are needed to relate temperature to the
other quantities (see e.g., Gough \& Kosovichev 1990; Shibahashi \& Takata 1996;
Antia \& Chitre 1998).

It turns out that most of the difference in frequencies between the Sun and
modern solar models are a result of  uncertainties in the near-surface layers. This makes 
it  difficult to test the internal structure of these models by directly
comparing the frequencies of the models with those of the Sun.
The frequency differences caused by the near-surface errors  are often larger
than the frequency differences caused by differences in the structure of the
inner regions, making the comparison ineffective. As a result, one resorts
to inverting the frequency differences in order to determine the differences
between solar models and the Sun as a function of radius.
A number of techniques have been
developed for solving the inverse problem (Gough \& Thompson 1991;
\jcd\ 2002 and references therein). 
Inverse problems are generally ill-conditioned since it is not possible
to infer a  function like sound speed in the solar interior
using only a finite and discrete set of observed modes. Thus additional assumptions have to
be made (e.g., solar sound-speed or density are positive quantities
and that the profiles are not discontinuous and do not vary sharply) in order to determine the structure
of the Sun. There are two classes of methods to determine $\delta c^2/c^2$ and 
$\delta\rho/\rho$ from Eq.~(\ref{eq:inv}): the  regularized least squares (RLS) method,
and the method of optimally localized averages (OLA).

In the RLS technique the unknown functions $\delta c^2/c^2$,
$\delta\rho/\rho$ and $F(\nu)$ are expanded in terms of a suitable
set of basis functions and the coefficients of expansion are determined
by fitting the given data. Noise in the data can cause
the solution to be highly oscillatory unless the result is smoothed by
applying regularization  (e.g., Craig \& Brown 1986).
Regularization is usually applied by ensuring that either the first or the second derivative
of the solution is also small. It is common to assume that the second derivative
is small, and this is achieved by minimizing the function
\be
\chi^2=\sum_{n,\ell}\left(d_{n,\ell}\over\sigma_{n,\ell}\right)^2+
\lambda_{c^2}\int_0^R\left(\d^2(\delta c^2/c^2)\over \d r^2\right)^2\d r +
\lambda_{\rho}\int_0^R\left(\d^2(\delta \rho/\rho)\over \d r^2\right)^2\d r\>,
\;\label{eq:rls}
\ee
where $d_{n,\ell}$ is the difference between LHS and RHS in Eq.~(\ref{eq:inv})
and $\sigma_{n,\ell}$ is the estimated error in the observed relative frequency
difference.
The smoothness is controlled by the regularization parameters $\lambda_{c^2}$ and
$\lambda_\rho$ which are adjusted to get the  required smoothness of the solutions
for $\delta c^2/c^2$ and $\delta\rho/\rho$. If these parameters are zero,
then the solution reduces to normal least squares approximation and the solution
is highly oscillatory, while for
very large values of these parameters, the solution will approach a
linear function in $r$.  Details
of how the regularization parameters can be determined in general has been described by Hansen (1992).
A description of how these parameters can be chosen for helioseismic inversions
is given in  Antia \& Basu (1994a) and Basu \& Thompson (1996).

\begin{figure} 
\begin{center}
\includegraphics[width=.99\textwidth]{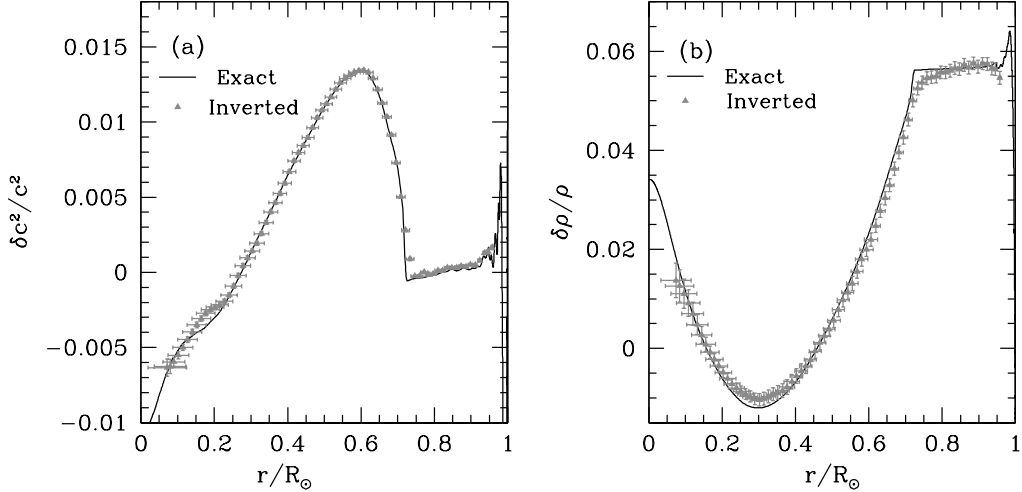}
\caption{The exact and inverted relative sound-speed (panel a) and
density (panel b) differences between two models. The mode set corresponding to
the first 360 days of MDI data (Schou et al.~1998a) was used for the inversion. Random errors
were added to the frequencies of the test model prior to the inversions. The vertical error bars show the
formal $1\sigma$ errors on the inversion results caused by errors in the MDI data. The
horizontal error bars are a measure of the width of the averaging kernels and hence, of the resolution of the inversions.
}
\smallskip
\label{fig:inv}
\end{center}
\smallskip
\end{figure}

In the OLA 
technique (Backus \& Gilbert 1968), a linear combination of the
kernels is obtained such that the combination is localized in space. The linear
combination is called the resolution kernel or the averaging kernel. The solution
obtained is then an average of the true solution weighted by the
averaging kernel. Very often a variant of the OLA  method
called ``Subtractive'' Optimally
Localized Averages  (SOLA; Pijpers \& Thompson 1992, 1994) 
is used. In either case, if $\mathcal K_{\rm av}$ is an
averaging kernel for an inversion, then
\be
<f>=\int {\mathcal K_{\rm av}} f(r)\;\d r
\label{eq:ave}
\ee
represents the average of the quantity $f$ over a sufficiently narrow
range in $r$.
Thus if
\be
{\mathcal K_{\rm av}}=\sum_{n,\ell} c_{n,\ell} K_{c^2,\rho}^{n\ell}\>,\quad\hbox{then,}\quad
\left<{\delta c^2\over c^2}\right > = \sum_{n,\ell} c_{n,\ell} {\delta\nu_{n\ell}\over
\nu_{n\ell}}\>.
\label{eq:csq}
\ee
From Eq.~(\ref{eq:inv}) we can see that this is
possible only if  $\int {\mathcal K_{\rm av}} \>\d r=1$,
and if ${\mathcal C}=\sum_{n,\ell} c_{n,\ell} K_{\rho,c^2}^{n\ell}$ and
${\mathcal F}=\sum_{n,\ell} c_{n,\ell} F(\nu_{n\ell})$ are small.
Details of how this is done for
solar structure inversions has been described by Rabello-Soares et al.~(1999).
Fig.~\ref{fig:inv} shows the result of inverting the frequency differences between
two models. As can be seen from the figure, we are able to invert for the differences in structure
very well.

It is difficult to determine precisely the structure of the solar core
through inversions   because of the dearth of
low-degree modes. The inversion results often have large errors and
poor resolution. As a result, it is also customary to compare the
cores of solar models and that of the Sun using combinations of
frequencies that are sensitive to the structure of the
core (e.g., Elsworth et al.~1990; Chaplin et al.~1997).
For example, the so-called  small frequency spacings of low
degree modes (Tassoul 1980; Gough 1986) is sensitive to the
sound-speed gradient in the core. The small frequency spacing (or separation)  is
given by
\be
\delta\nu_{n,\ell}=\nu_{n,\ell}-\nu_{n-1,\ell+2}\approx -(4\ell+6)
{\Delta\nu_{n,\ell}\over 4\pi^2\nu_{n,\ell}}\int_0^R{\d c\over\d r}\;{\d r\over c}\;,
\label{eq:nudif}
\ee
where $\Delta\nu_{n,\ell}=\nu_{n+1,\ell}-\nu_{n,\ell}$ is the large
frequency spacing. The approximation given in Eq.~(\ref{eq:nudif})  is
obtained from an asymptotic analysis of solar oscillation frequencies and
is valid for $n\gg\ell$.
The large frequency spacing is determined by the
sound travel time from center to surface and hence has large contribution
from near surface layers where sound speed is low. As a result, the
large frequency spacing for models is affected by our inability to
model the near-surface layers correctly. 
One way of reducing the effects of the near-surface errors is to use the 
so-called frequency separation ratios. The frequency separation ratios 
(Roxburgh \& Vorontsov 2003; Ot\'i Floranes et al.~2005; Roxburgh 2005) formed from the small 
frequency and large frequency spacings of the modes are given by
\be
 r_{0,2}(n) = \frac{\delta\nu_{n,0}}{\Delta\nu_{n,1}}\>,
 ~~~~~~~~r_{1,3}(n) = \frac{\delta\nu_{n,1}}{\Delta\nu_{n+1,0}}\>.
 \label{eq:rats}
\ee

\subsection{Determining the solar helium abundance}
\label{subsec:helmethod}

\begin{figure}[t]
\begin{center}
\includegraphics[width=.65\textwidth]{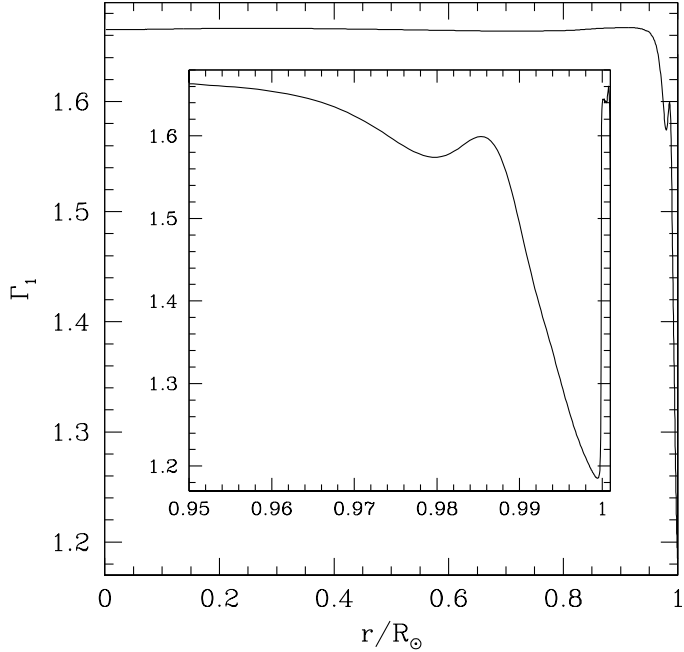}
\end{center}
\caption[]{The adiabatic index, $\Gamma_1$ in a solar model plotted as
a function of radial distance. The inset shows a blow-up of the outer
regions.}
\label{fig:gam}
\smallskip
\end{figure}

If one assumes that the equation of state of solar material is known well and that it is the
same as that used in solar models, Eq.~(\ref{eq:inv}) can be modified to determine
the helium abundance 
in the solar convection zone. Since helium
does not form lines at photospheric temperatures, it is difficult to
determine its abundance spectroscopically. Helioseismic techniques 
are the most accurate way to determine the solar helium abundance.
Helium abundance determinations are usually done through the adiabatic
index $\Gamma_1$. Although $\Gamma_1$ is generally close to the ideal-gas value of
5/3 in most of the solar interior, it deviates significantly from this
value in regions where elements undergo ionization. The adiabatic index
is smaller than 5/3 in the ionization zones, and the extent of reduction depends on the
abundance of the element undergoing ionization as well as the equation of state.
Thus $\Gamma_1$ is sensitive to the helium abundance in the layers
where helium ionizes, and this is exploited in several different ways to determine
$Y$ in those regions. Since the helium ionization zone lies in solar
convection zone, and since the convection zone is well mixed, the helium abundance
determined from $\Gamma_1$ is the abundance throughout the solar convection zone.
Fig.~\ref{fig:gam} shows $\Gamma_1$ in a solar
model as a function of radius. The dip near $r=0.98R_\odot$ is due
to the He\,{\small II} ionization zone, while the bigger dip closer to  the surface is due
to the ionization of H\,{\small I} and He\,{\small I}.  

D\"appen et al.~(1991) and Dziembowski et al.~(1990, 1991) showed that
Eq.~(\ref{eq:inv}) written in terms of $\Gamma_1$ and $\rho$ differences can be
modified to determine the helium abundance in the Sun because
\be
\!\!\!\frac{\delta \Gamma_1}{\Gamma_1} =
\left( \frac{\partial \ln \Gamma_1}{\partial \ln P} \right)_{Y,\rho}
\frac{\delta P}{P} +
\left( \frac{\partial \ln \Gamma_1}{\partial \ln \rho} \right)_{Y,P}
\frac{\delta \rho}{\rho} +
\left( \frac{\partial \ln \Gamma_1}{\partial Y} \right)_{P,\rho} \delta Y,
\label{eq:dy}
\ee
where the partial derivatives can be determined from the equation of state. This
equation ignores contributions of differences in the equation of state and  heavy element abundances
 to $\delta\Gamma_1$ between the models and the Sun.
Eq.~(\ref{eq:dy}) was then used by Dziembowski et al.~(1991) to rewrite Eq.~(\ref{eq:inv})
as
\be
{\delta \nu_{n\ell}\over \nu_{n\ell}}=\int_0^R \mathcal{K}_{u,Y}^{n\ell}(r)
{\delta u\over u}(r)\;\d r+
\int_0^R \mathcal{K}_{Y, u}^{n\ell}(r) \delta Y\;\d r+
{F(\nu_{n\ell})\over E_{n\ell}}\>,
\label{eq:invy}
\ee
where $u\equiv P/\rho$. They further assumed that since $\delta Y$ is constant in the
convection zone, and if one were only using modes trapped in the convection zone, $\delta Y$
could be brought out of the integral sign. This method has been used quite
extensively to determine solar helium abundance (e.g., Dziembowski et al.~1991;
Kosovichev et al.~1992), but suffers from extreme sensitivity 
to the equation of state of the reference model. 

\begin{figure}[t]
\begin{center}
\includegraphics[width=.65\textwidth]{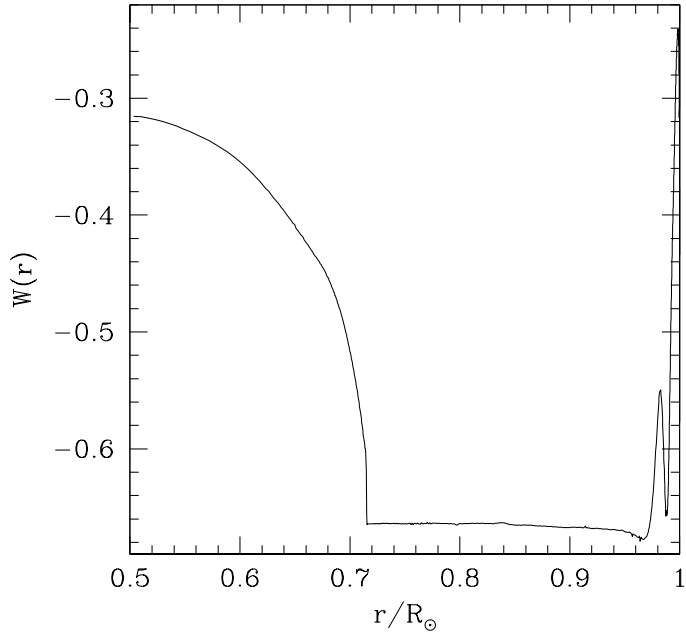}
\end{center}
\caption[]{The dimensionless gradient of sound speed, $W(r)$, of a solar model plotted as
a function of radius.}
\label{fig:wr}
\smallskip
\end{figure}

There are other methods to
determine the solar helium abundance that appear to be somewhat less sensitive to
the equation of state. This is done by calibrating the dip
in $\Gamma_1$ in the helium abundance zone (see Fig.~\ref{fig:gam}).
The reduction in $\Gamma_1$ in the ionization zones  also affects
the sound speed, but since  sound speed increases very rapidly with
depth, the variation may not be detectable in the sound-speed profile itself.
Gough (1984) suggested that this variation can be found  in the
dimensionless gradient of the sound speed, $W(r)$, where
\be
W(r)={1\over g}{\d c^2\over \d r}\;,
\label{eq:wr}
\ee
and $g$ is the acceleration due to gravity. In the lower part of the
convection zone, $\Gamma_1\approx 5/3$ and $W(r)\approx -2/3$. Figure~\ref{fig:wr}
shows $W(r)$ for a solar model.
The peak around $r=0.98R_\odot$ in this figure is due
to the He\,{\small II} ionization zone. This peak can be calibrated to find the
helium abundance (Gough 1984; D\"appen et al.~1991; Antia \& Basu 1994b). 
The prominent peak in $W(r)$
near the surface that can be seen from Fig.~\ref{fig:wr}, is due to the hydrogen and He\,{\small I}
ionization zones. Since this peak occurs in a region where inversions are
not very reliable, it is difficult to use this for any diagnostic purpose.
The solar helium abundance can also be determined by directly calibrating the
sound speed difference between the Sun and solar models with known $Y$ (Antia \&
Basu 1994b) around the He\,{\small II} ionization zone.

The solar helium abundance can also be determined by examining
the  surface term $F(\nu)$. The helium ionization zone is
relatively deep, but it is close enough to the surface to leave its signature
on that term. Although, $F(\nu)$ is a `smooth' function of frequency mainly
determined by surface effects, it is also influenced by discontinuities or
sharp variations in the sound speed profile in deeper regions. In general, these
discontinuities introduce a small oscillatory term in frequencies as a
function of $n$ and the scale of the oscillation is determined by the
acoustic depth (i.e., the sound-travel time) of the perturbation (Gough 1990), 
and its amplitude depends on the extent of variation around the discontinuity.
The oscillatory signature in the frequencies caused by the helium ionization
zone translates into a similar oscillatory signature in $F(\nu)$.
This signature of the 
helium ionization zone has been used in different ways by P\'erez Hern\'andez \& 
Christensen-Dalsgaard (1994), Antia \& Basu (1994b) etc., to determine the
solar helium abundance.

\subsection{Determining the depth of the solar convection zone}
\label{subsec:rb}

\begin{figure}[t]
\begin{center}
\includegraphics[width=.75\textwidth]{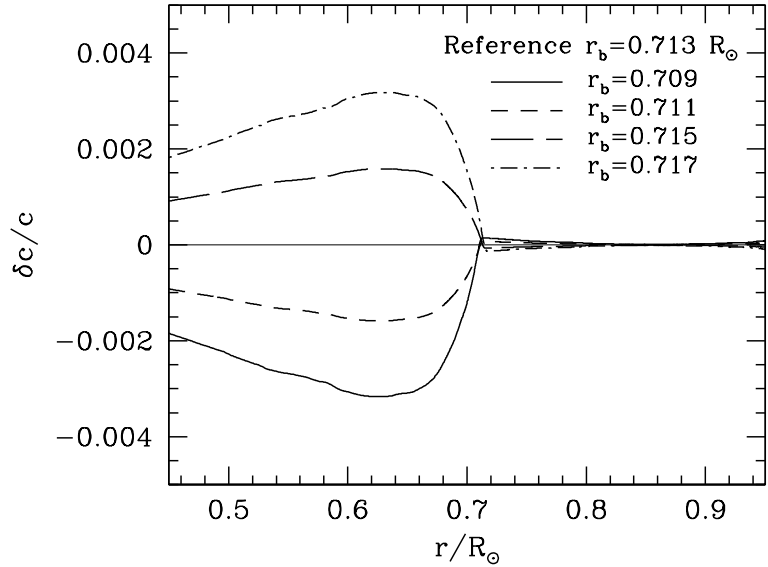}
\end{center}
\caption[]{The relative sound-speed difference between a model with a convection-zone base
at $r_b=0.713R_\odot$ and models with convection zone bases at different positions.}
\label{fig:cz}
\smallskip
\end{figure}

Solar oscillation frequencies can be used to determine $r_b$,  the position of the
convection-zone base. The function  $W(r)$ (Eq.~\ref{eq:wr}) obtained by inverting
solar oscillation frequencies  can be used
for this purpose.
At the base of the convection zone, the
temperature gradient switches from the adiabatic value inside the convection
zone to the radiative value below the convection zone. This introduces a
discontinuity in the second derivative of the temperature and hence, the
sound speed. This results in  a discontinuity in the gradient of $W(r)$, which
is clearly visible in Fig.~\ref{fig:wr}. The position of this discontinuity
can be determined to estimate the convection zone depth (\jcd\ et al.~1991).
The frequencies of solar oscillations are very
sensitive to this depth and hence, $r_b$ can be determined very accurately
from the observed  solar oscillations frequencies.  However, there is a 
more precise method to determine the base of the convection zone, and that is
by looking at the sound-speed difference between the Sun and models with
different position of the convection zone base. The shift between radiative and
adiabatic temperature gradients at the convection-zone base results in a large sound-speed
difference between two models or a model and the Sun when the depths of the
convection zones do not match, as can be seen in Fig.~\ref{fig:cz}.
 This difference can be calibrated to determine the position of the
solar convection-zone base (Basu \& Antia  1997; Basu  1998;  Basu \& Antia 2004 etc.).

\subsection{Testing equations of state}
\label{subsec:eos}

Equation~(\ref{eq:invy}) had been used extensively to determine both $\delta Y$
and $\delta u/u$. The equation, written in terms of $\delta \rho/\rho$ and
$\delta Y$ was used to determine density $\rho$. Basu \&
\jcd\ (1997) however, noticed that determining $\delta u/u$ and $\delta\rho/\rho$ using
that equation gives rise to systematic errors, which they found were related to 
differences in the equation of state. The problem lies in the fact that
Eq.~(\ref{eq:dy}) ignores the effect of equation of state on $\Gamma_1$.
Thus an extra term, 
$\delta\Gamma_{1,\rm int}/\Gamma_1$, needs to be added to Eq.~(\ref{eq:dy})
to account for the  differences in $\Gamma_1$  at fixed pressure, temperature,
and density. This term is often referred to as the `intrinsic' difference
in $\Gamma_1$. This quantity can be obtained from the frequencies by
rewriting the inversion equation as:
\begin{eqnarray}
\frac{\delta \nu_{n\ell}}{\nu_{n\ell}}&=&
\int_0^R \mathcal{K}^{n\ell}_{u, Y}(r) \frac{\delta u}{u}(r) \;\d r +
\int_0^R \mathcal{K}^{n\ell}_{Y, u}(r) \delta Y(r) \;\d r + \nonumber \\
& & \qquad\int_0^R \mathcal{K}^{n\ell}_{\Gamma_1, \rho}(r)
     \frac{\delta \Gamma_{1, {\rm int}}}{\Gamma_1}(r) \;\d r +
\frac{F(\nu_{n\ell})}{E_{n\ell}}\> ,
\label{eq:invdg1}
\end{eqnarray}
A non-zero $\delta\Gamma_{1,\rm int}/\Gamma_1$ would imply differences between the
equation of state used to construct a solar model and that of the Sun, and hence
can be used to test different equations of state.

\section{Helioseismic results}
\label{sec:res}

Early observations of high degree modes had provided significant
constraints on the solar interior, however, detailed results had to wait for
the availability of reliable frequencies of  low- and intermediate-degree modes,
as well as development of inversion techniques.
In this section we describe what we have learned about solar structure. We also
discuss the results of some tests of different inputs to solar models.
The discussion is mainly restricted to results that are relevant for this
review.

\subsection{Results about solar structure}
\label{subsec:solres}

One of the most important results obtained from the inversion of solar
oscillations frequencies
is the detailed knowledge of solar structure, in particular the solar sound-speed
and density profiles (Christensen-Dalsgaard et al.~1989; Dziembowski et al.~1990;
D\"appen et al.~1991; Antia \& Basu~1994a; Gough et al.~1996; Kosovichev et al.~1997;
Basu et al.~1997, 2000; etc.). With current frequency sets, such as those from
GONG or MDI,
the sound-speed, density and $\Gamma_1$ profiles of the Sun can be reliably  determined
from as close to the center as $0.05R_\odot$ to about $0.95R_\odot$. The dearth of
low-degree modes make it difficult to go deeper, while the lack of reliable high-degree
mode frequencies makes it difficult to investigate layers shallower than about $0.95R_\odot$.
A table listing solar sound speed and density as a function of radius can 
be found in the online material accompanying this review.
Sound speed is known in most of the solar
interior with a precision of better than 0.01\%, while $\Gamma_1$  is known
to a precision  of better than $0.1\%$. The density profile is
known less precisely, but nonetheless with a precision of  0.6\% in the
core and better (0.2\%) in the envelope. The basic inversion techniques
are described in \S~\ref{subsec:inv}.
Basu et al.~(2000) showed that the inferred solar structure depends minimally
on the reference model used for inversion, thereby justifying the linearization
used to obtain Eq.~(\ref{eq:inv}).

The position of the base of the solar convection zone is another important
structural parameter that has been determined. Comparisons of early
helioseismic data with models had shown that the convection zone was
deeper than what had been assumed (Ulrich \& Rhodes 1977),
and Rhodes et al.~(1977) showed that the base of the convection zone was
between $0.62$ and $0.75R_\odot$. By comparing the frequencies of a solar
model with observed ridges in the power spectrum Berthomieu et al.~(1980)
estimated the depth of the convection zone to be about 200 Mm.
More precise determinations had to
wait for inversions to determine the solar sound-speed profile.
As explained in \S~\ref{subsec:rb}, the solar  sound-speed profile can be 
used to measure
the depth of the solar convection zone. 
\jcd\ et al.~(1991) found that the base of
the convection zone is located at $r_b=(0.713\pm0.003)R_\odot$. Similar
results were also obtained by Kosovichev \& Fedorova (1991). Basu \&
Antia (1997) made a detailed study of systematic errors involved in this
measurement, and with better data from the GONG project determined the
base of the solar convection zone to be at $r_b=(0.713\pm0.001)R_\odot$. 
Basu (1998) studied the effect of small errors in the value of the 
solar radius in the estimating of the position of the solar convection-zone base,
and found that the effect was within current error-bars. Using data from both GONG and
MDI projects she determined the base to be at
$r_b=(0.7133\pm 0.0005)R_\odot$.
Basu \& Antia (2004) studied the influence of the heavy-element abundance on the
seismically estimated position of the convection zone base and found
the results to be insensitive to $Z/X$, and using GONG data they determined
the convection zone base to be at $r_b=(0.7133\pm0.0005)R_\odot$.
Furthermore, Basu \& Antia (2001) found that to within errors, the position of the base of convection 
zone is independent of latitude.
Analyses of observed solar frequencies have also been used to show that extent
of overshoot below the convection zone is very small, at least, when
a simple model of the thermal structure of the overshoot layer is assumed
(Basu et al.~1994; Monteiro et al.~1994; Roxburgh \& Vorontsov 1994;
Basu \& Antia 1994; Basu 1997).

Another important solar property that has been determined successfully using 
helioseismology is $Y_s$,  the helium abundance in the solar
convection zone.  Various techniques discussed
in \S~\ref{subsec:helmethod} have been used. The first attempts to determine the solar
helium abundance were made by D\"appen \& Gough (1986) and D\"appen et al.~(1988b).
They were, however, hampered by the lack of high-precision datasets.
D\"appen et al.~(1991), using a full
inversion technique (Eq.~\ref{eq:invy}), found 
$Y_s=0.268\pm 0.01$. Using a similar technique Dziembowski et al.~(1991)
found $Y_s=0.234\pm 0.005$. 
Kosovichev et al.~(1992), again through inversion of Eq.~(\ref{eq:invy}),  
found $Y_s=0.232\pm 0.006$.
They did a thorough analysis to determine the 
cause of the variation between different results about $Y_s$ obtained through inversions
and found that the major source of systematic error is the equation of state. 
The problem with the spread of results remained even when better data were
available.
Richard et al.~(1998) found  $Y_s=0.248\pm0.002$ using reference models
constructed with the MHD equation of state,
and
Di Mauro et al.~(2002) found $Y_s=0.2457\pm 0.0005$ using reference models
with the MHD equation of state and
$Y_s=0.2539\pm 0.0005$ using models with the OPAL equation of state.

There are of course other ways to determine the helium abundance and some
have been described in \S~\ref{subsec:helmethod}. Vorontsov et al.~(1992) 
using a different technique found $Y_s=0.25\pm 0.01$.
Guzik \& Cox (1992), by comparing frequencies of models with different $Y_s$ with
observations, found $Y_s=0.24\pm 0.005$ in the solar envelope.
Antia \& Basu (1994b) studied the sensitivity of the estimated $Y_s$ to the equation of
state for techniques based on calibrating $W(r)$ (Eq.~\ref{eq:wr}) or other equivalent functions.
When using
 the function $W(r)$ to determine $Y_s$,  they found that models constructed with equations of state
such as EFF  were inadequate to determine the solar helium
abundance, while models using the MHD equation of state 
yielded better
results  giving $Y_s=0.252\pm 0.003$ for the Sun. The OPAL equation of
state was not available at that time.
Basu \& Antia (1995) used both MHD and OPAL equations of state,
and found that the calibration methods of 
 finding $Y_s$ were less sensitive to the differences
between the two equations of state than results obtained by inverting 
Eq.~(\ref{eq:invy}).  They obtained $Y_s=0.246$ using calibration
models constructed with MHD equation of state and $Y_s=0.249$ for models with
OPAL equation of state. 
This work used frequencies from the Big Bear Solar Observatory (Libbrecht et al.~1990).
The errors due to
errors in frequencies in each case was very small, but taking into account systematic
errors in the calibration technique and some  uncertainty from the equation of state, 
they estimated the solar helium abundance to be $Y_s=0.249\pm0.003$.
Using better data from both GONG and MDI projects,
Basu (1998) found $Y_s=0.248$.

\begin{figure}[t]
\begin{center}
\includegraphics[width=.8\textwidth]{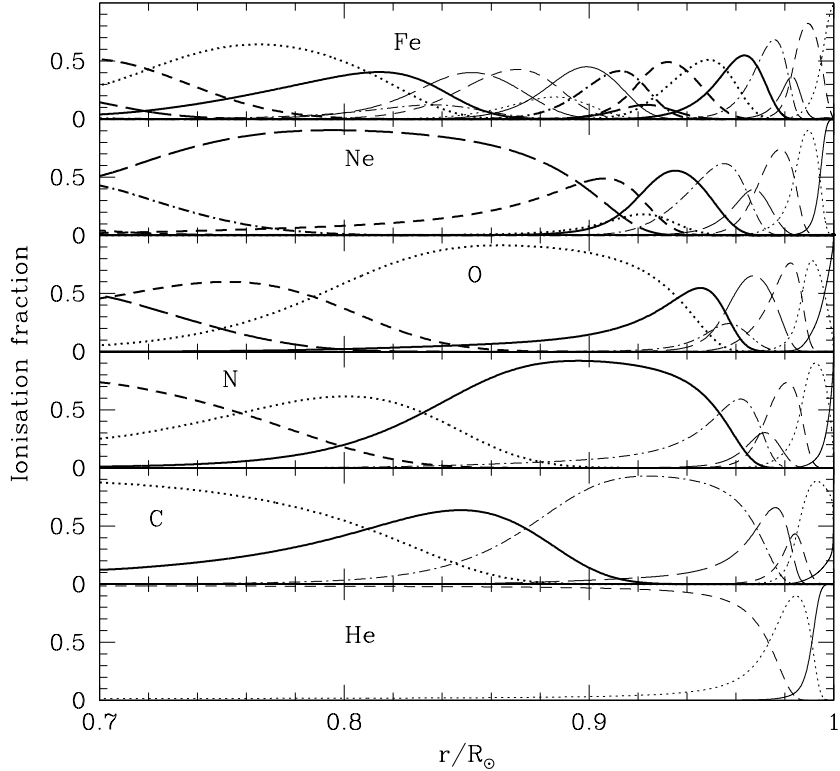}
\end{center}
\caption[]{The fractional abundances of different ionization states of 
some elements in a solar model plotted as a function of radius.
The elements are marked in each panel. Different lines represent the fractional
abundance of different ionization states, and the states  can be identified by
the fact that the peak in the abundance shifts inwards for each successive ionization state.
The CEFF equation of state was used in these calculations.}
\label{fig:ion}
\end{figure}

Basu \& Antia (2004) studied the effects
of heavy-element abundances on the estimated $Y_s$ to find
that $Y_s$ is not very sensitive to the value of $Z$ in the calibration models.
 They estimated
$Y_s=0.2485\pm0.0034$ using calibration models with
$Z/X=0.0171$. This is remarkably close to the estimate  Basu \& Antia (1995) obtained
using calibration models with $Z/X=0.0245$. This may be  a coincidence since there
is a small change in the value of $Y_s$ caused by improvements in input physics and
also because of the availability of better 
seismic data from GONG and MDI during
the intervening years. It turns out that the
estimated value of hydrogen abundance $X$ is remarkably insensitive
to $Z$. This would imply that $Y=1-X-Z$ will increase slightly when $Z$
decreases. Thus a decrease of the solar-envelope $Z$ by 0.0048 between the GS98 and AGS05 values
will lead to an increase in $Y_s$ by 0.0048. Trampedach et al.~(2006), while doing
a  detailed comparison of the effects of OPAL and MHD equations of state on solar
models, have pointed out that He\,{\small II} ionization zone also
coincides with the O\,{\small III} ionization zone (see Fig.~\ref{fig:ion}).  Because
of this the estimated helium abundance is affected and the effect is to
increase estimated $Y_s$ when $Z$  is reduced.
Using the MHD equation of state they estimate that $Y_s$ should
increase by 0.0039 when $Z$ is reduced from 0.018 to 0.011. This is slightly
less than the effect seen by Basu \& Antia (2004). The differences could be due to
differences in the equation of state --- Basu \& Antia (2004) used the OPAL equation
of state. Alternately, the difference could be due to the ionization
zones of other elements, not accounted for by Trampedach et al.~(2006), that also
coincide with the He\,{\small II} ionization zone. These other overlapping ionization
zones can also be seen in Fig.~\ref{fig:ion}.
The OPAL equation of state used in this study includes only the heavy elements C, N, O
and Ne in proportions that are different from  the AGS05
mixture. This introduces some systematic error in the seismic estimate of $Y_s$.
Basu \& Antia (2004) calculated $X$ using calibration models constructed with
different values of $X$ with fixed $Z$. By trying different heavy element
mixtures and $Z$ values, they found  that the estimated $X$ for the Sun appears to be
fairly independent of $Z$. When models with reduced $Z$ are used for calibration
they will have less contribution from heavy elements and as a result the
solar value of $Y_s$ could be slightly overestimated, as discussed by 
Trampedach et al.~(2006). Thus it would be preferable to use the seismic 
results for $X_s$, the surface value of $X$, which are less sensitive to $Z$.
Using  MDI and GONG data, Basu \& Antia (2004) found 
$X_s=0.7389\pm0.0034$ using calibration models that had $Z/X=0.0171$.

\subsection{Seismic tests of input physics}

As Eq.~(\ref{eq:inv}) shows, inverting the frequency differences
between models and the Sun immediately tells us how different the models are
compared with the Sun. This allows us to test different solar models. Additional tests
of solar models
are the position of the base of convection zone and the abundance
of helium in the convection zone. By examining differences between the Sun and
solar   models constructed with different input
physics  like the treatment of diffusion, equation of state,
opacity, nuclear energy generation rates, etc., one can test these inputs.

The analysis of
the solar sound-speed profile just below the base of the convection zone
demonstrated the need to include gravitational settling of helium and
heavy elements below the convection zone (\jcd\ et al.~1993).
The need for diffusion had earlier been discussed by Michaud et al.~(1984) 
as a means to solve the problem of lithium abundance in stars. 
Vauclair \& Vauclair (1982) proposed diffusion as the explanation for 
the large abundance variations in main sequence and horizontal branch stars,
as well as in white dwarfs.
Demarque \& Guenther (1988) and Cox et al.~(1989) constructed solar models
that included diffusion to find that the mode-frequencies of  models
with diffusion match observed solar frequencies better than those
of models without diffusion. 
Incorporation of diffusion of helium and heavy elements in solar models
led to significantly better agreement with seismic inversions.
The main reason for the improvement is that
models with diffusion have a deeper convection zone than models
without diffusion --- for models with the Grevesse \& Noel (1993) and
GS98 values of $Z/X$, the position of the
base of the convection zone almost matches that of the Sun when diffusion
is included.
Diffusion causes a buildup of helium and metals at the base of the convection zone
and this in turn results  in
an increase in opacity at relatively high temperatures,
resulting in a deeper convection zone in models with diffusion.
Incorporating diffusion also leads to a better agreement between
the convection-zone helium abundance of  the models and the Sun.
Models without diffusion have helium abundances of about $0.27$ to $0.28$,
much larger than the seismically derived value of $0.24$--$0.25$. 
The enormous improvements in  solar models caused by the inclusion of
diffusion has resulted in diffusion being considered as one of the
standard ingredients in the construction of solar models. All modern
standard solar models, such as those of \jcd\ et al.~(1996), Guzik \& Swenson (1997),
Guenther \& Demarque (1997),
Brun et al.~(1998), Neuforge-Verheecke et al.~(2001a,b), Couvidat et al.~(2003), 
Bahcall et al.~(2005a,b,c), etc., incorporate diffusion of helium and heavy
elements.

Since diffusion plays a major role in the Sun, it must be important in
other stars too. Diffusion moves helium from the outer layers to the core, which
 decreases the hydrogen abundance in the core, thereby decreasing the
main-sequence life-time of stars (Stringfellow et al.~1983). 
Proffitt \& Vandenberg (1991) found that including helium diffusion reduces globular 
cluster ages as determined from the main-sequence turn-off by 10\%. However,
ages determined from the difference in luminosity between the main-sequence turn-off
and the zero-age horizontal branch are less affected (Chaboyer et al.~1992a,b) except at
very low metallicities.
Not accounting for diffusion can also lead to
errors when determining the ages of individual stars by matching
their positions on the HR diagram  (Weiss \& Schlattl 1998).
The settling of heavy elements, such as lithium, below the
convection zone also means that diffusion must be taken into account
while deriving the primordial abundance of such elements (see e.g., Vauclair 1998).

\begin{figure} 
\begin{center}
\includegraphics[width=.99\textwidth]{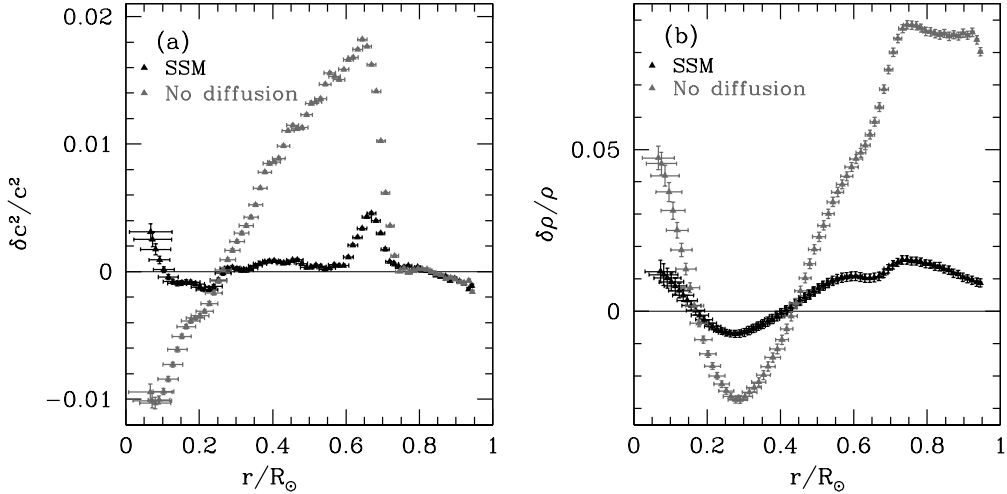}
\caption{The relative sound speed (panel a) and density (panel b) differences
between the Sun and two solar models as obtained by inverting the MDI 360-day
data set. One model is a standard solar model and includes diffusion, the
other does not. Other physics inputs are the same. The models have $Z/X=0.0245$
and are from Basu et al.~(2000).
}
\label{fig:inv1}
\end{center}
\end{figure} 

Modern solar models agree very well with the Sun.
Figure~\ref{fig:inv1} shows the relative difference in sound speed and density
between the Sun  and a standard solar
model as obtained by inverting Eq.~(\ref{eq:inv}). The model includes diffusion. For contrast, also shown in the
figure is the result for a model without diffusion which has the
same physics as the model with diffusion.
It is clear that in most of the  interior,
the sound speed in a solar model is within about 0.1\% of that in the Sun.
With the data set used, the surface layers cannot be resolved,
but for the rest we can see that the maximum difference in sound speed occurs just below
the base of the convection zone.  
Although,
we show only one standard solar model, all others give qualitatively similar results and
differ only in detail.
The strong peak in the
relative sound speed difference has been identified to
be the result of the  sharp gradient in the helium abundance just below the base of the
convection zone in solar models with diffusion (e.g., Basu \& Antia 1995;
\jcd\ et al.~1996).   
This discrepancy can be alleviated if some mixing is included in the
region just below the convection-zone base. Helioseismic tests show that the increase in
$Y$ below the solar convection-zone base is not as steep as in the models (Antia \& Chitre 1998).
Inversions  to obtain the rotation rate in the solar interior (e.g., Thompson et al.~1996;
Schou et al.~1998b) show the presence of a strong shear-layer around the
base of the convection zone, referred to as the tachocline
(Spiegel \& Zahn 1992). The strong shear in the tachocline may lead to
turbulence, which could mix these layers and smooth out the composition
profiles in this region.
Models which incorporate mixing in the tachocline region do show better agreement
with the Sun (Richard et al.~1996; Brun et al.~1999; Basu et al.~2000; etc.). 

The small difference between the Sun and the solar models,
particularly in the core,
led helioseismologists to believe that the solar neutrino problem was not caused by
deficiencies in  the solar models but by deficiencies in the standard model of particle physics
which postulates mass-less neutrinos
(e.g., Bahcall et al.~1997; Antia \& Chitre 1997; Takata \& Shibahashi 1998;
Watanabe \& Shibahashi 2001; etc.). Bahcall et al.~(1998)
showed that if a solar model that satisfied the observed solar neutrino
constraints provided by the Chlorine experiment was constructed,
the sound-speed difference between the Sun and the model will be about 10\% in the core --- much 
worse than that for a standard solar model.  Therefore, they claimed,  the
problem must be with our knowledge of neutrino physics.
Similar inferences were obtained by comparing observed neutrino fluxes from
different experiments (e.g., Hata et al.~1994; Haxton 1995; Castellani et al.~1997;
Heeger \& Robertson 1996) without involving solar models.
The helioseismic inference about solution of the solar neutrino problem has since been
confirmed by results from the Sudbury Neutrino Observatory (Ahmad et al.~2002).
With this the role of helioseismology in study of solar neutrinos has been reversed
since now the Sun can be used as a well calibrated source of neutrinos.
The observed neutrino fluxes at Earth are now used to study the properties
of neutrinos, such as mass differences and mixing angles between different
neutrino eigenstates (e.g., Bandyopadhyay et al.~2002; Bahcall \& Pe\~na-Garay
2004; etc.).

Except for a small region near the surface,  the convection zone is almost
adiabatically stratified and hence its structure is determined by the
equation of state and composition.  As discussed in \S~\ref{subsec:eos},
several methods can be used to test the equation of state used
in constructing solar models.
By comparing frequencies of models with those of the Sun,
\jcd\ \& D\"appen (1992) concluded that the older, simpler
EFF equation of state needed correction for Coulomb effects.
Antia \& Basu (1994b) 
 compared the function $W(r)$ (cf., Eq.~\ref{eq:wr})
for the Sun and different models and concluded that the MHD equation
of state was better than older equations of state such as the EFF
equation of state. 
Using the same function, Antia \& Basu (2006) also concluded that
the CEFF equation of state, i.e., the EFF equation of state
with Coulomb corrections %(\jcd\ \& D\"appen 1992; Guenther et al.~1992)
 is almost as good as the OPAL equation of
state throughout most of the convection zone.
Similarly, \jcd\ \& D\"appen (1992) also found that CEFF equation of state
gives a satisfactory agreement with seismic data.
Basu \& Antia (1995, 1997) examined the details of the sound-speed 
difference between models and the Sun in layers just below
the He\,{\small II} ionization zone to conclude that the OPAL equation of
state was a better description of the solar equation of state than the 
MHD equation of state.
Basu et al.~(1996) examined the sound-speed difference between models
constructed with MHD and OPAL equations of state throughout the
Sun to find that OPAL models  were 
in better agreement with the Sun at all radii compared to models with
MHD equation of state. This led them to conclude that
 that OPAL is better than MHD for the Sun. 

\begin{figure}
\begin{center}
\includegraphics[width=.65\textwidth]{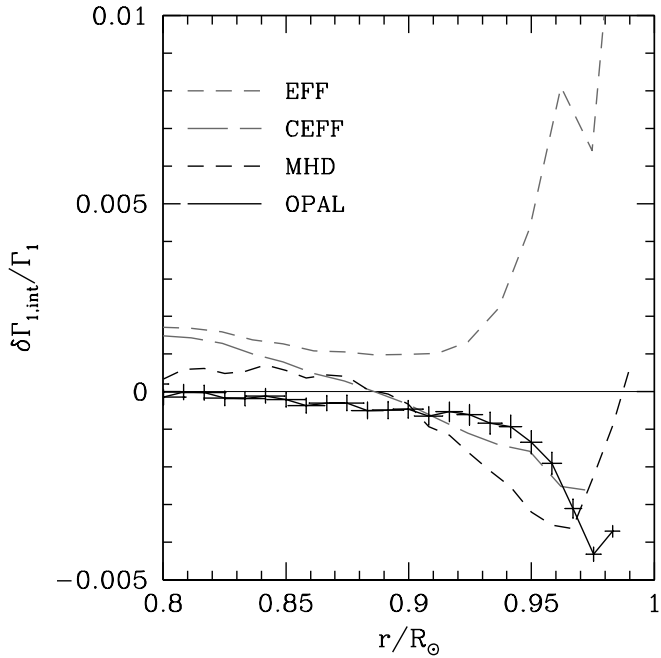}
\caption{The intrinsic $\Gamma_1$ difference between the Sun and solar
models constructed with different equations of state. All models
have $Z/X=0.0245$.
}
\label{fig:eos}
\end{center}
\end{figure}

There have also  been studies of the equation of state using the $\Gamma_1$ differences
between models and the Sun. By looking at the $\Gamma_1$ differences 
in the core, Elliott \& Kosovichev (1998) showed that 
both OPAL and MHD equations of state used
in the construction of models of solar-type stars were deficient 
at high-temperature and high-density
regimes. The cause of the discrepancy was identified as the use of the non-relativistic
approximation to describe partially degenerate electrons instead of the
relativistic Fermi--Dirac integrals.
The deficiency has since been corrected in both the OPAL equation
of state (Rogers \& Nayfonov 2002) and the MHD equation of state (Gong et al.~2001).
Since $\Gamma_1$ in the outer layers depends on the structure in addition to the
equation of state, Basu \& Christensen-Dalsgaard (1997) used the so-called
``intrinsic $\Gamma_1$'' difference (cf., Eq.~\ref{eq:invdg1}). They found that
the EFF equation of state was very obviously deficient, and that the CEFF, MHD, and OPAL
were reasonably similar, except that the MHD equation of state had problems
just below the He\,{\small II} ionization zone.
Basu et al.~(1999) did a more detailed study of the differences between
the MHD and OPAL equations of state and found that while indeed the
MHD equation of state fared poorly compared to OPAL below the He\,{\small II} 
ionization zone, it did better than OPAL much closer to the surface. The intrinsic
$\Gamma_1$ differences between the Sun and models constructed with 
different equations of state are shown in Fig.~\ref{fig:eos}.

Opacities used in constructing solar models can also be tested.
Guzik \& Cox (1991) compared frequencies of models with those of the Sun and
 suggested that the then used solar opacities
(Cox \& Tabor 1976) were too low. Somewhat more quantitative results
can be obtained by comparing the structure of models with that of the Sun.
Saio (1992), assuming that the  the sound-speed
difference between models and the Sun were completely due to
opacity errors, found that the discrepancy between models and the Sun can be reduced
below the convection zone if the Los Alamos opacities (Weiss et al.~1990)
were increased by 20-50\%.
The publication of the OPAL opacity tables by Rogers \& Iglesias (1992) resolved
this problem and confirmed that opacities needed to be increased substantially
in region near the base of the convection zone.
The Cox-Tabor and Los Alamos opacities are not in use these days,
but similar methods can be used to determine changes to OPAL opacities
used today.
Tripathy \& Christensen-Dalsgaard (1998), using the same assumption as 
that of Saio (1992),
calculated  kernels linking
the opacity changes to sound-speed changes. 
Tripathy \& \jcd~(1998) also used these kernels to determine the opacity
changes needed to explain the sound-speed differences between the Sun and  Model S of
\jcd\ et al.~(1996).
Basu \& Antia (1997) found that the relative density differences
between models and the Sun were  much more sensitive to opacities than the
sound-speed differences were, and using solar envelope models with the
correct position of the convection-zone base concluded that the OPAL
opacities were indeed consistent with seismic constraints.
Of course the result were based on models
that had the then accepted value of $Z/X$, i.e., 0.0245.

Assuming that opacities, equation of state,  and nuclear energy generation rates are
known, one can also infer the temperature and hydrogen-abundance profiles
of the Sun. For this one 
uses the equations of thermal equilibrium coupled with the sound-speed
and density profiles inferred through primary inversions, and a specified
heavy element abundance profile. Once all these
profiles are calculated, it is possible to calculate the total energy
generated through nuclear reactions. In general, this computed energy
does not agree with the observed solar luminosity, and the difference between
the two provides a test of the input physics as well as the heavy element abundance
profile used in the calculations. The nuclear energy generation rate in
the solar interior is primarily controlled by the reaction rate for the
p-p reaction, which has not been measured in laboratories. This reaction rate
is computed theoretically and seismic data allow us to test these calculations
(Antia \& Chitre 1998, 1999, 2002; Degl'Innocenti et al.~1998;
Schlattl et al.~1999). The current seismic estimate for this reaction rate
$S_{11}=(4.07\pm0.07)\times10^{-25}$ MeV barns
(Antia \& Chitre 2002; Brun et al.~2002),
is consistent with latest theoretical estimate of
$S_{11}=(4.00\pm0.03)\times10^{-25}$ MeV barns (Adelberger et al.~1998).

\section{Solar abundances}
\label{sec:solarz}

The Sun is unique among stars in that there are multiple ways to determine its heavy element
abundance. Spectroscopy allows us to determine the heavy element abundance of the 
solar photosphere, chromosphere and corona, and sometimes in sunspots. Solar abundances can also be determined
from solar winds (cf., Bochsler 2007a) and solar energetic particles (cf.,
Stone 1989; Reames 1994, 1998). And another important source of solar
abundance information are C1 chondrites --- i.e., meteorites that have never been heated to melting
temperatures. These meteorites are expected to have preserved the original
composition of the solar nebula, except for the most volatile elements like,
H, C, N, O and the noble gases. Meteoritic abundances are usually given
relative to silicon, which is used as a bridge to determine the photospheric
abundances of meteoritic elements. For most of the non-volatile elements, the agreement between
meteoritic and photospheric abundances determined through spectroscopy is
fairly good.
The meteoritic abundances may be expected to represent the initial solar
abundance, which is different from the current photospheric abundances
because of diffusion of helium and heavy elements below the convection zone.
If the diffusion rates of various heavy elements are different,  their
relative abundance could also  be different in  the  meteorites compared to the
present-day solar photosphere. The accuracy of current photospheric abundance determinations
is probably not sufficient to detect these small differences.
The abundances of elements is usually expressed in a logarithmic
scale.  This scale is normalized with respect to the hydrogen abundance, which is 
assumed to have a logarithmic abundance of 12. Thus if $N_{\rm A}/N_{\rm H}$ is the ratio of the
number densities of element A and hydrogen respectively, then 
the `abundance' of A is written as
$\log_{10}(N_{\rm A}/N_{\rm H})+12$ dex, this is often denoted by $\log\epsilon({\rm A})$.

The method of determining solar abundances from spectroscopy is the same as that for other stars
and can be broken down into four steps: (i) identifying the lines to be used to
measure the abundance, (ii) construction of a model atmosphere, (iii) use the model
 atmosphere to do a line-formation calculation, and (iv) determine the
abundance either using a curve-of-growth analysis  or through spectral synthesis.
In principle, this process is iterative since the model atmosphere in step (ii)
also depends on abundances obtained in step (iv). Hence, ideally these steps
should be repeated till model becomes self-consistent, though in practice
this is often not done.
Techniques to analyze stellar spectra are described by Gray (2005).
 The calculations are usually done
under assumption of local thermodynamic equilibrium (LTE), though non-LTE calculations
are gaining ground since departures from LTE affect many lines.
Usually, a 1D photospheric model with a given
 $T$--$\tau$ relation is used. 
Since these
models lack dynamical information, free parameters are used to represent
the line-broadening effects of turbulence on the spectral lines in either the form of
micro-turbulence or macro-turbulence. 
The microturbulence parameters is supposed to account for 
effects of turbulent elements that are smaller than
unit optical depth. Macroturbulence is the effect of  turbulent elements
larger than unit optical depth. Microturbulence is brought into
the radiative-transfer calculations by increasing the thermal velocity in the
absorption coefficients. Calculations to estimate the effect of  macroturbulence on lines
 bypass radiative transfer calculations altogether. These effects are usually determined by convolving the 
line intensity function calculated without macroturbulence to the velocity distribution
of the macroturbulent cells. In either
case only kinematic effects are included.
These effects, however,  cannot  reproduce the line-bisectors of stellar spectral lines
and hence, cannot reproduce the spectral lines accurately.

Modern solar abundance calculations usually rely on semi-empirical  photospheric models
such as those by Vernazza et al.~(1973), Holweger \& M\"uller (1974), Maltby et al.~(1986), or
on photospheric models obtained from ab initio calculations, such as
those obtained by using the MARCS code (Gustafsson et al.~1975), the atmospheric
model of Kurucz%
\footnote{Atmospheric models of Kurucz can be downloaded from http://kurucz.harvard.edu.}
(Kurucz 1970; Sbordone et al.~2004; Kurucz 2005a,b, etc.).
The semi-empirical atmospheric models rely on some form of inversion procedure to infer
the stratification of the solar atmosphere using spectral lines formed at
different heights. The atmospheric models  keep improving with advances
in input physics.
In addition to the model atmosphere, there is one important input to the whole
process of determining abundances from spectroscopy, and that is the transition probability
(or alternately the oscillator strength) of the line. Improvements in the
measurement of oscillator strengths can lead to sudden changes in the estimated 
abundance, an interesting case being that of the abundance of iron in the
late 1960s and early 1970s. Garz \& Kock (1969)
measured the oscillator strengths of thirty iron lines
leading to a downward revision of the oscillator strength. This led to a nearly
ten-fold increase in the estimated solar iron abundance (Garz et al.~1969;
Baschek et al.~1970; etc.). In addition, including line-broadening effects (other than
turbulence already mentioned earlier)  is important. Errors and
uncertainties in spectroscopic techniques to determine stellar spectra have been
discussed by Johnson (2002).

Solar abundance determinations started with spectroscopy. Russell (1929) estimated the 
line intensities of several lines by eye and derived the abundances of more than
fifty elements. An important conclusion from his work was that the Sun was
largely made of hydrogen. The abundance of helium was not known at that time.
Uns\"old (1948) had much better data and techniques at his disposal and
determined the abundances of twenty five elements. Surprisingly, Russell's estimates were 
not very different from those of Uns\"old (1948). Perhaps the first `modern' abundance
analysis is that by Goldberg et al.~(1960). They did a careful curve-of-growth analysis
with improved photospheric models and oscillator strengths to determine the abundances
of more than forty elements. Since the work of Goldberg et al.~(1960), many groups have
been engaged in determining solar abundances through spectroscopy. The abundance of
carbon, nitrogen and oxygen, 
elements that are important for solar structure, are usually derived
from photospheric lines.

The solar photospheric spectrum however, does not contain lines of the noble gases.
As a result, the  abundances of these elements  are determined from coronal lines,
solar energetic particles or solar wind. This includes determining  the abundances of helium and neon. Helium, of course,
is the second most abundant element, and its abundance determines the evolution of
the Sun, and neon is an important source of opacity in the solar interior. 
Because of their very low abundances, other
noble gases do not affect solar structure or evolution significantly.
Initial estimates of the solar helium abundance were  obtained through
theoretical solar models (e.g., Str\"omgren 1935; Schwarzschild 1946) as explained
in \S~\ref{subsec:ssm}. This continued to be the most reliable technique
until seismic techniques were developed (cf., \S~\ref{subsec:helmethod}).
Although helium was discovered in a prominence spectrum, it is difficult
to use prominence spectra for abundance determinations because these regions are
more difficult to model than the photosphere and equilibrium conditions do not apply.
Values of $N_{\rm He}/N_{\rm H}$ ranging from 0.065 to 0.16 were obtained from
prominence and flare spectra (e.g., Hirayama 1979; Heasley \& Milkey 1978;
Feldman et al.~2005). Observations made in 1985 by the Coronal
Helium Abundance Spacelab Experiment (CHASE) on the space shuttle Challenger
  yielded an abundance of $0.070\pm0.011$ (Gabriel et al.~1995b).
Similar results
were obtained from solar winds and solar energetic particles. Furthermore, the
helium abundance found in the slow solar wind was found to vary with solar cycle. It 
varies from 0.02 during minimum activity to 0.04 during maximum activity
(Aellig et al.~2001).
Because of these difficulties,  attempts
were made to measure the helium abundance in giant planets that may be expected
to preserve the abundances at the time of formation of solar system.
The Voyager spacecraft was used to
measure the abundance ratio of helium to hydrogen in Jupiter and Saturn 
and it was found to be $0.057\pm0.013$ and $0.016\pm0.013$ respectively
(Gautier et al.~1981; Conrath
et al.~1984),  lower than expected value.  Uranus and Neptune, however,  yielded a value
of $0.092\pm0.017$ (Conrath et al.~1987) which is  close to the now accepted value.
Later measurements by the Galileo spacecraft have resulted in  a higher value, 
0.078, for Jupiter (von Zahn \& Hunten 1996), while a reanalysis of Voyager
data for Saturn (Conrath \& Gautier 2000) has also yielded higher value of
$Y=0.18$--0.25 (by mass).
As a result of these discrepant results, most of the early estimates of the solar
helium abundance  were obtained form
extrasolar sources, such as planetary nebulae,
H\,{\small II} regions, or hot stars in solar neighborhood (e.g., Bowen 1935).

The major problem in determining the abundance of elements using coronal
lines or solar wind is that the
abundances in the outers layers of the Sun show evidence of fractionation.
This occurs at 
low chromospheric levels leading to the so-called FIP (first ionization potential) effect. 
Elements with low first ionization potential ($\le 10$ eV) show much higher abundances
in the corona than in the photosphere. The FIP effect also varies strongly with
coronal structure (e.g., Reames 1999). 
Abundances in the solar wind also vary tremendously. Elements with a low FIP are overabundant by a 
factor of 3--5 relative to high-FIP elements in the slow solar wind, but only by a factor of 1.5--2 in the 
fast streams that emanate  from coronal holes (e.g., von Steiger et al.~1995). Thus
these sources are not very useful for determining abundances required for modeling the
solar interior. An estimate of the extent of fractionation can be obtained by
studying  elements which have both   photospheric and coronal lines. However, these elements
have comparative low first ionization potential compared to the noble gases, e.g., the FIP
of  C, N and O 
are 11.3, 14.5 and 13.6 eV, respectively, while those for He and Ne are
24.6 and 21.6 eV respectively. Thus it is difficult to get a direct calibration
of fractionation for noble gases because of the lack of  elements with
comparable first ionization potentials. Even if the normal
correction for fractionation is applied, the estimated helium abundance from coronal
studies is about 0.2 dex
lower than the currently accepted value (Anders \& Grevesse 1989).
It turns out that currently the most 
precise method for determining the solar helium abundance is 
 through helioseismology as  has been discussed in  \S~\ref{subsec:helmethod}\
and \S~\ref{subsec:solres}. 
While the issue of the solar helium abundance has been resolved,
Neon poses a difficulty: because there are no independent estimates of the solar neon 
abundance, it is difficult
to estimate the systematic error in its abundance determined from coronal lines.
The abundance of neon is
still determined from the outer layers of the Sun and hence uncertainties
are large. These are generally based on the relative abundance of neon with respect to
either oxygen or magnesium. From the ratio Ne/O or Ne/Mg one can determine
the absolute abundance of neon by using the photospheric value of O or Mg
abundances. Appropriate corrections need to be applied for the FIP effect,
particularly for Ne/Mg since Mg is a low FIP element.
In view of the FIP effect, it is not clear if this  gives
the correct photospheric abundance of neon, and 
it should be noted that the same technique failed to give reliable abundance estimates
of helium. Recently Drake \& Ercolano (2007) have suggested that the solar neon
abundance can be measured by observing the Ne K$\alpha$ line that is fluoresced
by coronal X-rays and is emitted near the temperature minimum region of the
solar atmosphere. They have done  Monte Carlo simulations to show
that it would be feasible to use this line if good quality spectra of the relevant
spectral region are available.
This uncertainty in neon abundance
plays a role  in the current controversy about solar abundances and is
 discussed later in \S~\ref{subsec:neon}.

For many elements, particularly refractory elements, meteoritic abundances are more precise than 
spectroscopic abundances. 
Meteoritic abundances have been measured extensively since the 1920s and 1930s. Early meteoritic-abundance estimates
were systematically tabulated by Goldschmidt (1937).
Brown (1949) and Suess \& Urey (1956) put together tables of ``cosmic abundances'' by choosing
among different source of data. Among the more recent tables of meteoritic abundances
are those by Cameron (1982) who based it on his previous works, and those of
Anders \& Ebihara (1982), Palme \& Beer (1993), and Lodders (2003).
The abundance of refractory elements in the C1 chondrites and
the Sun is now believed to be the same (Grevesse \& Noels 1993; Palme \& Jones 2005), 
and even abundances of moderately
volatile elements, such as sodium, zinc and sulfur, are in agreement
with the Sun (Palme \& Jones 2005). 

There have been significant improvements in abundance determinations during
the last few decades. Solar photospheric spectra with very high resolution
and very high signal to noise ratio, covering a large wavelength
range from ultra-violet to far infra-red, have been obtained from the ground and
from space (e.g., Kurucz 1995). Solar atmospheric models have also improved
significantly. Finally, accurate atomic and molecular data including the
transition probabilities have been obtained. As a result of all these
improvements, discrepancies between the photospheric abundances determined
spectroscopically and the abundances in C1 meteorites has steadily decreased.

Solar structure calculations during the last two decades have generally
been done using solar abundance tables compiled by Anders \& Grevesse (1989) and the
updates to that by Grevesse \& Noels (1993). This was later superseded by the
compilation of GS98. The default OPAL opacity tables are calculated for
Grevesse \& Noels (1993) mixture, though recently it has become possible
to calculate OPAL opacity tables for any specified mixture of 19 heavy
elements listed in Table~\ref{tab:abund}\footnote{http://www-pat.llnl.gov/Research/OPAL/}.
The abundance table of Anders \& Grevesse (1989)  was an update of the meteoritic
data of Anders \& Ebihara (1982) and Grevesse (1984a,b). They obtain  $Z/X=0.0274$ for the
Sun and C, N, O abundances
of $\log \epsilon ({\rm C)}= 8.56\pm 0.04$ dex, $\log \epsilon ({\rm N)}= 8.05\pm 0.04$,
and  $\log \epsilon ({\rm O)}= 8.93\pm 0.035$. They have a photospheric Fe abundance
of $\log \epsilon ({\rm Fe)}= 7.67\pm 0.03$, and Ne abundance of $\log \epsilon ({\rm Ne)}= 8.09\pm 0.10$
dex.
Grevesse \& Noels (1993) updated these abundances. In the process, abundances of
C, N, O, Ne and Fe were lowered to $8.55\pm 0.05$, $7.97\pm 0.07$, $8.87\pm 0.07$, $8.07\pm0.06$ and
$7.51\pm 0.01$ respectively. This lowered the total $Z$ to give $Z/X=0.0245$. 
The photospheric Fe abundance was lowered because of work by different groups with more
accurate transition probabilities. This photospheric Fe abundance matches the meteoritic abundance, 
while the earlier values did not.
The updates for C, N and O were
a result of new infra-red data  observed by the Atmospheric Trace Molecule Spectroscopy (ATMOS)
 experiment, new transition probabilities for 
C\,{\small I}, N\,{\small I} and O\,{\small I} lines, and the change in the gas and electron pressures
in the Holweger-M\"uller model atmosphere caused by the change in the Fe abundance. The new neon abundance
was a weighted mean of the abundance derived from the analysis of impulsive flare spectra and
from solar energetic particles. 

The next revision of these table was that by GS98.
This revision lowered solar $Z$ further to $Z/X=0.023$. GS98 redid some of the
computations for C, N and O abundances with different photospheric models based on the
empirical photospheric model of Grevesse \& Sauval (1999). The abundances of C, N and O in
their table are $8.52\pm 0.06$, $7.92\pm 0.06$ and $8.83\pm 0.06$ respectively. The iron abundance
was lowered slightly to $7.50\pm 0.05$, and they changed the neon abundance estimate to
$8.08\pm 0.06$ based on the measurements of the Ne/Mg ratio by  Widing (1997).

The compilations of GS98 were changed dramatically by AGS05. Most of the
updates were a results of the new abundance determinations by Allende-Prieto et al.~(2001, 2002)
and Asplund et al.~(2004, 2005a). All these calculations were based on 3D atmospheric models
obtained from simulations of the upper convection zone and lower photosphere of the Sun.
Since this change is the main
effect responsible in decreasing the total heavy element abundances of the Sun, we
discuss these works, and others that followed, in some detail.
A comparison of the abundances of some of the elements
in the GS98, AGS05 and earlier tables is given in Table~\ref{tab:abund}. 
This table includes all elements used in OPAL opacity calculations.

%%%%%%-------------- TABLE -------------
\begin{table}[t]
\caption{A comparison of solar abundances of some
elements in tables of Anders \& Grevesse (1989, AG89), Grevesse \& Noels
(1993, GN93), Grevesse \& Sauval (1998, GS98) and Asplund, Grevesse \&
Sauval (2005b, AGS05). Abundances are in units
of $\log_{10}(A/{\rm H})+12$.\label{tab:abund}}
\begin{center}
\begin{tabular}{lrcccc}
\noalign{\smallskip}
\hline
{\hbox{Element}}&$Z$&AG89&GN93&{\hbox{GS98}}&{\hbox{AGS05}}\\
\hline
C &  6 & $8.56\pm 0.04$ &  $8.55\pm 0.05$ & $8.52\pm 0.06$ & $8.39\pm 0.05$\\
N &  7 & $8.05\pm 0.04$ & $7.97\pm 0.05$ & $7.92\pm 0.06$ & $7.78\pm 0.06$\\
O &  8 & $8.93\pm 0.04$ & $8.87\pm 0.04$ & $8.83\pm 0.06$ & $8.66\pm 0.05$\\
Ne & 10 & $8.09\pm 0.10$ & $8.07\pm 0.06$ & $8.08\pm 0.06$ & $7.84\pm 0.06$\\
Na & 11 & $6.33\pm 0.03$ & $6.33\pm 0.03$ & $6.33\pm 0.03$ & $6.17\pm 0.04$\\
Mg & 12 & $7.58\pm 0.05$ & $7.58\pm 0.05$ & $7.58\pm 0.05$ & $7.53\pm 0.09$\\
Al & 13 & $6.47\pm 0.07$ & $6.47\pm 0.07$ & $6.47\pm 0.07$ & $6.37\pm 0.06$\\
Si & 14 & $7.55\pm 0.05$ & $7.55\pm 0.05$ & $7.55\pm 0.05$ & $7.51\pm 0.04$\\
P  & 15 & $5.45\pm 0.04$ & $5.45\pm 0.04$ & $5.45\pm 0.04$ & $5.36\pm 0.04$\\
S &  16 & $7.21\pm 0.06$ & $7.21\pm 0.06$ & $7.33\pm 0.11$ & $7.14\pm 0.05$\\
Cl & 17 & $5.50\pm 0.30$ & $5.50\pm 0.30$ & $5.50\pm 0.30$ & $5.50\pm 0.30$\\
Ar & 18 & $6.56\pm 0.10$ & $6.60\pm 0.14$ & $6.40\pm 0.06$ & $6.18\pm 0.08$\\
K  & 19 & $5.12\pm 0.13$ & $5.12\pm 0.13$ & $5.12\pm 0.13$ & $5.08\pm 0.07$\\
Ca & 20 & $6.36\pm 0.02$ & $6.36\pm 0.02$ & $6.36\pm 0.02$ & $6.31\pm 0.04$\\
Ti & 22 & $4.99\pm 0.02$ & $5.04\pm 0.02$ & $5.02\pm 0.06$ & $4.90\pm 0.06$\\
Cr & 24 & $5.67\pm 0.03$ & $5.67\pm 0.03$ & $5.67\pm 0.03$ & $5.64\pm 0.10$\\
Mn & 25 & $5.39\pm 0.03$ & $5.39\pm 0.03$ & $5.39\pm 0.03$ & $5.39\pm 0.03$\\
Fe & 26 & $7.67\pm 0.03$ & $7.51\pm 0.01$ & $7.50\pm 0.05$ & $7.45\pm 0.05$\\
Ni & 28 & $6.25\pm 0.04$ & $6.25\pm 0.04$ & $6.25\pm 0.04$ & $6.23\pm 0.04$\\
\hline
$Z/X$& & $.0274\pm.0016$ & $.0244\pm.0014$ & $.0231\pm.0018$ & $.0165\pm.0011$ \\
\hline
\noalign{\smallskip}
\end{tabular}
\end{center}
\end{table}

Although oxygen is the most abundant of all heavy elements, determining its abundance 
spectroscopically is problematic since the relevant spectral
lines are difficult to model. The forbidden lines at 630 nm and 636.3 nm 
are affected by blends, while the O\,{\small I} triplets at 777.3 nm and 844.6 nm 
require extensive NLTE corrections. Additionally, OH and CO lines are also
difficult to model and are very sensitive to atmospheric models. As a result,
the solar oxygen abundance determined from various lines do not generally agree with each
other (e.g., Reetz 1999). It should be noted that  1D atmospheric model can
only represent an average atmospheric structure. Such models cannot
represent thermal inhomogeneities and velocity fields that are present in the solar
atmosphere. Kiselman \& Nordlund (1995)
attempted to determine the solar oxygen abundance using a 3D model with NLTE treatment.
Although, they found some improvement in the sense that the abundances
determined from different lines were somewhat closer to each other, the
differences were still significant. They pointed out that an oxygen abundance
of less than 8.8 dex may be justified from their studies.

Asplund et al.~(1999, 2000b) started a systematic investigation using 3D radiative 
hydrodynamical simulations of 
solar convection instead of 1D model atmospheres
to determine solar abundances from spectroscopy. They first applied the 3D models
 to some Fe\,{\small I} and Fe\,{\small II} lines. 
 These model atmospheres include dynamic
information and hence free parameters for turbulence are not required to fit spectral lines.
These models are, however, the result of  a simulation, and hence, whether or not they
really represent the solar atmosphere depends on the input physics and correct treatment
of radiative transfer in the simulation.
The 3D simulation used by Asplund et al.~(2000b) was obtained with the 3D compressible,
radiative-hydrodynamic code developed to study solar surface convection and granulation developed
by Nordlund \& Stein (1990), and  Stein \& Nordlund (1989, 1998). This  is a
large eddy simulation (LES) code. The equations of mass, momentum,
and energy conservation coupled to the radiative transfer equation along a ray are solved.
The simulations have  periodic boundary conditions in the horizontal direction and transmitting
boundary conditions at the top and bottom. The simulation covers a small part
of the solar atmosphere and the convection zone with a horizontal extent
of $6\times 6$ Mm and a vertical extent of 3.8 Mm of which about 1.0 Mm is
above the level at which optical depth for the continuum is unity. They had $200\times 200\times 82$
grid points in their simulations which typically covered 2 solar hours.
The code uses the hyper-viscosity 
diffusion algorithm described by Stein \& Nordlund (1998) for  numerical viscosity.
Radiative transfer is assumed to happen in LTE.
The 3D equation of radiative transfer is solved along 8 inclined rays  using the opacity binning 
technique (Nordlund 1982), and four bins were used.  
The code used the MHD equation of state.
Continuum opacities of Gustafson et al.~(1975 and updates) were used, and line opacities
were privately obtained from Kurucz. The simulations were done using the GS98 solar heavy element
abundances. The simulation was then used as the model atmosphere to do line formation
calculations using a spectral line formation code. 
To do the spectral line calculation, a 50-minute subset of the
simulations was chosen with snapshots stored at every 30s. They were also interpolated to a finer
depth scale to improve vertical sampling. Asplund et al.~(2000b) showed that they could get
better fits to the Fe lines than conventional 1D calculations even without using free parameters for
micro- and macro-turbulence, and also that they could fit the line bisectors. They did not
need any adjustable parameters to fit the lines.

Allende Prieto et al.~(2001)  used the same simulation to
determine the solar oxygen abundance. They analyzed the forbidden
[O\,{\small I}] line at 630~nm. Since this a forbidden line, it is not affected by
non-LTE effects, however, this line is blended with a Ni\,{\small I} line.  Allende Prieto et al.~(2001)
used the Uppsala synthesis package of Gustafsson et al.~(1975, and updates) for inputs needed
for the line formation calculation, and included natural broadening using data from the
Vienna Atomic Line Database (Kupka et al.~1999), and solar line fluxes from
Kurucz et al.~(1984). The oscillator strength of the Ni\,{\small I}
line was not well known and was considered to be a free parameter. Allende Prieto et al.~(2001)
got a very good fit to the [O\,{\small I}] line for an abundance of  $\log\epsilon({\rm O})=8.69\pm0.05$.
However,  since this work, the oscillator strength of the Ni\,{\small I} has been measured
(Johansson et al.~2003)
and  the fit to the line is not as good if that value is used (Koesterke et al.~2007a).
Asplund et al.~(2004) investigated the solar oxygen abundance further using the same simulation as
the Asplund et al.~(2000b) work. They used a mixture of [O\,{\small I}], O\,{\small I}, OH vibration-rotation
and OH pure rotation lines. They used the MULTI3D code (Botnen 1997; Botnen \& Carlsson 1999;
Asplund et al.~2003) to do non-LTE calculations. Instead of the 50 snapshots from the simulations, as
was used by Asplund et al.~(2000b) and Allende Prieto et al.~(2001), they used two. They
derived an oxygen abundance of $\log \epsilon ({\rm O)}= 8.66\pm 0.05$, again much smaller than the
GS98 value of $8.83 \pm 0.06$. 

Allende Prieto et al.~(2002) examined the solar carbon abundance  using the 
[C\,{\small I}] line at 872.7 nm,  again using the 3D simulation of Asplund et al.~(2000b),
and did a calculation similar to what Allende Prieto et al.~(2001) did for the [O\,{\small I}] 
line at 630~nm. They find $\log \epsilon ({\rm C)}= 8.39\pm 0.04$ dex, which is
lower than the GS98 value of $8.52\pm 0.05$. Asplund et al.~(2005a) 
investigated the solar photospheric carbon abundance using [C\,{\small I}], C\,{\small I},
CH vibration-rotation, CH A-X electronic and C$_2$ Swan electronic lines using the
same simulations and a non-LTE line calculation using the
code MULTI (Carlsson 1986). 
They get the same abundance as Allende Prieto et al.~(2002).

AGS05 used the same simulation to study the solar nitrogen abundance
using N\,{\small I} and NH vibration-rotation lines. They obtained an abundance
of $\log\epsilon({\rm N})=7.88\pm0.08$ using LTE analysis of N\,{\small I} lines.
Since there is no NLTE study of these lines for 3D models they applied the NLTE
corrections from Rentzsch-Holm (1996) to get an abundance of $7.85\pm0.08$.
They argue that this may require a small downward correction. The NH vibration-rotation
lines are more difficult to treat as they are very sensitive to temperature
and hence there is a large difference between results obtained from  1D and 3D models. They obtain
an abundance of $\log\epsilon({\rm N})=7.73\pm0.05$ from these lines, which
is significantly lower than the value obtained from N\,{\small I} lines.

Earlier, Asplund (2000) used the same 3D simulations to get $\log \epsilon ({\rm Si)}= 7.51\pm 0.04$.
This is slightly smaller than the GS98 value of $7.55\pm 0.04$ and
led AGS05 to the lower abundances of meteoritic elements listed
in Lodders (2003).
AGS05 have also applied the 3D model to study abundances of other elements
and in all cases the abundances are lower than those obtained using 1D models.
The Na abundance is obtained using six weak Na\,{\small I} lines to get
a value of $6.17\pm0.04$, which is about 0.1 dex lower than the meteoritic
value and also lower than the GS98 value. The cause of this discrepancy is
not understood. The Al, P and S abundances are also reduced by about 0.1 dex
when 3D models are used.

The investigations  mentioned in this section resulted in a net lowering of the solar
heavy element abundance to $Z/X=0.0165$ as compiled by AGS05. The largest effects are due to 
C, N, O and Ne abundances, these are elements that  play important roles in determining solar structure.
Since Ne and Ar abundances cannot be determined from photospheric spectra due to the lack
of  spectral lines, their values were estimated by AGS05 using measured
abundance ratios in the solar corona and solar energetic particles
along with the photospheric abundance for oxygen. Thus the reduction in the
O abundance, along with refined coronal abundance ratio from Reames (1999),
led to the reduction in the abundances of Ne and Ar. If the Ne/Mg ratio
from Feldman \& Widing (2003) were used instead,  a much higher Ne abundance of
$8.06\pm0.10$, a value similar to the GS98 one, would be obtained.
Before the AGS05 revision, the
two ratios (Ne/O or Mg/O) gave mutually consistent results.
This is also true for the abundance of argon.
Other changes in abundances,
predominantly caused by the lowering of the Si abundance,
have been smaller. 
It should be noted however, that the
underlying 3D photospheric model used in each case had the higher GS98 abundances, and the
calculations have not been repeated with simulations that have the AGS05 background abundances. Nor
have the investigations been re-done with
 other simulations with different input physics (such as equation of
state or opacities), or even different numerical viscosity parameters to check for the
sensitivity of the results to the background 3D model.

The main reason for confidence in the abundances obtained using these 3D models is the good agreement
with observed line profiles and line bisectors, and the fact that different
lines give similar abundances. 
While this is true for oxygen, the nitrogen abundance obtained using N\,{\small I}
and NH lines are significantly different even when 3D models are used.
Another argument in favor of the reduced
abundances is that old abundances suggested that the Sun has much higher
metallicity as compared to galactic neighborhood and Magellanic
clouds (Meyer 1989; Turck-Chi\`eze et al.~1993, 2004). With the reduced
heavy element abundances the Sun now appears to be naturally enriched
in oxygen in comparison with extragalactic H\,{\small II} regions, Magellanic
clouds, other clusters and neighbors. 
One measure of enrichment is the ratio $\Delta Y/\Delta {\rm O}$, when 
$\Delta Y$ and $\Delta {\rm O}$ are respectively the increase in the helium and
oxygen abundances over the same epoch. The starting epoch is 
assumed to be just after the Big Bang nucleosynthesis.
Peimbert et al.~(2007) estimate the ratio to be $3.3\pm0.7$.
Using the primordial helium abundance, $Y_p=0.2482\pm0.0004$ from WMAP
(Spergel et al.~2007) we find the ratio of the helium to oxygen enrichment to be 2.78 for a solar
model with GS98 abundances and 1.90 for a model with AGS05 abundance.
Of course, if a lower primordial helium abundance is used the numbers may
change in favor of the AGS05 model. It is not clear if these arguments can
be used to constrain solar abundances.
Cunha \& Lambert (1992)
found that the mean oxygen abundance in 18 B type stars in Orion association
is $\log\epsilon({\rm O})=8.65\pm0.12$, which is consistent with the AGS05 value.
However, Edvardsson et al.~(1993) had done a detailed study of abundances of
various elements in 189 F and G type nearby stars, and concluded that
solar abundances are consistent with those in other similar stars after accounting
for the dispersion in abundances between different stars.
Similar conclusions have been drawn by Gustafsson (1998).
Furthermore, there is evidence that stars with planetary
systems have higher heavy element abundances as compared to stars without
planets (e.g., Santos et al.~2003, 2005). Thus we would expect the Sun to have a higher metallicity
compared to other, average, stars.
It should also be kept in mind that comparison of solar abundances with 
interstellar medium and H\,{\small II} region abundances
is affected by dust formation, which  traps some elements.
Additionally, comparison of solar abundances with that of   nearby stars  needs to account for
diffusion of elements below the outer convection zone. The amount of depletion caused by 
diffusion  depends on the 
star's age and the thickness of its outer convection zone, and hence a proper comparison
is not possible without first determining the amount by which
diffusion has depleted the atmosphere of any star. It should be kept in mind that 
the abundances of other stars are determined by using 1D
stellar atmosphere models and these too could change if abundance calculations with
3D models are done.

\section{Consequences of the new abundances}
\label{sec:abund}

Chemical composition enters the equations of stellar structure
through input physics like the opacity, equation of state and nuclear energy generation
rate. In studies of stellar evolution, it has been found that stars that have lower 
heavy-element abundances  are bluer (higher effective
temperature)  and
more luminous than higher metallicity stars of the same mass and helium
abundance (see e.g., Salaris \& Cassisi 2005). The Sun, however, has a known effective temperature, luminosity
and age, and hence whatever the metallicity, a solar model has to
have the same effective temperature and luminosity as the Sun at the same age. This
usually implies a change in  the helium abundance (one of the unknown parameters
of the system as described in \S~\ref{subsec:ssm}). Low-metallicity solar models,
therefore, generally have lower helium abundances in order to have the correct luminosity
at the solar age.
Thus solar models
with different metallicities  have different evolutionary histories that 
intersect in the HR diagram at the solar age, and the models also have 
different structures. 
In this section we examine the consequences of the lowered solar heavy-element
abundance on the structure of solar models. Prior to that we discuss
why and how heavy element abundances
affect solar structure.
It should be noted that spectroscopy measures the photospheric
abundance of the Sun.
 However, 
the convection-zone abundances are believed to be the same as the
photospheric  abundances because of constant convective overshoot into the
photosphere. Moreover, the convection zone is believed to be
chemically homogeneous because convective turn-over time-scales are
much shorter than evolutionary and diffusion time-scales.
We should keep in mind that the
 composition of the solar surface today is not  the composition
with which the Sun was formed since helium and heavy elements diffuse out of the
convection zone to the layers below.

Even before the current revision of heavy element abundances, solar models
with low $Z$ were examined to explore the possibility of reducing neutrino
fluxes. Although helioseismic data were not very precise at that point, they did not favor 
these models
(e.g., \jcd\ \& Gough 1980). Much more seismic
data have become available since this early work and solar models have also improved through advances in
in our understanding of input physics.
However, the conclusion still remains valid. Of course, as we know today, the solution of
the solar neutrino problem lies in nonstandard properties of neutrinos rather than nonstandard
processes in solar models.

The primary effect of the heavy-element abundance on solar structure is
through its effect on opacity. Opacity  mainly affects the structure of the radiative interior
and the position of the base of the convection zone. 
The structure of the convection zone itself is
independent of opacity, except for a thin layer near the surface where
convection is inefficient.  Of course,
there are some indirect effects of $Z$ on the convection zone since  conditions at 
the convection-zone base determine the adiabat that defines the structure of 
the lower convection zone.
While hydrogen and helium, the main constituents of the Sun, are
completely ionized in the deeper parts of the solar interior, some of the heavy elements, such
as iron, silicon, sulfur, etc.  are only partially ionized. These elements
contribute to opacity in these regions  through different atomic transitions.

The effect of heavy-element abundances on the equation of state is smaller.
The equation of state  mainly depends on the number of particles per unit
volume. The number density of  heavy elements is two orders of
magnitude smaller than that of   hydrogen or helium, as a
result they  only have  a  small effect on pressure. However, the process
of ionization of the elements causes the
adiabatic index $\Gamma_1$ to decrease in the ionization zones. The magnitude of the
dip in $\Gamma_1$, though small for heavy-element ionization, depends 
on the abundance of these elements. 
The structure of most of the convection zone is essentially independent
of opacity and  is dominated by the equation of state. Thus the effect
of heavy element abundances on the structure of the solar convection zone
is due to their effect on the equation of state.
The convection zone is also the region where helioseismic inversions
are very reliable and hence the equation-of-state effect is detectable.

Nuclear energy generation in the solar core occurs
primarily through the $p$-$p$ chain, which depends only
on the abundance of hydrogen and not that of heavy elements.
However, a small fraction of the energy in the Sun
is produced by the CNO cycle, which does depend on the
abundance of C, N and O. Thus there is some effect of the
new abundances on the nuclear energy generation rate too.
Using the updated reaction rates for the ${}^{14}{\rm N}(p,\gamma){}^{15}{\rm O}$
reaction (Runkle et al.~2005), it was found that models with the
GS98
abundances produce 0.8\% of their energy through the CNO cycle,
while models with the AGS05 abundances
produce 0.5\% of their energy through  the CNO cycle (Bahcall et al.~2006).
Nuclear energy generation through the CNO cycle is effective only
in the inner core.
It is not clear whether the effect of the heavy elements on the structure
of core is predominantly through
their effect on the CNO reaction rate or through their effect on opacities. Most of
the heavy elements are fully ionized in these regions, and hence their
effect on opacity is small, but the CNO reactions contribution to the total energy generated is
small too.   Thus it is not completely
clear which effect dominates in the inner core. 
A related effect of heavy-element abundances on the CNO cycle
is in the production of neutrinos. Since the nuclear reaction rates change
with change of C, N, and O abundances, the neutrino fluxes from these
reactions  change too.

\begin{figure} 
\begin{center}
\includegraphics[width=.99\textwidth]{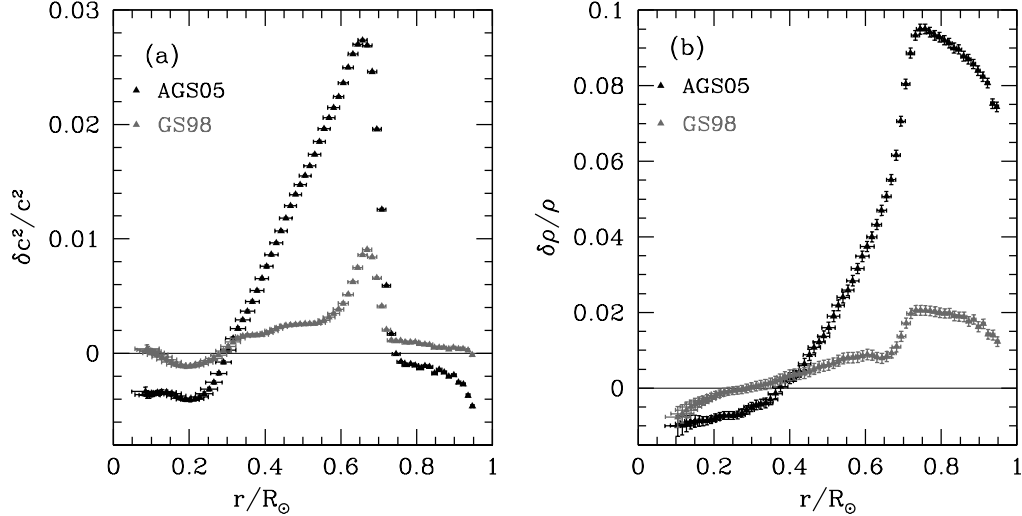}
\caption{The relative sound-speed (panel a) and density differences (panel b)
between the Sun and a model constructed with the AGS05 abundances
(Bahcall et al.~2005c). For comparison
we also show the results for a models constructed with the GS98 abundances.
MDI 360 day data have been used for the inversions.
}
\label{fig:asp}
\end{center}
\end{figure} 

In Fig.~\ref{fig:asp} we  show the sound-speed and
density differences between a standard solar model constructed with the
new AGS05 abundances and the Sun. For 
comparison, we also show the results for a comparable model constructed
with the GS98 abundances. As can be seen from the
figure, the new abundances result in a solar model that is much
worse than the model with the older abundances in terms of agreement
with seismically inferred solar sound-speed and density profiles. 
Thus it is clear that while solar models with
old abundances are closer to the seismically
inferred profiles of the Sun, the one using newer
abundances differs significantly. The maximum difference in sound-speed
and density increases by more than a factor of five
when AGS05 abundances
are used instead of the GS98 abundances. 
The root mean-square deviation
in squared sound speed, $\langle (\delta c^2/c^2)^2\rangle^{1/2}$, 
between the solar models and the Sun
increases from 0.0014 for the GS98 model to 0.0053 for the model with AGS05 abundances.
Similarly the rms density difference, $\langle (\delta \rho/\rho)^2\rangle^{1/2}$, increases from
0.011 to 0.047 when AGS05 abundances are used instead of GS98 (Bahcall
et al.~2005a).
The model with the AGS05 abundances has
a convection-zone helium abundance of $Y_s=0.230$, much lower than the
helioseismically determined value of $Y_s$. The position of the
convection-zone base in that model is $r_b=0.7289R_\odot$, again very
different from the seismically determined value of $r_b$.
We describe below in detail the changes in various parts of the
solar model caused by the lowering of $Z/X$.

\subsection{The base of the convection zone}

The most easily detectable effect of the reduction of heavy-element abundances is a
change in the position of the base of the convection zone.
The temperature gradient in the radiative interior is determined by
opacity, and hence, its structure is affected by the heavy element abundances.
The base of the convection zone occurs at a point where opacity is just small
enough to allow the entire heat flux to be transported by radiation, and  thus
the location of this point depends on the abundance of those  heavy elements that are
the predominant sources of opacity in that region.
If these abundances are reduced, opacity reduces, and  the depth of
the convection zone also reduces. 
Since the depth of the convection
zone has been measured very accurately, it is the most sensitive indicator
of opacity or heavy element abundances. 

After the revision of the solar
heavy-element abundances was announced by 
Allende-Prieto et al.~(2001, 2002) and
Asplund et al.~(2004),  Bahcall \& Pinsonneault (2004) constructed
a solar model with the reduced abundances and found
that the depth of the convection zone in the model is reduced significantly
with respect to a similar model with the GS98 abundances.
The convection-zone base of that model was at $r_b=0.726R_\odot$, which
differs  from the seismically determined value by
$26\sigma$. Bahcall et al.~(2005a) also found a similar value.
This solar
model used $Z/X=0.0176$ since some of the downward revisions of abundances  found
in the tables of AGS05
had  not been determined at that stage. Using the same $Z/X$,  Basu \& Antia~(2004)
found $r_b=0.732R_\odot$.  Using a  slightly different 
composition ($Z/X=0.0172$), Turck-Chi\`eze et al.~(2004) found
$r_b=0.7285R_\odot$ for a standard model and $r_b=0.7312R_\odot$ for 
a model that incorporates mixing below the convection zone.
There is considerable variation in the location of $r_b$ in different
models and  these arise  because of differences in input physics used in calculating the
models, but all these values are significantly different from the seismically
determined value of $r_b$ when low $Z/X$ is used. The further lowering of $Z/X$ to 0.0165 by
AGS05 changed the position of the
convection-zone base  further, and Bahcall et al.~(2005c)
found that models constructed with these abundances have $r_b=0.729R_\odot$,
a difference of about $32\sigma$  from the seismically determined value.
Table~\ref{tab:cz} lists $r_b$ for several solar models, standard
and non-standard, that have been
published recently.

Bahcall et al.~(2006) studied the consequences of the
lowering of $Z/X$ through a detailed Monte-Carlo simulation of
solar models where parameters controlling input physics were 
randomly drawn for each model from separate probability distributions for every parameter.
The parameters varied include the rates of some of the nuclear reactions, age of
the Sun, diffusion coefficients, luminosity, and most importantly the
individual abundances of the elements that are used to calculate OPAL opacities.
Two types of models were constructed, ones where
the central value of the heavy-element abundances were obtained 
from the GS98  abundances, and another where 
they were drawn from the AGS05 abundances.
Two types of error-distributions were considered for the heavy elements. The
so-called ``optimistic'' uncertainties corresponded to the error-bars
in the composition tables of GS98  or 
AGS05; and the so-called ``conservative''
uncertainties, which assumes that the uncertainty in the abundance of a 
particular element is the difference in its abundance between the
GS98  and the AGS05 tables.
For all cases, Bahcall et al.~(2006)  found
that the position of the base of the convection zone had a Gaussian
distribution. The ``conservative'' distribution of the
GS98  models show that the convection-zone base
at $r_b=(0.7154\pm0.0102)R_\odot$, while for the AGS05 models
$r_b=(0.7296\pm0.0105)R_\odot$.
The `optimistic'  AGS05 models fare worse, with the
convection-zone base at $r_b=(0.7280\pm0.0037)R_\odot$,
 showing that
changing other solar inputs within their errors do not 
bring  the convection-zone base of any of these low-$Z/X$ models in agreement with the Sun. 
For the 'optimistic' GS98 models on the other hand,  $r_b=(0.7133\pm0.0035)R_\odot$,
which is consistent with the seismic value.

\begin{table}[p]
\renewcommand{\baselinestretch}{0.8}
\caption{The position of the base of the convection zone ($r_b$) and the helium abundance $Y_s$ in the convection zone  for different solar models. Unless mentioned otherwise, the models were calculated with OPAL opacities.\label{tab:cz}}
\begin{tabular}{lcccl}
\hline
Reference& $Z/X$& $r_b$& $Y_s$& Remarks\\
\hline
Basu et al.~(2000) & 0.0245 & 0.7123 & 0.2453 & GN93\\
Bahcall et al.~(2001)  & 0.0229 & 0.7140 & 0.2437 & GS98 \\
Montalb\'an et al.~(2004)& 0.0245 & 0.714\phantom{0} & 0.246\phantom{0} & GN93\\
Montalb\'an et al.~(2004)& 0.0177 & 0.727\phantom{0} & 0.243\phantom{0}\\
Montalb\'an et al.~(2004)& 0.0177 & 0.723\phantom{0} & 0.248\phantom{0} & enhanced opacity\\
Montalb\'an et al.~(2004)& 0.0177 & 0.718\phantom{0} & 0.249\phantom{0} & enhanced opacity\\
Montalb\'an et al.~(2004)& 0.0177 & 0.714\phantom{0} & 0.226\phantom{0} & enhanced diffusion\\
Montalb\'an et al.~(2004)& 0.0177 & 0.717\phantom{0} & 0.239\phantom{0} & enhanced diffusion \& opacity\\
Turck-Chi\`eze et al.~(2004)& 0.0172 & 0.7285 & 0.2353\\
Turck-Chi\`eze et al.~(2004)& 0.0172 & 0.7312 & 0.2407 & mixing in tachocline\\
Bahcall et al.~(2005a)& 0.0176& 0.7259 & 0.238\phantom{0}  \\
Bahcall et al.~(2005a)& 0.0176& 0.7133 & 0.239\phantom{0} & 21\% increase in opacity \\
Bahcall et al.~(2005a)& 0.0176& 0.7162 & 0.243\phantom{0} & 11\% increase in opacity \\
Bahcall et al.~(2005b)& 0.0192& 0.7174 & 0.2411 & OP, increased Ne \\
Bahcall et al.~(2005b)& 0.0207& 0.7146 & 0.2439 & OP, increased Ne, CNO \\
Bahcall et al.~(2005c)& 0.0229& 0.7138 & 0.243\phantom{0}   & GS98, OP opacity \\
Bahcall et al.~(2005c)& 0.0165& 0.7280 & 0.229\phantom{0} & AGS05, OP opacity\\
Guzik et al.~(2005)& 0.0244 & 0.7133 & 0.2419 & GN93 \\
Guzik et al.~(2005)& 0.0196 & 0.7022 & 0.1926 & enhanced diffusion\\
Guzik et al.~(2005)& 0.0186 & 0.7283 & 0.2339 & enhanced Z diffusion\\
Guzik et al.~(2005)& 0.0206 & 0.7175 & 0.2269 & enhanced diffusion\\
Guzik et al.~(2005)& 0.0173 & 0.7406 & 0.2541 & enhanced diffusion\\
Yang \& Bi~(2007) & 0.0174 & 0.7335 & 0.2294 & \\
Yang \& Bi~(2007) & 0.0176 & 0.7168 & 0.2225 & enhanced diffusion\\
Castro et al.~(2007) & 0.0164 & 0.730\phantom{0} & 0.223\phantom{0} \\
Castro et al.~(2007) & 0.0165 & 0.732\phantom{0} & 0.240\phantom{0} & GS98+low-$Z$ accretion\\
Castro et al.~(2007) & 0.0165 & 0.712\phantom{0} & 0.249\phantom{0} & GS98+low-$Z$ accretion  \\
& & & &  \& mixing \& overshoot  \\
\hline
\end{tabular}
\end{table}

The large difference in the sound speed between the Sun and the low-$Z/X$ solar
model shown in Fig.~\ref{fig:asp} is caused mainly by the difference in
the position of the base of the convection zone. Since convection zones
are adiabatically stratified, and thus have larger temperature gradients than
radiatively stratified regions, a deeper convection zone results in higher
temperatures at the base of the convection zone. The sudden increase in
the sound-speed difference in  Fig.~\ref{fig:asp} at the radius corresponding
to the convection-zone base indicates that the Sun has a higher temperature
than the model in that region which is a result of the deeper convection zone in
the Sun. This sharp increase in the sound-speed differences with respect to the
Sun is seen for other low-$Z$ models too (see, e.g.,
Basu \& Antia 2004;
Turck-Chi\`eze et al.~2004; Montalb\'an et al.~2004, Bahcall et al.~2005a, Guzik
et al.~2005, etc.).
In their Monte-Carlo study, Bahcall et al.~(2006) too found 
that average rms sound-speed and density
differences between the models and the Sun are much worse for the 
AGS05 models than the GS98 
models.

\begin{figure}[t]
\begin{center}
\includegraphics[width=.8\textwidth]{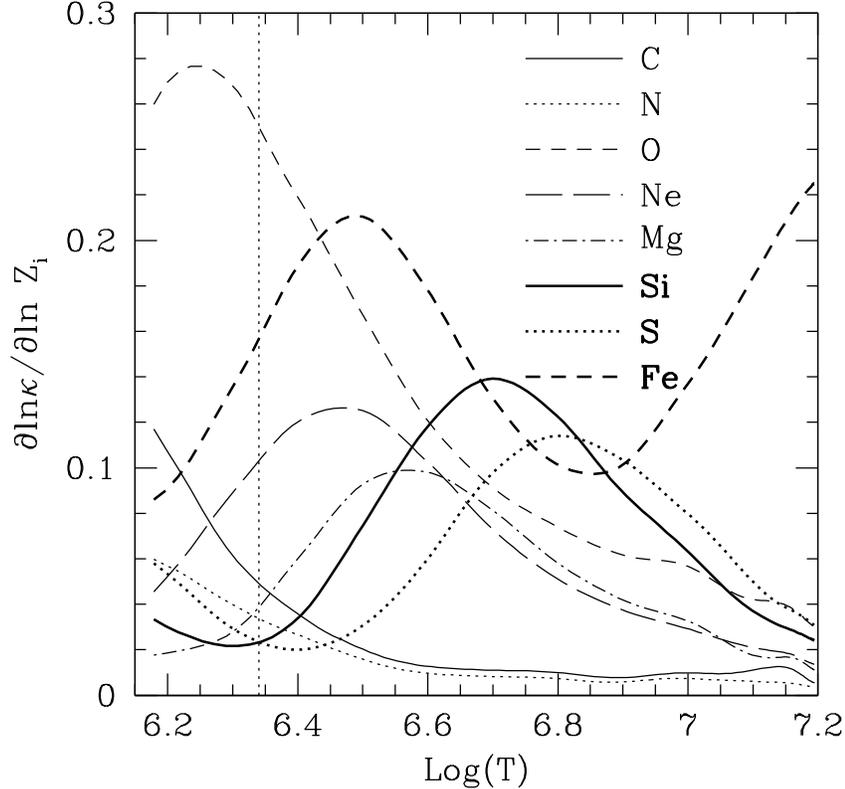}
\caption{The logarithmic gradient of opacity with respect to abundances of
different heavy elements plotted as a function of temperature. The gradient
is calculated at different points in a solar model, using the temperature,
density and composition profiles of the solar model. The dotted vertical line
in the figure marks the position of the base of the convection zone.}
\label{fig:opac}
\end{center}
\smallskip
\end{figure}

Since the primary source of discrepancy in solar models is the reduction in opacity due to
the lowering of  $Z/X$, it is of interest to identify the  elements that are
responsible for the opacities in the radiative interior. Figure~\ref{fig:opac}
shows the logarithmic derivative of opacity with respect to abundances
of individual elements. These derivatives are calculated for OPAL opacities
 at the GS98   abundances. Although the results are plotted as a function
of temperature, the derivatives were calculated at temperature
and density values for a  solar model.
The dashed vertical line in the figure marks the temperature
corresponding to the  base of the convection zone. 
It can be seen that elements that have low atomic number, such as C and N, that
are more or less completely ionized inside the convection zone
do not make a significant contribution to the opacity in the radiative interior.
Oxygen, which is the most abundant among the heavy elements in the Sun,
is the dominant source of opacity in the region near the base of the
convection zone. In addition to O, Fe and Ne  also contribute significantly
to the opacities in this region. In the solar core, Fe, S, Si and  O are the
dominant sources of opacity. Thus it is not surprising that a reduction
in oxygen abundance has a strong effect on the depth of the convection
zone in a solar model.

\subsection{The convection-zone Helium abundance}

The largest effect that the lowering of the heavy-elements has within
the convection zone is a change in the helium abundance $Y_s$. As
mentioned earlier, the  helium abundance of solar models  needs to be changed in
order to ensure that the models satisfy solar constraints at the
current age of the Sun. The reduction of $Z/X$ causes a reduction of
$Y_s$ in the calibrated solar models. This also implies that the
initial helium abundance of the low-$Z$ models at the zero-age main sequence is
lower than the initial helium abundance for the higher-$Z$ models.
For a $Z/X=0.0176$ model Basu
\& Antia (2004) found that helium abundance in the convection zone
is 0.237, a difference of about
$3\sigma$ from the seismically determined value. Similarly, for their
$Z/X=0.0176$ model, Bahcall et al.~(2005a) found $Y_s=0.238$.
Turck-Chi\`eze et al.~(2004) found $Y_s=0.235$ for their
standard model, and $Y_s=0.241$ for their model that incorporated
mixing below the convection-zone base. All these results are lower than
the seismic estimates, though as is evident, the exact value of $Y_s$ in the models
also depends on the other physical inputs of the model. 
The convection-zone helium abundance
is lowered further when the AGS05 value
of $Z/X=0.0165$ is used,
and Bahcall et al.~(2005c) found that their
models had $Y_s=0.230$ when OPAL opacities were used, and $Y_s=0.229$ when
OP opacities were used.
Table~\ref{tab:cz} also lists $Y_s$ for several solar models that have been
recently published.

From the results of their  Monte-Carlo simulations, Bahcall et al.~(2006)
found that the average $Y_s$ for models with GS98  abundances
with the `conservative' errors is $Y_s=0.2420\pm0.0072$, while for the
AGS05 models with conservative errors is
$Y_s=0.2285\pm 0.0067$,
i.e., marginally inconsistent with the seismic value. However, with
optimistic errors, AGS05 models have
$Y_s=0.2292\pm0.0037$, which is more inconsistent. It should be noted
that the optimistic uncertainties correspond to the uncertainties 
in the AGS05 table. The GS98 models with `optimistic' errors have
$Y_s=0.2425\pm 0.0042$.

\begin{figure} 
\begin{center}
\includegraphics[width=.99\textwidth]{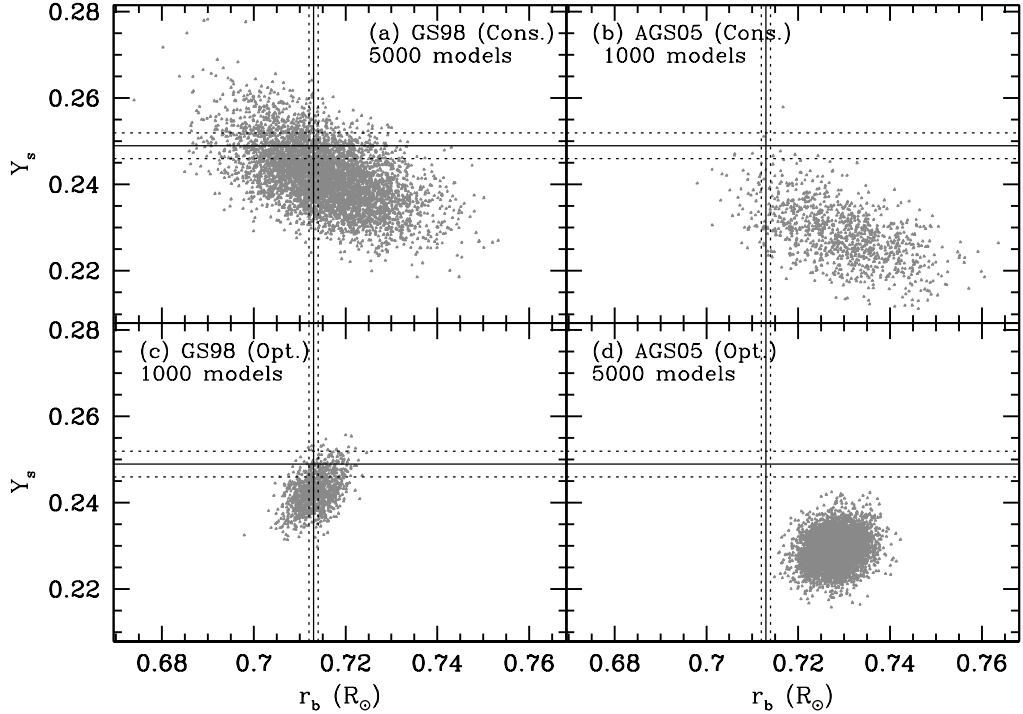}
\caption{The helium abundance in the convection zone, $Y_s$, plotted
against $r_b$, the  position of base of the convection zone,
for the solar models from the Monte Carlo simulations of
Bahcall et al.~(2006). The horizontal and vertical lines respectively, show the
seismically determined values of $r_b$ and $Y_s$ with dotted lines showing
the $1\sigma$ error bars. The labels `Cons.' and `Opt.' refer to
conservative and optimistic errors respectively.}
\label{fig:rby}
\end{center}
\end{figure}

Figure~\ref{fig:rby} shows the models from Monte Carlo simulations in
the $r_b$--$Y_s$ plane. It can be seen that many of the GS98 models
fall in the region that is consistent with seismically
determined values, while almost all the AGS05 models are outside the
seismically accepted range.
From a detailed analysis of the seismic constraints on $Y_s$ and $r_b$, Delahaye \& Pinsonneault
(2006) estimate that solar models with revised abundances disagree
at the level of $15\sigma$ with seismic data, while models constructed
with older solar abundances are consistent within $2\sigma$ with seismic
data. Their analysis was mainly based on the seismic value of the
depth of the convection zone and the helium abundance in convection zone. 
Using these constraints they also attempted to estimate the solar heavy element 
abundances and obtained values close to the GS98 ones.

\subsection{The radiative interior}
\label{subsec:radint}

The effect of the heavy elements on opacities implies that the
structure of the radiative zone changes when the heavy-element
abundance is changed. Both sound-speed and density profiles
are significantly different, as can be seen from Fig.~\ref{fig:asp}.
The main difference between
models using different composition is seen in regions just below the
base of the convection zone. 
Since the equation of state is not very important in determining the structure
of the solar radiative zone, the effect of $Z$ on the equation of state
does not affect the structure of the radiative zone significantly. 
Although, we can
invert for $\Gamma_{1,int}$ differences between models and the Sun in order
to investigate signatures of differences in abundances,
the errors in the results are large enough in these 
regions to make it difficult to detect any signatures of ionization of
elements in the radiative zone.

Since solar models are constructed to 
produce the observed luminosity
irrespective of composition, the structure of the inner core is less
sensitive to opacity and $Z$ variations than the rest of the
radiative zone.
Most of the reduction in abundances is for lighter elements like
C, N, O, and Ne that are fully ionized in the core and hence, do not contribute
significantly to the opacities (cf., Fig.~\ref{fig:opac}).
The main source of opacities in the core are the heavier elements like Fe, S, Si.
Of these, only sulfur abundance has been reduced by a substantial amount (about
a factor of 1.55),  however, the small differences in these 
abundances do produce small differences in opacities.
Although, the C, N and O abundances in the core do not affect opacities
significantly, they  affect the rates of the
nuclear reactions involved in the CNO cycle.
There are, therefore,
noticeable differences between the models. 
For instance, the central temperature of model with
GS98  abundances is $15.67\times 10^6$~K, but that
of a model with AGS05 abundances is
$15.48\times10^6$~K (Bahcall et al.~2006). 
Also, for the
models in the Monte Carlo study of (Bahcall et al.~2006), it is found that the
average mean molecular weight in the inner 2\% of the Sun by radius is
$0.7203 \pm 0.0029$  models with GS98  composition
and $0.7088 \pm 0.0029$ for models with the AGS05
composition (Chaplin et al.~2007b). Unfortunately, there are no direct
seismic determinations of these quantities, though Chaplin et al.~(2007b)
have attempted to determine $\mu$ for the solar core (see \S~\ref{subsec:zcore}).
These differences affect the  sound speed profile of the core and hence, the 
frequency separations.
Basu et al.~(2007) examined the scaled small-frequency spacing and
separation ratios of the  
low-degree modes, which are sensitive to conditions in the solar core and
hence provide an independent test of solar models. They find that models
with AGS05  abundances are not consistent 
with the observed frequency spacings and separation ratios as obtained 
from observations carried out by the BiSON network. 

The change in the structure of the solar core caused by changes
in the heavy-element abundance also affect the 
neutrino fluxes expected from a solar model. However, as explained above, the
major changes are in the outer radiative zone, where the temperature is
too low for nuclear reactions to take place. 
As a result, the neutrino fluxes are comparatively less affected by the
changes in the abundances
(e.g., Bahcall \& Pinsonneault 2004; Turck-Chi\`eze et al.~2004;
Bahcall et al.~2005c). Neutrino fluxes for two models,
one with GS98 and the other with AGS05 abundances are listed in Table~\ref{tab:neutrino}.
Bahcall \& Serenelli (2005) have studied the
effect of the abundance of each element separately on the neutrino fluxes
to get a better estimate of uncertainties due to the uncertainties in the
 abundances of heavy
elements. They find that the calculated uncertainties in neutrino fluxes
are overestimated if $Z$, the total abundance of all elements is considered.
 This is a result of the fact that heavier elements like Fe and Si
have the most  effect on neutrino fluxes, but they do not contribute much
to errors in $Z$ since their abundances are believed
to be more reliably determined. On the other hand, lighter elements like
C, N, O, Ne have little effect on neutrino fluxes (except for those emitted
by the  CNO reactions),
but their abundances are more uncertain in view of recent developments, and affect $Z$ changes
more. Thus errors in solar $Z$ are more a result of changes in elements that do not affect
neutrino emissions by much, rather than a result of changes in those elements that have a
larger influence on the emitted neutrinos.
Hence, Bahcall \& Serenelli (2005)  argue that more realistic estimate of uncertainties due to
heavy element abundances can only be calculated by examining the effect of each
element separately. They also include the effects of correlations between the
abundances of different elements.
The flux of
CNO neutrinos is significantly affected by the difference in GS98 and
AGS05 abundances, but it is difficult to isolate these neutrino fluxes.

\begin{table}[t]
\caption{Neutrino fluxes at the Earth (in cm$^{-2}$ s$^{-1}$) from solar models with
GS98 and AGS05 abundances. Models are from Bahcall et al.~(2006).\label{tab:neutrino}}
\begin{center}
\begin{tabular}{lcc}
\noalign{\smallskip}
\hline
{Reaction/Detector}& GS98 & AGS05\\
\hline
pp & $5.99\times10^{10}$ & $6.06\times10^{10}$ \\
pep & $1.42\times10^8$ & $1.45\times10^8$ \\
hep & $7.93\times10^3$ & $8.25\times10^3$ \\
$^7$Be & $4.84\times10^9$ & $4.34\times10^9$ \\
$^8$B & $5.69\times10^6$ & $4.51\times10^6$ \\
$^{13}$N & $3.05\times10^8$ & $2.00\times10^8$ \\
$^{15}$O & $2.31\times10^8$ & $1.44\times10^8$ \\
$^{17}$F & $5.83\times10^6$ & $3.25\times10^6$ \\
Cl (SNU) & 8.12 & 6.58 \\
Ga (SNU) & 126.08 & 118.88\\
\hline
\noalign{\smallskip}
\end{tabular}
\end{center}
\end{table}

\subsection{The ionization zones}
\label{subsec:ion}

There are other, more subtle, effects of the change in heavy-element abundances
on the structure of the convection zone. These are caused by the effects heavy elements
have on  the equation of state, which in turn determines the structure of
the convection zone. The convection zone essentially has an adiabatic
temperature gradient, and that is determined by the equation of state.
The effect of heavy-element abundances on the structure of the convection zone
is usually too small to be seen directly in the sound speed and density
profiles. The effect is seen more clearly in the dimensionless sound-speed
gradient $W(r)$ (cf., Eq.~\ref{eq:wr}) or the adiabatic index
$\Gamma_{1,int}$.

Both $\Gamma_1$ and $W(r)$ show dips or peaks at the ionization
zones of different elements. In Fig.~\ref{fig:ion} we
showed the fractional abundances  of different ionization stages of various elements
in a solar model. These ionization fractions were calculated for a solar model
using the CEFF equation of state. The CEFF equation of state allows us to
explicitly include different elements and  to follow the ionization
stages of the different elements. Since hydrogen and helium, which form the bulk of 
solar material, are ionized in the outer 5\% of the Sun,  $\Gamma_{1, int}$ and $W(r)$
in these layers are dominated by these elements, making it difficult to
discern the signatures of heavy elements.
Below this depth one can expect to
see features due to ionization of heavy elements. High ionization
states of light elements like  C, N, O and Ne have broad ionization regions, in particular
C\,{\small V}, N\,{\small VI}, O\,{\small VII}, and Ne\,{\small IX} have
ionization zones
around  (0.90--0.95)$R_\odot$, (0.87--0.93)$R_\odot$,
(0.83--0.90)$R_\odot$ and (0.75--0.85)$R_\odot$ respectively.
 There is
considerable overlap between the different ionization zones of different elements
and hence, unlike in the case of helium, it is difficult to isolate the signal from these
elements in either $\Gamma_1$ or $W(r)$, but we should expect to see the cumulative
effect of all the heavy elements. 

\begin{figure}[t]
\begin{center}
\includegraphics[width=.7\textwidth]{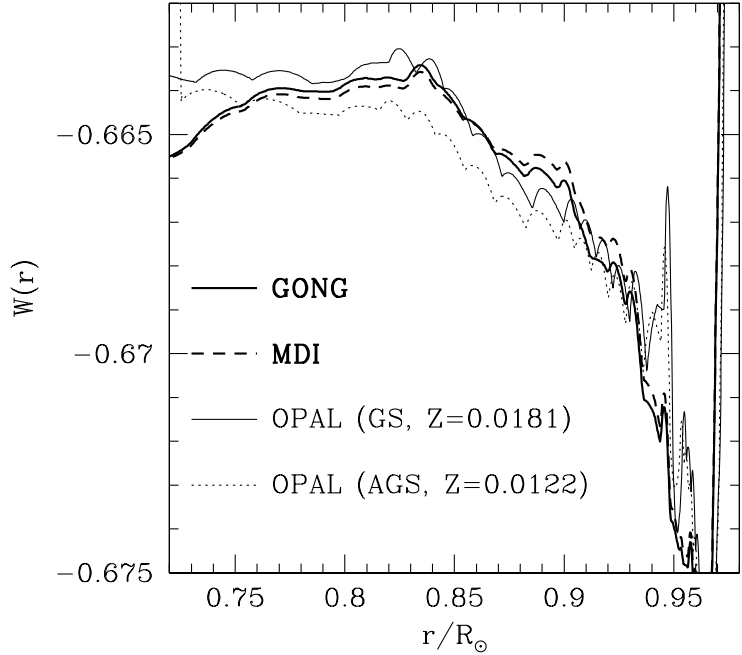}
\end{center}
\caption[]{The dimensionless sound speed gradient, $W(r)$ in solar models
with GS98 and AGS05 abundances are compared with those inferred from
inverted sound speed profiles obtained using GONG and MDI data.}
\label{fig:wrsun}
\end{figure}
\begin{figure}
\begin{center}
\includegraphics[width=.7\textwidth]{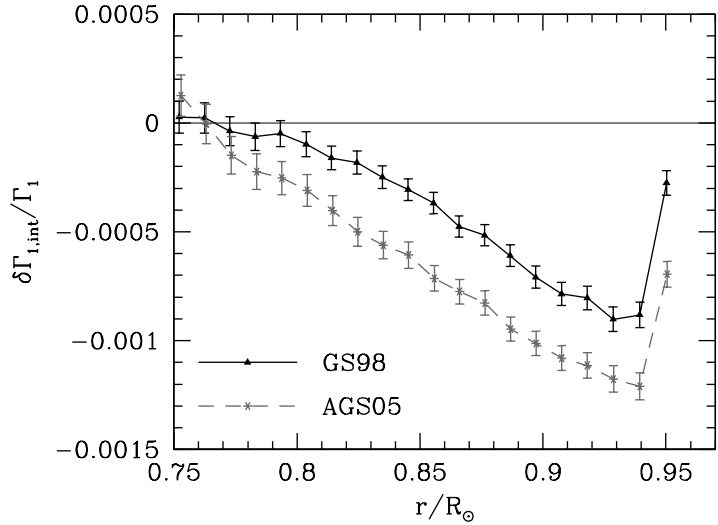}
\end{center}
\caption[]{The intrinsic $\Gamma_1$ differences between the Sun and
two solar models, one with AGS05 abundances, the other with GS98 abundances.
}
\label{fig:gam1}
\end{figure}

Fig.~\ref{fig:wrsun} shows the function $W(r)$ for the Sun and two models,
one with the GS98  abundances, and one with the 
AGS05 abundances.
As can be
seen,  $W(r)$ for the low-$Z$ model is consistently lower
than the solar value of $W(r)$, while the higher $Z$ model matches the solar value
very well, at least below $0.9R_\odot$, where errors in the equation of
state have a small effect.
Fig.~\ref{fig:gam1} shows the relative difference in $\Gamma_{1, int}$ between two
models with different $Z$ and the Sun. 
It can be seen that although none of the models
agree perfectly with the Sun, the low-$Z$ model has a larger disagreement.
Thus although the largest difference between models constructed with the
AGS05 models and the Sun is at the base of the
convection zone, the structure of the convection zone also
shows considerable disagreement.

\subsection{Some consequences for stellar models}
\label{subsec:stellar}

The revision in solar metallicity also affects  models for other
stars. This is because the abundances of stars are often expressed relative to  solar
abundances. Very often stellar metallicity  is expressed as [Fe/H],
and translating this to $Z$ assumes that solar metallicity is known.  Furthermore, most evolutionary tracks and isochrones for population I stars
are calculated assuming that they have the solar mixture of heavy elements.
Thus using AGS05 mixture instead of GS98 would result
in a lower value of $Z$, and also a different relative abundances of elements.
Stellar models with lower-$Z$ but the same helium abundance are bluer
and more luminous than their high-$Z$ counterparts (Salaris \& Cassisi 2005).
There have been a number of recent investigations examining the effects of the
lowered solar abundances on models for other stars. The results do not clearly demonstrate whether or
not the lower solar $Z$ values are better for other stars.
Some of these results are discussed below.

Degl'Innocenti et al.~(2006) examined the effects of heavy-element abundances
on the determination of the stellar cluster ages. They found that there 
are two ways in which
the revised heavy element abundances can affect the age calibration for
globular clusters. First is the effect of the change in the heavy element
mixture, and second is the effect of the reduction of $Z$  for the same value of [Fe/H].
They have examined the
effect of updating the heavy-element mixture on theoretical evolutionary tracks and
isochrones to find a maximum variation of the estimated age of order
of 10\%.  They also found that when the $Z$ values are adjusted from the observed
[Fe/H] for solar like stars, the revised solar $Z/X$ value leads to
a disagreement between theoretical isochrones and the color-magnitude
diagram of the Hyades cluster,  although,
the discrepancy is not large enough to rule out the revised solar abundances.

Piersanti et al.~(2007) studied the effect of heavy element abundance
on stellar evolution calculations for low mass stars. They found that the
evolutionary tracks on the HR diagram are significantly affected when
the lower abundances from Lodders~(2003) were used instead of the
GS98 abundances. They estimate that this leads to an increase in the
estimated age of globular clusters by up to 0.7 Gyr.

VandenBerg et al.~(2007) have studied the effect of heavy element abundances 
on fits of isochrones to the color-magnitude diagram of M67. The open
cluster M67 has an age of about 4 Gyr and was chosen since high-resolution
spectroscopy indicates that this cluster has almost same heavy element
abundance as the Sun and its turnoff stars have masses just around the
lower limit for sustained core convection on the main sequence.
Using MARCS model atmospheres as boundary conditions, they constructed
evolutionary models of stars in range of 0.6--1.4 $M_\odot$ to find that
isochrones formed from models with AGS05 abundances predict a turnoff gap
that does not match the
observations. This difficulty does not arise when GS98 abundances are used.

Montalb\'an \& D'Antona (2006) studied  lithium depletion in pre-main
sequence stellar models with 2D radiative-hydrodynamical convection models.
They found that with higher $Z$,  lithium depletion in  models is
much higher than that expected from the observations of lithium in young
open clusters. Reduction of heavy element abundance as suggested by AGS05
significantly reduces the lithium depletion in pre-main sequence stellar
models, thus bringing them in better agreement with observations.
Sestito et al.~(2006) also found that lithium depletion increases
with increase in $Z$ in the pre-main sequence stage, however, they
conclude that none of the adopted solar mixtures and variations in 
the chemical composition appear to explain the very small amount of 
pre-main sequence lithium  depletion observed for solar-type stars.

Miglio et al.~(2007a,b) have examined the effect of opacity and heavy-element
abundances on the stability of slowly pulsating B stars and $\beta$ Cephei stars.
They have used both OP and OPAL opacities, and  have used heavy element
mixtures from Grevesse \& Noels (1993)  or from AGS05  after enhancing the
abundance of neon. The instability in these
stars is believed to be due to an opacity bump which occurs at lower
temperatures than those at the base of the solar convection zone. At these temperatures
the difference between OP and OPAL opacities, as well as the contribution of
heavy elements,  is much larger and they find significant shifts
in the instability strip due to both opacity and abundances. 
By comparing these models with observations it may be possible to test the
opacities and heavy-element abundances.
Unfortunately, they did not use the AGS05 mixture without neon enhancement
and hence that mixture can not be tested. 
Pamyatnykh \& Ziomek (2007) find that the
$\beta$ Cephei instability domain in the HR diagram when computed with
AGS05 abundances and OP opacities is very similar to the instability
domain calculated earlier with OPAL opacities and the higher abundances.
Looking at the stability of beat Cepheids, 
 Buchler \& Szab\'o (2007) favor the lower solar abundances. They found that
the stability of beat Cepheid models  depend sensitively
on $Z$ as well as the mixture of heavy elements. They showed that with
reduced $Z$ the galactic beat Cepheid models are in better agreement with
observations as compared to models with Grevesse \& Noels (1993) mixture.

Alecian et al.~(2007) have investigated the impact of the new solar
abundances on the calibration of the pre-main sequence binary system RS Cha
to find that it is possible to reproduce the observational data of the
RS Cha stars with models based on AGS05 mixture using standard input
physics. On the other hand, using GN93 mixture it is not possible to find
such models. The difference presumably arises from the relative abundance
of Fe and C, N, O in the two mixtures.
They have not used GS98 abundances.

\section{Attempts to reconcile low-$Z$ solar models with the Sun}
\label{sec:recon}

The large discrepancies between models constructed with the 
AGS05 abundances and the Sun have resulted in numerous attempts to
change the inputs to solar models in order to  reconcile the
models with the helioseismic data.
The proposals generally fall into four categories  or combinations
thereof.
These are (1) increasing input opacities, (2) increasing the heavy element
abundance of the radiative interior through increased diffusion, or alternatively, decreasing
the heavy-element abundance of the convection zone through the accretion of low-$Z$
material,
(3) increasing the abundance of other elements to compensate for the reduction in oxygen
abundance, (4) other processes such as mixing, or the deposition
of energy by gravity waves which effectively changes opacity.
 We consider each of these in the following subsections.
It should be noted that all these proposals only address the effect of
abundances on the structure of solar models through  opacity in the radiative interior. The discrepancy, if
any, in equation of state is not addressed by these proposals. 
Guzik (2006) has reviewed some of the proposed modifications.
Of course, increasing the oxygen
abundance to its old value will resolve all discrepancies. 
As it happens,
none of the proposed changes resolve the discrepancy fully, but it is
possible that some combination of these will fare better depending on
how much the inputs  are changed compared to their expected values.

\subsection{Increasing input  opacities}

Since the primary effect of reducing $Z/X$ is to reduce opacity,
an obvious possibility is to increase the opacities. 
Near the base of the
convection zone $\partial \ln\kappa/\partial \ln Z$ is about $0.7$ (Bahcall et al.~2004) and
hence, opacity  reduces by about 25\% when $Z/X$ is changed from
the GS98 value to the AGS05 value. The actual change is slightly different
since  the relative abundances  of the heavy elements in the AGS05 table is different from
that in the GS98 tables.
If the available opacity tables  underestimate the opacity of solar
material, then the effect of the reduction of $Z/X$ on the opacities may
be compensated to some extent.
Opacities are
theoretically calculated for a given composition, temperature and density,
and hence, it is possible that some opacity sources, such as weak lines, are 
missed.
The fact that there are no laboratory measurements of opacities
also makes it  difficult to estimate the errors in these computations. 
The only way that is available for estimating the errors in the 
computed opacities is through comparison with other, independent calculations.
However, the extent of opacity modifications required to restore agreement
between solar convection zone model and seismic data can be estimated by 
looking at the changes in the position of the convection zone base in models constructed
with different opacities.

Using low-$Z$ solar envelope
models constructed with specified helium abundance and convection-zone depth, Basu \&
Antia (2004) estimated that the opacity near the convection-zone base needs to be
increased by about 20\% to restore agreement with the seismically inferred
density profile.  Basu \& Antia (2004) also calculated the extent of  the
opacity increase required to match the convection zone model for different
mixtures of heavy elements as a function of $Z/X$.
From that they could 
estimate the value of $Z/X$, or alternatively the opacity, required to match the seismic model
of the convection zone.  Since the
contribution to opacity from oxygen and neon fall off with temperature
above a temperature of about $3\times10^6$ K (Fig.~\ref{fig:opac}),
it would be necessary to
taper off the required increase in opacity beyond this point.

\begin{figure}[t]
\begin{center}
\includegraphics[width=.95\textwidth]{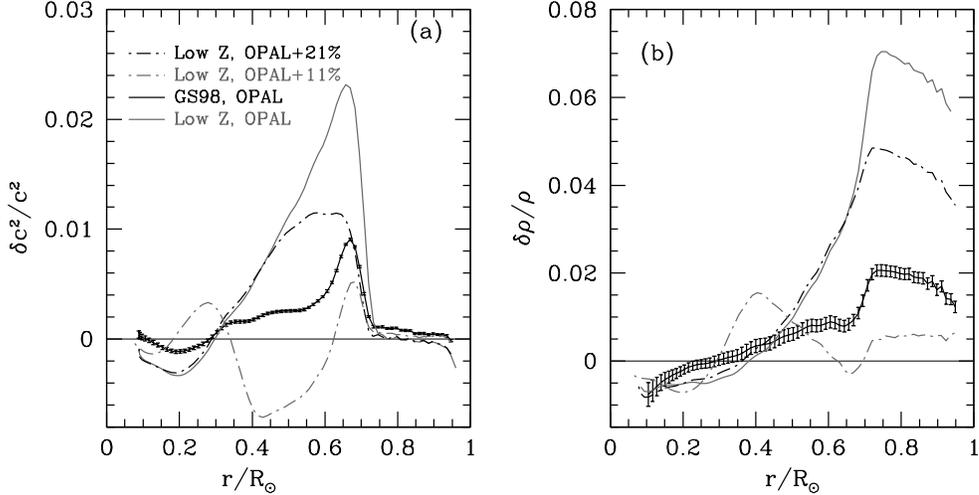}
\caption{The relative sound-speed and density differences between the Sun and
several models with modified opacities. The models marked ``Low Z'' have $Z/X=0.0176$.
The model ``Low Z'' was constructed with unmodified OPAL opacities.
Model ``Low Z+21\%'' had OPAL opacities increased by 21\% near the base of the convection zone,
and model ``Low Z+11\%'' had OPAL opacities increased by 11\% for temperatures ranging from $2\times  10^6$
 to $5 \times  10^6$ K. For comparison a model with GS98 abundances and unmodified OPAL
opacities is also shown. Details of these models can be found in Bahcall et al.~(2005a).
}
\label{fig:opacmod}
\end{center}
\end{figure}

Montalb\'an et al.~(2004) constructed a model in fairly good agreement
with the solar sound-speed profile with a 14\% opacity increase in region near
the convection-zone base.
Bahcall et al.~(2005a) 
 estimated that increases in opacity
ranging from 11\%--21\% may be needed to reconcile the 
low-$Z$ models with helioseismology, the exact increase depends on the
temperature-dependence of the opacity increase. 
They found that a localized increase in opacity near the base of the convection
zone improves the agreement with the depth of the convection zone, but the sound speed
profile in deeper layers is significantly different (see Fig.~\ref{fig:opacmod}). Hence it is necessary
to modify the opacity over a wide range of temperature to compensate for
the reduction in $Z$. They find that an increase in opacity by 11\% in
the temperature range of 2--5$\times10^6$ K, which in a solar models
corresponds to a radius range of (0.4--0.7)$R_\odot$, can restore agreement
in sound speed and density profiles of solar models. The helium abundance
in this model is 0.243, which is  within $2\sigma$ of the seismically
estimated value. From these experiments it appears that an opacity
increase of at least 10\% is required to resolve the discrepancy in
solar models caused by reduced $Z$.
Bahcall et al.~(2005a) also compared
the opacities in solar models constructed using GS98 and AGS05 abundances.
After accounting  for density differences between the models, they find an
opacity difference of 15\% near the base of the convection zone, but the difference
falls sharply to about 5\% around $T=5\times10^6$~K. For $T>9\times10^6$~K
or $r<0.2R_\odot$, the opacity difference is less than 3\%. This is 
expected from what is seen in Fig.~\ref{fig:opac}, and is a result of the fact that  most of the elements 
are fully
ionized in this temperature range, and as a result the difference in opacities
due to difference in abundances is not significant.

Earlier, Bahcall et al.~(2004) had
done 
a detailed study of uncertainties in determining opacities for conditions
in stars using   stellar evolution
codes, in particular they looked at what happens to position of the
base of the convection zone. They
found that different interpolation techniques used to calculate opacities
from OPAL tables give results which can differ by up to 4\%. They stressed
the need for improved opacity tables with finer grid spacings to reduce
 interpolation errors. Nevertheless, these errors do not explain
the discrepancy caused by the reduction of $Z/X$. 

\begin{figure}
\begin{center}
\includegraphics[width=.95\textwidth]{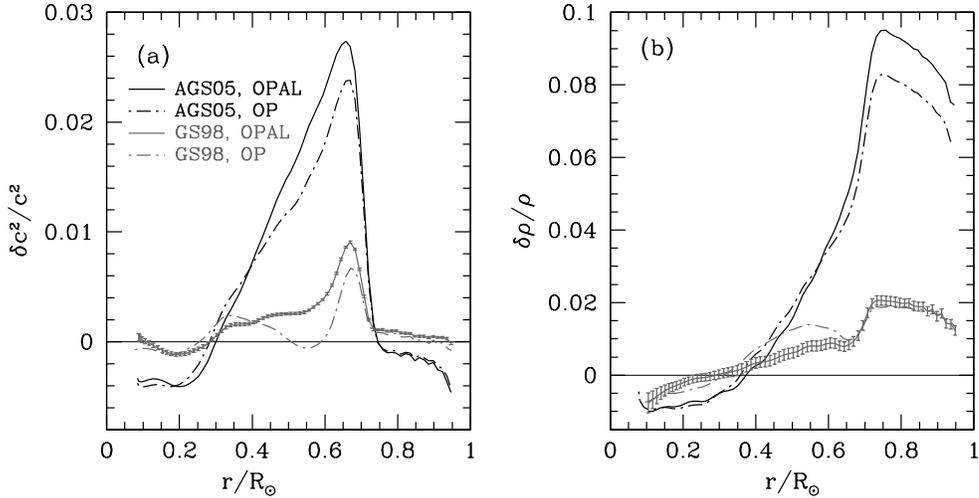}
\caption{
The relative sound-speed and density differences between the Sun and
models constructed with OPAL and OP opacities. Models with both GS98 and
AGS05 abundances are shown.
These models are described by Bahcall et al.~(2005c). 
}
\label{fig:opal_op}
\end{center}
\end{figure} 
At around the  time that the implications of the low-$Z/X$ solar models
were being discussed, 
Seaton \& Badnell (2004) presented the results of independent opacity
calculations made by the OP project, and showed that OP opacities were larger than
OPAL values by about 5\% in the region near the base of the convection zone.
This comparison was for opacities calculated with  a mixture of only six elements,
H, He, C, O, S and Fe. These initial results raised  hopes that opacity
errors could, at least partially, account for the discrepancy caused by the
reduction in $Z$.  The expectations, however, were not fulfilled.
Badnell et al.~(2005) calculated OP opacities with more elements
in the heavy-element mixture, and a comparison 
between models constructed with these OP opacities and those
constructed with OPAL opacities showed that
the difference between OP and OPAL
opacities near the base of the convection zone is less than 2.5\%.
As can be seen from Fig.~\ref{fig:opal_op}, OP opacities  do not solve the problem 
with the low-$Z/X$ models
(Bahcall et al.~2005c).
Guzik et al.~(2005) also compared different opacity
calculations, including the LANL LEDCOP opacities (Neuforge-Verheecke et al.~2001a), 
to find that these agree within about 3\%.
Thus the estimated error in opacities near the base of the convection zone is 2--4\%.
As the temperature increases, the difference between different opacity estimates
generally reduces.
Thus, although improvements in low-$Z$ solar models can be made
by ad hoc opacity increases, there appears to be little justification
for doing so, at least if OP, OPAL and LEDCOP opacities are compared.
The difference between OP, OPAL and LEDCOP opacities could be considered to
be an estimate of uncertainties in the  computed opacities, and this does
not seem to be large enough. Thus
 increasing opacities
within reasonable limits is unlikely to explain the discrepancy between
 low-$Z$ solar models and helioseismology.

Primary inversions for sound-speed and density are independent of
opacity, however estimating temperature and composition profile in radiative
interior needs the  equations of thermal equilibrium, which requires
opacity. Antia \& Basu (2005) have investigated the effect of opacity on
these inversions using models with  both OPAL and OP opacity for calculations.
They find that the difference between OPAL and OP opacity does not affect
the results for temperature and hydrogen-abundance profiles significantly.
Using reduced heavy-element abundances in these calculations will certainly
have significant effects on  the result, but since these abundances are
not consistent with the primary seismic constraints, the solar $T$ and $X$ profiles
obtained with these abundances will not be meaningful.

It may be possible to get better
agreement between solar models and helioseismology by combining a modest
increase in opacity with an increase in $Z$ at the convection-base base
by increasing diffusion.
This option is discussed in the next subsection.

\subsection{Increasing diffusion}

Since only  convection-zone abundances are measured, one
way to improve the agreement of the models with helioseismic data is
to somehow increase the heavy-element abundance at the base of the 
convection zone and the radiative interior. This can be done relatively
easily if it is assumed that diffusion and gravitational settling of helium
and heavy elements is more efficient than what is normally assumed. This was
indeed suggested as a possibility by Asplund et al.~(2004).
Increased diffusion would lead to higher helium and heavy element
abundance in the radiative interior, thereby increasing opacities
and  thus bringing the structure of radiative interior closer to that in the Sun. 

Basu \& Antia (2004) constructed models with
increased  diffusion to find that although,
the convection zone became deeper, the helium abundance in the convection zone
reduced further to 0.225. Thus increasing diffusion in the radiative
interior increases the discrepancy with the seismically inferred
helium abundance. This has also been reported by  Montalb\'an et al.~(2004) and Guzik et al.~(2005).
Guzik et al.~(2005) also
tried to enhance diffusion separately for helium and heavy elements to find
that although the agreement between solar models and seismically inferred
profiles improves to some extent, the required increases in
thermal diffusion rates are unphysically large. None of the variations that they
tried completely restored the good agreement obtained using earlier abundances.
They found that depending on how the diffusion rate is enhanced, it may be
possible to improve the agreement with either the convection zone depth or the helium
abundance in the convection zone, but it is difficult to restore both to the seismically
measured values. They did find that a combination of modest opacity increases,
diffusion enhancements and abundance increases at the level of their respective
uncertainties may be able to restore agreement with helioseismology, however, they
admit  that the solution is rather contrived. Montalb\'an et al.~(2004) too tried
a combination of increased opacity and increased diffusion. They found
that they were able to restore most of the sound speed
profile agreement by  e.g., a
50\% increase in diffusion velocities combined with a 7\% increase in
opacities. However,  the helium abundance in the convection zone in this model, $Y_s=0.239$,
is still significantly lower than the seismic value.
Yang \& Bi~(2007) also tried to resolve the discrepancy between low-abundance models
and seismology by enhancing diffusion. They too find that diffusion coefficients 
have to be increased by a factor of two in order to construct  models with convection-zone depths
that are closer to the seismic value.
They found that the inclusion of mixing below the convection-zone base  helps in improving the agreement
with helium abundance in the convection zone.
However, Yang \& Bi~(2007) do not give  any physical justification of why
diffusion needs to be  enhanced by such a large amount.
Moreover, the sound-speed and density profiles of the resulting models still
disagree considerably with the seismically obtained solar profiles, 
unless $Z$ is increased to 0.0154, a  value that lies  between the AGS05 and GS98 
abundances.

Guzik et al.~(2005) suggested that considering
later accretion of low-metallicity material onto the solar surface that keeps a
high-metallicity interior may resolve the discrepancy.
They assumed that the Sun had 98\% of its current
mass in its pre-main sequence stage and had metallicities like the GS98 metallicities
and accreted the last 2\% of material after it was no longer fully convective. This
material was assumed to be low-$Z$.  The accretion model
has an acceptable value of $Y_s$ (0.2407), but it still has
a shallow convection zone ($r_b=0.7235R_\odot$).
Castro et al.~(2007) too invoked accretion of low-metallicity material in
the early main sequence phase  of the Sun to explain the
observed low abundances of heavy elements in the current solar atmosphere.
They too found that these models have
better agreement with helioseismology, as compared to normal AGS05 models, but they
find  a spike in  $\delta c^2/c^2$ just below the
convection zone that is as large as 3.4\%. Furthermore, both the convection-zone depth 
and the helium abundance in this model
are  lower than the corresponding seismically determined values.
Castro et al.~(2007) also found that the agreement with the solar sound-speed
profile could be improved
by invoking mixing in the region just below the convection zone, however, the
sound-speed profile remains significantly different.

\subsection{Increasing the abundance of  neon and other elements}
\label{subsec:neon}

Since the errors in opacity calculations are unlikely
to be large enough to resolve the discrepancy caused by reduction in oxygen
abundance, and since increasing diffusion does not help either, the next
possibility that was explored was increasing the abundance of 
some other element to compensate for the decrease in the oxygen 
abundance.

Antia \& Basu (2005) examined this issue and found that the convection
zone depth is sensitive to abundances of O, Fe and Ne. Other elements
have little influence on convection-zone models. They suggested that the neon abundance may be increased
to compensate for the reduction in oxygen abundance. Iron would work as
well as neon in providing opacity at the base of the convection zone,
however, it is believed that the solar iron abundance is known reasonably well.
Besides, iron is also an important source of opacity in the solar core
and thus increasing iron abundance could have adverse effects on the structure there.
Neon is an ideal candidate. It cannot be detected in the solar photosphere
since it does not form lines at photospheric temperature. Thus usually
neon abundance is determined from lines formed in the solar corona. It is
assumed that the abundance ratio 
of neon to oxygen in the solar corona and the photosphere is the same, and
hence the photospheric oxygen abundance, along with the coronal Ne/O ratio (0.15) is
often used to determine the photospheric abundance of neon. Since coronal
models are uncertain, the abundance of  neon is quite uncertain,  justifying
models with increased neon abundance.

It can be seen from Fig.~\ref{fig:opac} that the logarithmic opacity gradient 
with respect to  the abundance of oxygen is about 2.5 times that with neon
 near the base of the convection zone.
This implies that to compensate for a reduction in the  oxygen abundance
by a factor of 1.48 we need to increase the neon abundance by almost
a factor of 4 (an additional factor of 1.48 was included to
take into account the fact that the reduction of the oxygen abundance
in the AGS05 table automatically led to the reduction of the
neon abundance).
Since this increase is probably larger than the expected
uncertainties and hence  may be unacceptable,
Antia \& Basu (2005) also explored
the option of increasing the abundances of C, N, and O by $1\sigma$ 
of their tabulated
uncertainty, and then increasing the Ne abundance  to restore
agreement with the seismically determined structure of the
convection zone. They found that in this case the neon abundance needs to
be increased by a factor of 2.5, which is more acceptable.

\begin{figure}[t]
\begin{center}
\includegraphics[width=.99\textwidth]{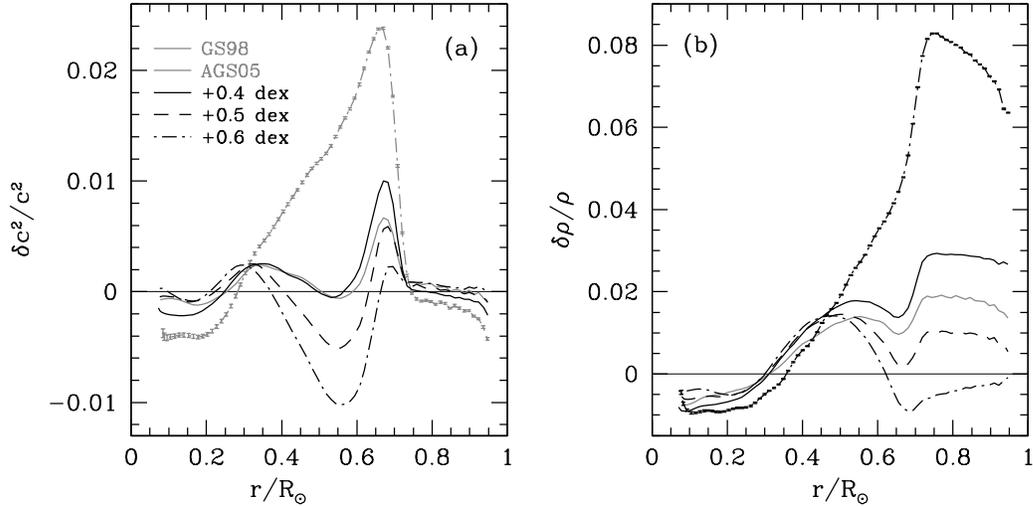}
\caption{The relative sound-speed and density differences between the Sun and
several neon-enhanced models. The amount by which the abundance of neon is
enhanced over the AGS05 values is mentioned in the legend. For comparison
a  model with normal AGS05 abundances and one with GS98 abundances are also 
shown. All models are from Bahcall et al.~(2005b) and  were constructed
with OP opacities.}
\label{fig:purene}
\end{center}
\end{figure}
\begin{figure}
\begin{center}
\includegraphics[width=.99\textwidth]{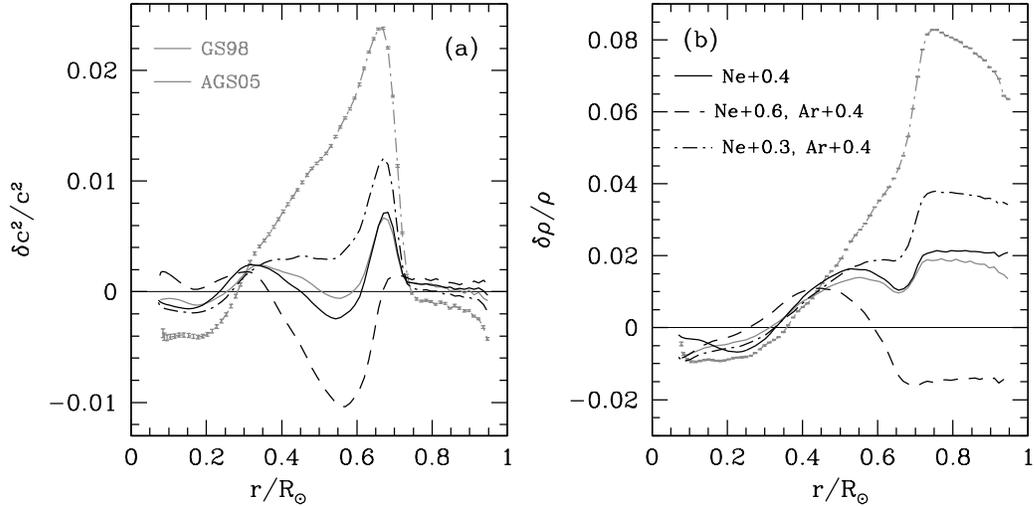}
\caption{The relative sound-speed and density differences between the Sun and
several models with neon and other abundances increased over the AGS05 values.
The amount by which the neon abundance is increased is shown in the figure.
The CNO abundances in all models have been increased by 0.05 dex, and the
abundances of meteoritic elements have been increased by 0.02 dex. In addition, two
models have the argon abundance increased by 0.4 dex. For comparison
a  model with normal AGS05 abundances and one with GS98 abundances are also
shown.
All models are from Bahcall et al.~(2005b) and  were constructed
with OP opacities.}
\label{fig:neplus}
\end{center}
\end{figure}

 
Fig.~\ref{fig:opac} shows
that the contributions of oxygen and neon to opacity have rather different
temperature dependences and hence just increasing the neon abundance cannot
compensate for the reduction of the oxygen abundance throughout the radiative
interior. It may also be necessary to adjust the diffusion coefficient or
other factors to construct models whose structure is in
 better agreement with the  seismically inferred solar structure.
 Antia \& Basu (2005)
used models of only  the solar envelope and hence, could not examine the
effects of neon enhancement in the deep interior.
Bahcall et al.~(2005b) examined the issue of neon
in more detail using
full solar models evolved from the zero-age main sequence stage. They found
that increasing the neon abundance by 0.4--0.5 dex (i.e., factors of 
about 2.5 to 3)  over the AGS05 values
improves agreement with helioseismology in that the position of the
convection-zone base and the convection-zone helium abundance are then
within a few $\sigma$ (rather that $26\sigma$) of the respective helioseismic
estimates. However, the sound-speed and density profiles do not match the
seismic estimates very well (see Fig.~\ref{fig:purene}). Nevertheless, they did get an overall good match
to the seismic profiles when they considered simultaneous increases in
Ne and Ar, as well as a $1\sigma$ increase in C, N, O and some of the
heavy elements whose meteoritic abundances have been well determined. The
sound speed and density results for some of these models are shown in 
Fig.~\ref{fig:neplus}.

Zaatri et al.~(2007) examined the effect that  increasing the abundance of  neon  has  on the core
of
solar models. They compared the scaled small-frequency spacings and the
separation ratio (cf., Eqs.~\ref{eq:nudif},\ref{eq:rats}) on the
low degree modes of the models with the observed
small-frequency spacing derived from GOLF data  (Gelly et al.~2002)
to test the models, and like Basu et al.~(2007) found that the
small frequency separations are extremely sensitive to the metallicity of the
models. They found that increasing neon can reduce the discrepancy between
the  frequency spacings of the solar models and those of the Sun.
In addition to examining frequency spacings, they also did the usual helioseismic
tests of comparing the position of the convection-zone base,  the
convection-zone helium abundance, and the sound-speed profile.
They found that an increase in neon abundance by $0.5\pm0.05$ dex can bring
the model in reasonably good agreement with 
observations, particularly if in addition to enhancing the neon abundance, they
slightly modified the solar age  and the CNO abundances. 
Zaatri et al.~(2007) also pointed out that  the abundance of sulfur
 also plays a role in the discrepancy. Sulfur primarily contributes
to opacity in core and its abundance affects the helium abundance required by a
solar model
to match solar luminosity and radius constraints.

It is by no means clear that increasing the solar neon abundance is
justified. Soon after Antia \& Basu (2005) and Bahcall et al.~(2005b) 
suggested  neon-abundance enhancement as a means of reducing the
discrepancy between the AGS05 models and the Sun,
 Drake \& Testa (2005) claimed that twenty one solar neighborhood 
almost  Sun-like stars seem to have a Ne/O ratio of 0.41
which is much higher than 0.15,  the accepted value of the Ne/O ratio
for the Sun which is also what was used by AGS05.
These results were based on the study of X-ray spectra from Chandra. If this higher
ratio 
is adopted for the Sun, then the solar neon abundance will increase by a
factor of 2.7, and  thus, would  be of some help in resolving the discrepancy between solar models
and helioseismology.
However, Schmelz et al.~(2005) and Young (2005)
reanalyzed solar X-ray and ultraviolet data respectively to find that the Ne/O ratio of the 
Sun is indeed
consistent with the old lower value. 
Young (2005) used an emission-measure method
applied to extreme ultraviolet emission lines of 
Ne\,{\small IV}--{\small VI} and O\,{\small III}--{\small V} ions
observed by the Coronal Diagnostic Spectrometer on board SOHO to find the
abundance ratio Ne/O to be $0.17\pm0.05$ which is consistent with the value
used by AGS05. Schmelz et al.~(2005) analyzed solar active-region spectra from the
archives of the Flat Crystal Spectrometer on the Solar Maximum Mission
to find that the data are consistent with the standard Ne/O abundance
ratio of 0.15. 
They explain the
higher ratio found by Drake \& Testa~(2004) as being due to the higher activity
level in the stars studied. 

Bochsler et al.~(2006) examined the abundances in solar winds and  found that 
the neon abundance is $\log\epsilon({\rm Ne})=8.08\pm0.12$, 
i.e., consistent with the value in the GS98 tables,
but  lower by about a factor of $1.5$  than the neon abundance of
$\log\epsilon({\rm Ne})=8.29\pm0.05$  proposed by 
Bahcall et al.~(2005c) to bring the AGS05 solar models back into agreement
with the Sun.
Bochsler (2007b) extended this work to determine the absolute abundances of  Ne and O.
 After accounting for fractionation in the solar wind,  he finds
$\log\epsilon({\rm O})=8.87\pm0.11$ and $\log\epsilon({\rm Ne})=7.96\pm0.13$. These
values are closer to those listed in the GS98 table, though their large
errors mean that they are also  within $2\sigma$ of the
AGS05 values.
Landi et al.~(2007) used the ultraviolet spectrum  of a solar flare obtained with the SUMER instrument
on board SOHO to measure the absolute abundance of neon in the solar
atmosphere. They  find  $\log\epsilon({\rm Ne})=8.11\pm0.12$, which is much 
higher than the neon abundance in the AGS05 tables ($7.84\pm0.06$), but
somewhat lower than the abundance proposed by  Bahcall et al.~(2005c)  

The situation of  the solar neon abundance remains  confusing, particularly
since measurements of other stars seems to yield high Ne/O ratios.
Cunha et al.~(2006) measured the abundance of neon in B stars in the Orion association
and found $\log\epsilon({\rm Ne}) = 8.11\pm0.04$, about twice the value of AGS05, and they
found Ne/O=0.25, higher than what is assumed for the Sun.
Using X-ray spectra obtained with XMM-Newton,
Liefke et al.~(2006) determined  Ne/O for $\alpha$ Cen, and
found a value that is twice the usual   adopted value  for the Sun.
Using Chandra X-ray spectra again, 
Maggio et al.~(2007) studied the chemical
composition of  146 X-ray-bright pre-main-sequence
stars in the Orion nebula cluster, and found an Ne/O ratio which is a factor
of 2.2 higher than that used by AGS05. However,  their  error-bars  are large.
Planetary nebulae at solar galactic distances appear to have high neon abundances too.
Pottasch \& Bernard-Salas~(2006) determined abundances in planetary
nebulae using infra-red lines. They find a higher (by about a factor of 2)
neon abundance at solar
distances from the galactic center. Their neon-abundance result is consistent with
that of Feldman \& Widing (2003) who used the Ne/Mg ratio to
determine the solar  neon abundance rather than the  Ne/O ratio used by 
GS98 and AGS05. 
Stanghellini et al.~(2006) also examined  neon abundances in planetary
nebulae and obtained Ne/O=0.26.
Ignace et al.~(2007) attempted to estimate Ne abundances from a Spitzer/IRAS
survey of Wolf-Rayet stars and obtained  results that are consistent with the old
solar neon abundances listed by Cox~(2000).
Wang \& Liu (2007) studied elemental abundances of Galactic bulge planetary
nebulae from optical recombination lines and found a value of Ne/O which
is much higher than that used by AGS05.
Similarly, Lanz et al.~(2007) studied Argon abundance in a sample of B
main-sequence stars in the Orion association to find a value
$\log\epsilon_{\rm Ar}=6.66\pm0.06$ which is about a factor of 3 higher than that obtained
by AGS05. Lodders (2007) has tried to estimate solar Argon abundance using
variety of techniques, including solar wind, solar flares, solar energetic
particles as well as from Jupiter, B stars, H\,{\small II} regions etc.
to find that all these are higher than the estimated value by AGS05.
She has proposed a mean value of $\log\epsilon_{\rm Ar}=6.50\pm0.10$ for the
current photospheric abundance and $6.57\pm0.10$ for the protosolar
abundance after taking into account diffusion of Argon in solar interior.

\begin{figure}
\begin{center}
\includegraphics[width=.7\textwidth]{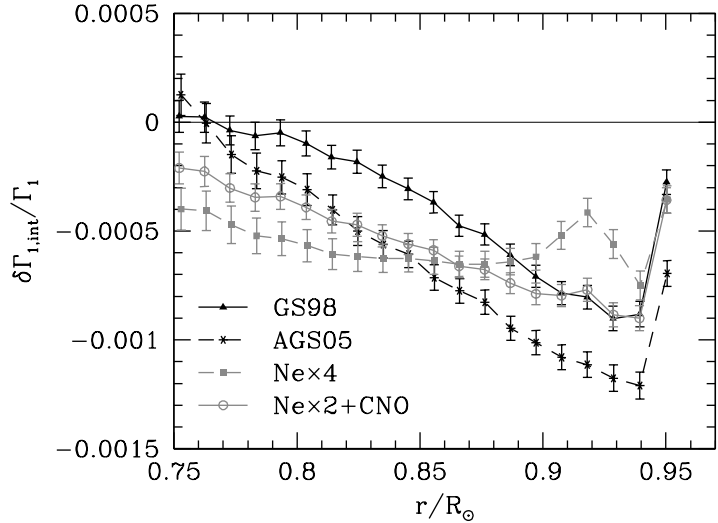}
\end{center}
\caption[]{The intrinsic $\Gamma_1$ differences between the Sun and neon
enhanced solar models. Unmodified GS98 and AGS05 models are also shown
for comparison. Model Ne$\times4$ has neon increased by a factor of
4 over the AGS05 abundances, and Ne$\times2+$CNO has neon abundance
increased by a factor of 2 and CNO abundances increased by 1$\sigma$
of observed errors.
}
\smallskip
\label{fig:gam1ne}
\end{figure}

As is evident from the discussion above, it is not clear whether  an increase in the
adopted solar neon abundance is justified. Increasing the neon abundance
does bring models with low C, N, and O abundances in better agreement with the Sun. However,
the main reason for this is the increased opacities at the base of the convection zone.
Lin et al.~(2007) investigated the consequences of increased neon abundances
in the ionization zones by looking at $\Gamma_{1, int}$ in those regions. They found that 
increasing neon abundances alone does not help in resolving the discrepancy
in the ionization zones, it actually increases the discrepancy in the region
0.75--0.9 $R_\odot$. Better results are obtained only if
C, N and O abundances are raised simultaneously. Thus, the improvement appears to
be a result of increased $Z/X$ rather than a result  of the increased neon abundance.
Although  they could not rule out increased neon
abundances, they showed that an increased neon abundance does not resolve the discrepancy 
between solar models with AGS05 abundances and seismic data. Thus while
increasing neon abundance reduces the discrepancy at the
convection-zone base, it does not reduce the discrepancy in the ionization
zones within the convection zone. This can be seen from Fig.~\ref{fig:gam1ne}.

\subsection{Other processes}

Several other processes have been tried to bring  low-$Z/X$ solar models in
agreement with helioseismic data. Given that there is evidence of mixing below
the convection-zone base, Turck-Chi\`eze et al.~(2004) tried models with mixing
in the tachocline region.
They found that mixing raised the convection-zone helium abundance to
acceptable levels, however, the convection zone itself became shallower. 
Montalb\'an et al.~(2006) 
explored the role of overshoot below the convection-zone base. They assumed
an overshoot of order $0.15H_p$, and by  increasing the opacity by $\approx 7$\% they
were able to reproduce the position of the convection-zone base and the convection-zone
helium abundance. However, the discrepancy of the sound-speed profile in the radiative
zone remained quite large.

Arnett et al.~(2005) and Young \& Arnett (2005) 
following Press (1981) and Press \& Rybicki (1981) 
have tried models in which internal
gravity waves  deposit  energy in the radiative zone.
These waves are assumed to be excited and launched inwards at the
convection-zone base. The radiative damping of these waves, as they travel inwards,
deposits energy and changes the structure in a manner similar to 
opacity enhancements. 
 Young \& Arnett (2005) use  inertial
wave-driven mixing as formulated by Young et al.~(2001).
Their models have mixed success with some being able to reproduce the position
of the convection-zone base and convection-zone helium abundance. However,
none of the models has a radius that is exactly $1R_\odot$ and in that sense
these are not really solar models.  
An alternate mechanism for energy transfer in the solar interior is the
role of dark matter particles like the Weakly Interacting Massive Particles (WIMP).
Although there are no published studied  investigating role of these particles
in low-$Z$ models, solar models with dark matter have been studied
in other contexts (e.g., D\"appen et al.~1986; Faulkner et al.~1986;
Giraud-Heraud et al.~1990; \jcd~1992; Lopes et al.~2002). These models
generally show large effects in the core, and hence may not explain
the departures between the Sun and low-$Z$ solar models in the region below the base of 
the convection zone.
 Whether or not WIMPs with different
characteristics can give the required differences in the right region has not been studied.

\section{Seismic estimates of solar abundances}
\label{sec:seisz}

Even before the current controversy over solar abundances started,
there had been attempts to infer the heavy-element abundance
of the Sun using helioseismic data, as was done for determining  the
abundance of helium, as an independent test of solar
abundances. The efforts gained urgency once it was clear that
solar models with the AGS05 abundances did not match helioseismic
constraints, and that none of the changes made to bring the
models back in concordance with helioseismology worked.

Like  helium,  heavy elements affect
oscillation frequencies indirectly, and hence, one cannot determine
$Z$ directly through inversions. One depends on the effect of
$Z$ on different inputs to the solar models. Unlike the helium
abundance in the Sun, the abundance of heavy elements is small,
as a result, attempts to `invert' for $Z$ in a manner similar to
that for $Y_s$ (cf., Eq.~\ref{eq:invy}, \ref{eq:invdg1}) have not been very successful and
have given results that have very large errors and are somewhat
unstable (e.g., Takata \& Shibahashi 2001).

Attempts to determine the solar heavy-element abundance can be 
grouped into three categories: (1) Studying the regions
where opacity plays a role and depends
on the heavy element abundance. The base of the convection
zone is the most obvious region. 
(2) Using information about core structure, which depends predominantly on the
mean molecular weight of material there (which in turn depends on
the abundances) and to a smaller extent on opacities. 
(3) Looking at the ionization
zone where ionization causes changes in the adiabatic index and
sound-speed. The equation of state plays a role here.  
We describe below
how different regions of the Sun have been used in attempts to determine
the heavy-element abundance of the Sun.

\subsection{Results that depend on the $Z$-dependence of opacity}
\label{subsec:zopac}

\begin{figure}
\begin{center}
\includegraphics[width=.75\textwidth]{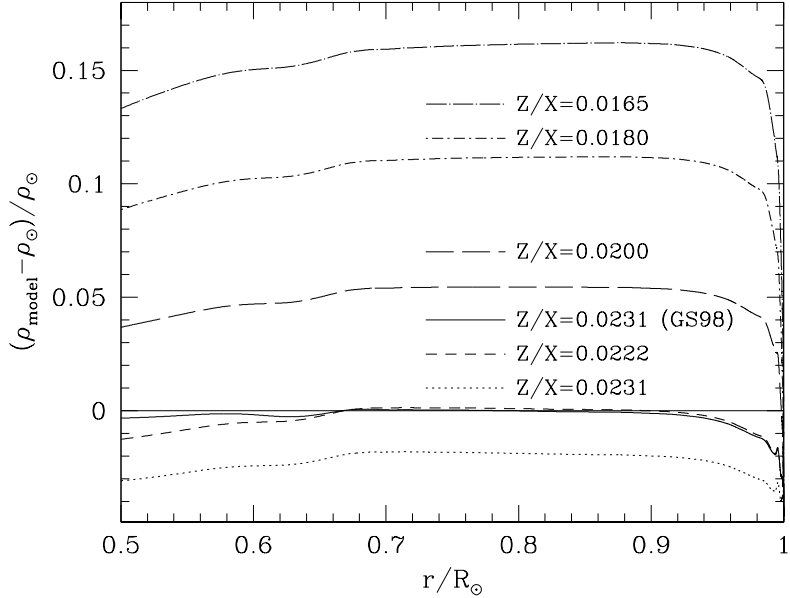}
\end{center}
\caption[]{The relative density difference between the Sun and solar envelope models constructed
with different values of $Z/X$. All models, except the one marked GS98, have
AGS05 relative abundances. All models were constructed with OPAL opacities, and
have the seismic values of $r_b$ and $Y_s$. }
\label{fig:rhozbx}
\smallskip
\end{figure}

\begin{figure}
\begin{center}
\includegraphics[width=.8\textwidth]{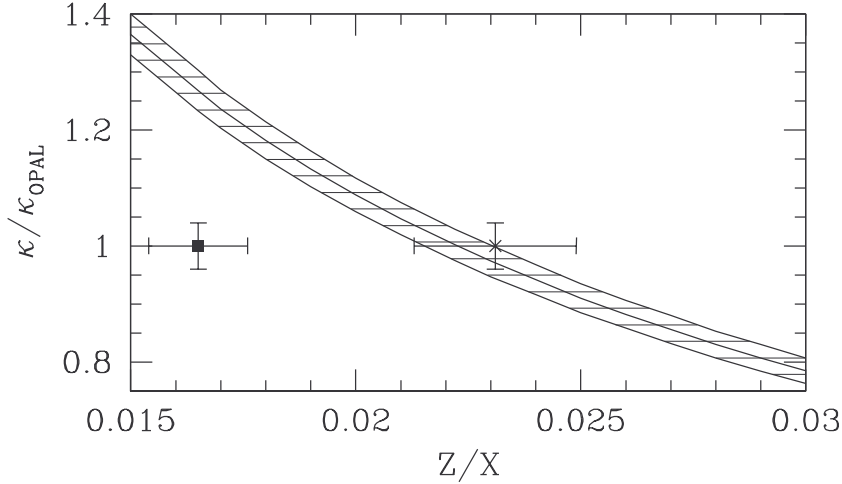}
\end{center}
\caption[]{The shaded area shows the allowed region in the $Z/X$--opacity plane
that is consistent with seismic constraints when OPAL opacities with AGS05
mixture is used. The points with error bars show
current value of opacity and abundances from AGS05 and GS98. The solid
square on the left side marks the AGS05 abundances, while the cross on
the right side marks the $Z/X$ value in GS98.}
\label{fig:zop}
\smallskip
\end{figure}

The depth of the solar convection zone is determined by the
opacity at that position, which in turn depends on the solar heavy-element 
abundances. Basu \& Antia (1997) found that the density profiles of  solar envelope models
constructed to have the correct convection-zone depth and helium abundance is
extremely sensitive to the heavy-element abundance.  They looked at
the possible sources of systematic error in the determination
of the position of the convection-zone base, and concluded that
if the OPAL opacity tables constructed with the then standard
heavy-element abundances  (i.e., those of
Grevesse \& Noels 1993) were valid, then  the heavy element abundance
in the solar envelope should be $Z/X=0.0245\pm0.0008$, consistent with
the then accepted value of $Z/X$.  Basu (1998) repeated the
exercise with different data and obtained the same result.
Subsequently, GS98 
revised the value of $Z/X$ to 0.0231, which is within $2\sigma$ of the
seismic value. However, the latest revision of $Z/X$ by AGS05
makes it $10\sigma$ different from the seismically estimated value. 
Of course, the seismic estimate of $Z/X$ obviously depends on the mixture
of heavy element used to calculate the opacity.  Basu \& Antia (2004) 
repeated the exercise for a  mixture 
similar to that of AGS05 to get a seismic estimate of $Z/X=0.0214$,
 which  still differs by about $6\sigma$ from the AGS05 value.
This exercise can be repeated using the AGS05 mixture.
 Fig.~\ref{fig:rhozbx} shows 
density differences between the Sun and solar envelope models  that have the seismically
estimated values of the convection-zone helium abundance and the 
position of the convection-zone base, but have different values of $Z/X$. 
Most of the models were constructed with the AGS05 mixture. An analysis of
the figure will show that for $Z/X=0.0222\pm0.0008$ the models with AGS05
mixture are consistent with the seismic density profile.
It is possible to estimate the extent of opacity
modification required to get the correct density profile in the lower part
of the convection zone for other values of $Z/X$.
The results are shown in Fig.~\ref{fig:zop}, where the shaded regions show
the allowed region in the opacity--$Z/X$ plane where the envelope
models are consistent with seismic data within estimated errors.
It is clear that the
AGS05 abundance with OPAL opacity is well outside the allowed region, while
the $Z/X$ value of GS98 is consistent with OPAL opacities. For $Z/X=0.0165$,
the value obtained by AGS05, the required opacity modification is
$26.8\pm 3.5$\%.
It is also clear that the seismic estimate of $Z/X$ depends rather sensitively
on the relative mixture of heavy elements, though the AGS05 abundances
are well below the seismic estimates using any of the mixtures.
OP opacities were not available when the work of Basu \& Antia (2004)  was done.
Since the OP opacities are somewhat larger than OPAL opacities at conditions
near the base of the solar convection zone, the seismic estimate 
of $Z/X$ will decrease slightly. If OP opacities are used instead of OPAL,
then the allowed region in the Fig.~\ref{fig:zop} shifts downwards by about
1.7\% and for AGS05 mixture, the models are consistent with seismic data
when $Z/X=0.0218\pm0.0008$.

 Delahaye \& Pinsonneault (2006) did a detailed analysis of seismic constraints,
in particular the convection-zone depth and the convection-zone helium abundance,
in order to try and estimate the heavy-element abundance in the convection zone.
For their study they assumed  reasonable errors on input physics and other parameters
 such as  opacities, diffusion coefficients, nuclear reaction rates,
and the  equation of state. They estimated that the theoretical error in 
the values of $r_b$ in solar models is about $0.0027R_\odot$ and that in
$Y_s$ is about 0.0031. These values are consistent with those found by Bahcall
et al.~(2006).
They found that the depth
of the convection zone  is primarily determined by abundances of lighter elements
like C, N, O and Ne, while the helium abundance is more sensitive to
abundances of heavier elements like Fe. Thus, they
separated the heavy elements in two groups, one consisting of the lighter
elements (C, N, O, Ne) which they refer to as `photospheric' since the abundances
of these elements are determined spectroscopically;
and another group with heavier elements like Fe, which they refer to as
`meteoritic' since the abundances of these elements are primarily 
determined from meteorites.
Since these two groups of elements have a different influences on the
depth of convection zone  and its helium abundance, Delahaye \& Pinsonneault (2006)
argue that  it is, in principle,
possible to determine their abundances independently.
Using the seismic constraints on $r_b$ and $Y_s$, they attempted to 
estimate  solar heavy element 
abundances and found that 
$\log\epsilon({\rm O})=8.87$ dex and $\log\epsilon({\rm Fe})=7.51$ dex, which
are values very similar to those in the GS98 tables.

\subsection{Results from the core}
\label{subsec:zcore}

Basu et al.~(2007) used very precise frequencies of low-degree solar-oscillation modes
measured from 4752 days of data collected by the Birmingham
Solar-Oscillations Network (BiSON)  to determine the degree of agreement between
the cores of solar models constructed  with different values of $Z$ and the Sun.
They examined the scaled small-frequency spacings (cf.,~Eq.~\ref{eq:nudif})
 and separation ratios (cf., Eq.~\ref{eq:rats}) of the  
low-degree modes --- these are sensitive to conditions in the solar core and
hence provide an independent test of solar models. 
Basu et al.~(2007) showed that the small spacings and separation ratios
of models constructed with the old GS98 composition match the
observed BiSON spacings and ratios much more closely than do the
spacings and ratios of models with the lower AGS05 composition. In
short, models constructed with higher metallicities compare better
with the BiSON data than do models constructed with lower
metallicities. The level of agreement deteriorates when the
metallicity becomes very large, indicating that  one should be able
to determine solar metallicity using the spacing and ratio data. They showed
that the separation ratios depend predominantly on the mean-molecular weight of the
core.

\begin{figure}
\begin{center}
\includegraphics[width=.99\textwidth]{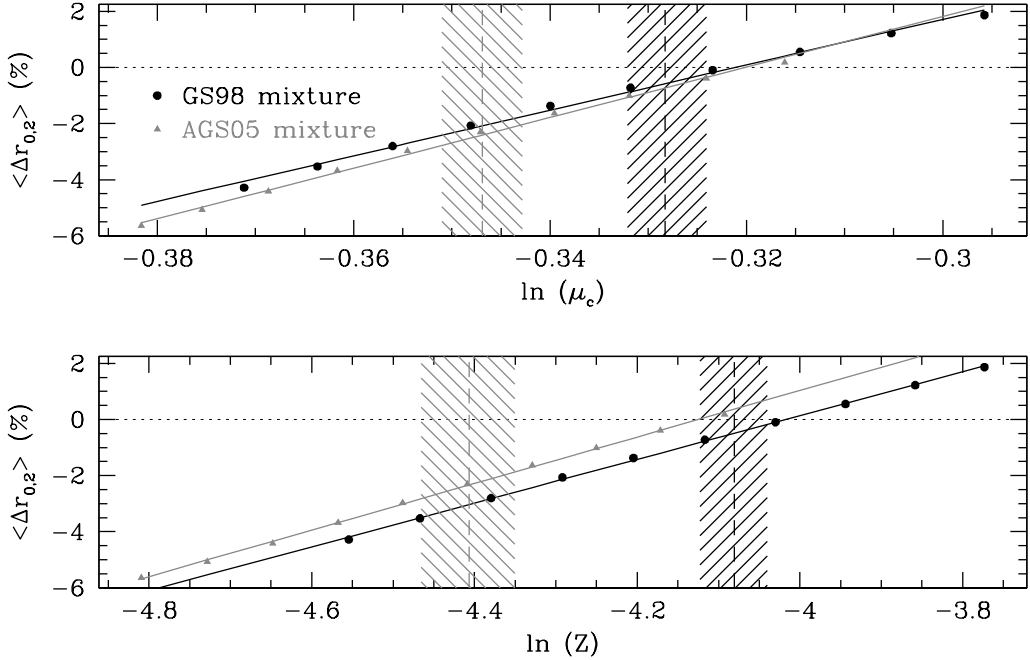}
\caption{The relation between the averaged difference of the
frequency separation ratios between a solar model and that obtained from BiSON data, and the
average mean molecular weight in the core of solar models  as well
as $Z$ for the models. Only results for the (0,2) separation ratios are shown, results for
the (1,3) ratios are similar.
The best-fit straight lines are also
shown. 
Results are shown for models with both GS98 
relative heavy element abundances and AGS05 relative heavy element abundances.
The two vertical lines mark the models with GS98 and AGS05 abundances.
The hashed region around these lines represent uncertainties in $\mu_c$ and $Z$. 
Errors in $<\Delta r_{0,2}> $ are of the size of the points. In the case of the
$\mu_c$ plot, the errors represent the range of $\mu_c$ for a given $Z$
caused by differences in relative composition, errors in the
diffusion rate, etc. The errors were obtained from the Monte Carlo simulations
of Bahcall et al.~(2006). 
}
\label{fig:mulin}
\smallskip
\end{center}
\end{figure}

Chaplin et al.~(2007b) expanded on the work of  Basu et al.~(2007)
to try and determine the average mean molecular weight of the 
core of the Sun. Since the mean molecular weight of the core is related
to the metallicity of the outer layers, they also put constraints on
$Z$. They used the average difference in separation ratios between models
and the Sun. Thus if the separation ratios of the BiSON data are 
 $r_{\ell, \ell+2}(n)$ and
model ratios are  $r'_{\ell, \ell+2}(n)$, then the difference of the two is 
 \begin{equation}
 \Delta r_{\ell,\ell+2}(n) = r_{\ell,\ell+2}(n) - r'_{\ell,\ell+2}(n).
 \label{eq:diff1}
 \end{equation}
These differences were then averaged over $n$, for each of the $\Delta
r_{02}(n)$ and $\Delta r_{13}(n)$, to yield weighted mean
differences, $\langle \Delta r_{\ell, \ell+2} \rangle$:
 \begin{equation}
 \langle \Delta r_{\ell,\ell+2} \rangle =
 \frac{\displaystyle\sum_{n}^{} \Delta r_{\ell,\ell+2}(n) / \sigma^2_{r_{\ell,\ell+2}}(n)}{
       \displaystyle\sum_{n}^{} 1/\sigma^2_{r_{\ell,\ell+2}}(n)}.
 \label{eq:diff2}
 \end{equation}
The formal uncertainties of the BiSON spacings,
$\sigma_{r_{\ell,\ell+2}}(n)$, were used to weigh the averages. 
The data were averaged over the radial-order ranges where good determinations of the separation
ratios were available, which for the data set used was $n=9$ to 25.
While Basu et al.~(2007) had shown that the separation ratios depend on the
molecular weight, they did not determine the exact dependence of
the separation ratios on the average mean molecular weight of the
core. To do so, Chaplin et al.~(2007b)
constructed two sets of test models. The models in each set had different
values of $Z/X$, but one sequence of models was
constructed with the relative heavy element abundances of GS98, while
the second sequence was made with the relative heavy-element abundances of
AGS05.  To fix the $Z/X$ of a given model in either sequence, the
individual relative heavy element abundances of GS98 (or AGS05) were
multiplied by the same constant factor. The models otherwise had identical
physical inputs. Chaplin et al.~(2007b) 
found that $ \langle \Delta r_{\ell,\ell+2} \rangle$,
the mean difference of the separation ratios between the Sun and the models
was a linear function of $\ln \mu_c$ and $\ln Z$, where
$\mu_c$ is the mean molecular weight in the inner 20\% (by radius) of the Sun, and
$Z$ is the metallicity in the convection zone. This relation can be seen in Fig.~\ref{fig:mulin}.
This monotonic relationship led the authors to argue that  $\ln \mu_c$ and $\ln Z$
for the model that lead to perfect match between the separation ratios of the
model and the Sun would be a good estimate of  $\ln \mu_c$ and $\ln Z$ of the
Sun.  The quantities $\mu_{\rm c}$ and $Z$ are
related --- higher $Z$ results in a higher $\mu_{\rm c}$. Two
models with the same $Z$, nuclear reaction rates, opacities and
equation of state can have different values of $\mu_{\rm c}$ only if
the diffusion rates are different in the two models.  It is, however,
not surprising that the dependence of the separation ratios on the two
parameters is somewhat different given that $\mu_{\rm c}$ also depends
heavily on the helium abundance in the core. Since all the models were
calibrated to have the same radius and luminosity at the solar age,
differences in $Z$ generally gave rise to differences in the
core helium abundance.

The value  of $\mu_c$ for which $ \langle \Delta r_{\ell,\ell+2} \rangle=0$
is  almost independent of
the heavy element mixture used in the model, but the value of  $Z$
for which $ \langle \Delta r_{\ell,\ell+2} \rangle=0$  depends on
the mixture. This is not difficult to understand. For the calibrated solar
models used by Chaplin et al.~(2007b), the dominant contribution to the $Z$ and
$\mu_{\rm c}$ of each model comes from different elements. For $Z$,
the dominant elements, in order of importance, are oxygen, carbon,
neon and iron.  The value of $\mu_{\rm c}$ is determined by the mass
fractions of helium and hydrogen in the core. The abundances of
hydrogen and helium in the core depend strongly on the abundances of
heavy elements that contribute to the opacity in the core. These elements,
again in order of importance, are iron, sulfur, silicon and
oxygen. The difference between the GS98 and AGS05 mixture lies
predominantly in the relative abundances of oxygen, carbon, nitrogen
and neon, less so in sulfur, and much less so in the abundances of iron and silicon.
This explains why for the same $Z$, $\mu_{\rm c}$ is different
for the GS98 and AGS05 models. 
From the calibration curves shown in Fig.~\ref{fig:mulin}, Chaplin et al.~(2007b)
determined the solar $\mu_c$ and $Z$ to be $0.7248\pm 0.0008$  and $0.01785\pm 0.00007$
respectively using GS98 models, and $\mu_c=0.7258\pm 0.0008$ and 
$Z=0.01611\pm 0.00008$ using  AGS05 models. The errors here are caused by 
errors in the BiSON frequencies alone and do not have any contribution from
errors in inputs to the two sets of models.

Since $\mu_c$ for a model depends on other input physics
Chaplin et al.~(2007b) 
used the large set of solar models 
of Bahcall et al.~(2006) that were obtained from
a Monte-Carlo simulation (see \S~\ref{sec:abund})
 to actually
determine $\mu_c$ and $Z$ for which $ \langle \Delta r_{\ell,\ell+2} \rangle=0$.
 These models
take into account the relevant uncertainties in standard solar model calculations, and
hence,
the spread in the results will give the uncertainty in the results.
Since the Monte Carlo results did not take into account the error
due to the input opacity tables, these were determined separately and
added to the uncertainties. 
From the results obtained using the Monte Carlo models, Chaplin et al~(2007b)
found that the error on $\mu_c$ is about 0.5\%. The uncertainty in $Z$ is 
larger and is in the range 12--19\%.
Because of the indirect dependence of the
separation ratios on $Z$, it is difficult to obtain a more precise result.
The main source of 
uncertainty in the $Z$ results is the relative abundance of the heavy elements,
and not any of the other physical inputs used to make solar models.
The solar results indicate that the AGS05 models do not satisfy seismic constraints
even in the core.
The average $\mu_c$ 
for the AGS05 models in the Monte Carlo simulation
was  $0.7088\pm 0.0029$.  The GS98 models had an average $\mu_c$ of  $0.7203\pm 0.0029$.
Thus the Chaplin et al.~(2007b)  results
indicate that the discrepancies between solar models constructed with
low metallicity and the helioseismic observations extend to the solar
core and thus cannot be attributed to deficiencies in the modelling of
the solar convection zone.

\subsection{Results that depend on the $Z$-dependence of the Equation of State}
\label{subsec:zeos}

The results mentioned in \S~\ref{subsec:zopac} and \S~\ref{subsec:zcore} exploited 
the effect of heavy element abundances
on opacity. Such estimates are sensitive to other uncertainties in
input physics, like errors in the opacity table, diffusion coefficients, etc., and it is
quite difficult to estimate the effect of these on the derived abundances.
As a result, it is necessary to use seismic techniques that do
not depend on opacity, but on some other input physics. 
In the case of determining the helium abundance, one used the
dimensionless sound speed gradient $W(r)$ (see Eq.~\ref{eq:wr}, and also
\S~\ref{subsec:helmethod}, \S~\ref{subsec:solres})
 or the adiabatic index, $\Gamma_1$. These two functions
are determined by the equation of state  rather than opacity.
There have been attempts to determine $Z$ using $W(r)$ and $\Gamma_1$ (e.g., Elliott 1996).
The main problem with these techniques is that the effect of heavy element
abundances on $W(r)$ or $\Gamma_1$ is rather small because of the low
abundances of heavy elements. The typical ratio of an heavy element to
hydrogen abundance by number  is of the order of $10^{-4}$. It is this quantity, rather than
the mass ratio,  that is the relevant measure needed to quantify the effect of heavy
elements
on the equation of state. Hence the expected  effect of the
heavy-element abundance on  $W(r)$ or $\Gamma_1$  in the
lower convection zone is of this order.  This is just about the accuracy of the inversion
techniques in this region. 

The dimensionless gradient of squared sound speed, $W(r)$ shows peaks in
the ionization zones of various elements. As mentioned in
\S~\ref{subsec:solres}, The peak due to the He\,{\small II} ionization
zone around $r=0.98R_\odot$ has been successfully used to determine
the helium abundance. The height
of this peak is about 0.1, which is roughly the relative abundance of He by number.
Thus  the bumps due to heavy elements are expected  to be of the order of
$10^{-4}$. Furthermore, as can be seen from Fig.~\ref{fig:ion} the ionization
zones of different elements overlap and hence it is difficult
to isolate the signal from each heavy element, making it  difficult
to measure the abundances of individual elements using this technique.
However, Antia \& Basu (2006) showed that it is possible to
determine the total heavy element abundance, $Z$, using $W(r)$. 

Antia \& Basu (2006)  used the RLS
inversion technique to estimate the solar sound-speed profile using  observed frequencies.
The RLS technique was used because it gives the sound speed in a form that
can be easily differentiated to calculate $W(r)$. The acceleration due
to gravity, $g(r)$, was estimated from the inverted density profile.
Before applying this technique to observed frequencies, they tested it by
using frequencies of  known solar models to infer the
 $Z$ for these models. They found that it is indeed possible to reliably
determine $W(r)$ for the Sun to sufficient accuracy and precision to
determine $Z$, and obtained $Z=0.0172\pm 0.002$ in the solar
convection zone.

The main source of uncertainty in the determination of $Z$ from $W(r)$ is the equation
of state, since  in addition to abundances, $W(r)$ depends on the equation of
state.
 The more sophisticated
equations of state like OPAL and MHD are tabulated for a fixed mixture of heavy elements
that is quite different from the solar abundances mixture.
For example, the OPAL equation of state is calculated for a mixture consisting of
only C, N, O and Ne, while the MHD equation of state only  uses 
C, N, O and Fe. 
Hence, it is not
possible to use these equations of state for determining solar $Z$.
As a result, Antia \& Basu (2006) had to use the CEFF equation of state for their work.
They calculated this equation of state using a mixture of 
20 elements (which includes almost
all elements used in OPAL or OP opacity calculations).  The
advantage of using CEFF equation of state was that they could change the mixture
of heavy elements (and $Z$) as per requirements. 

\begin{figure}[t]
\begin{center}
\includegraphics[width=.8\textwidth]{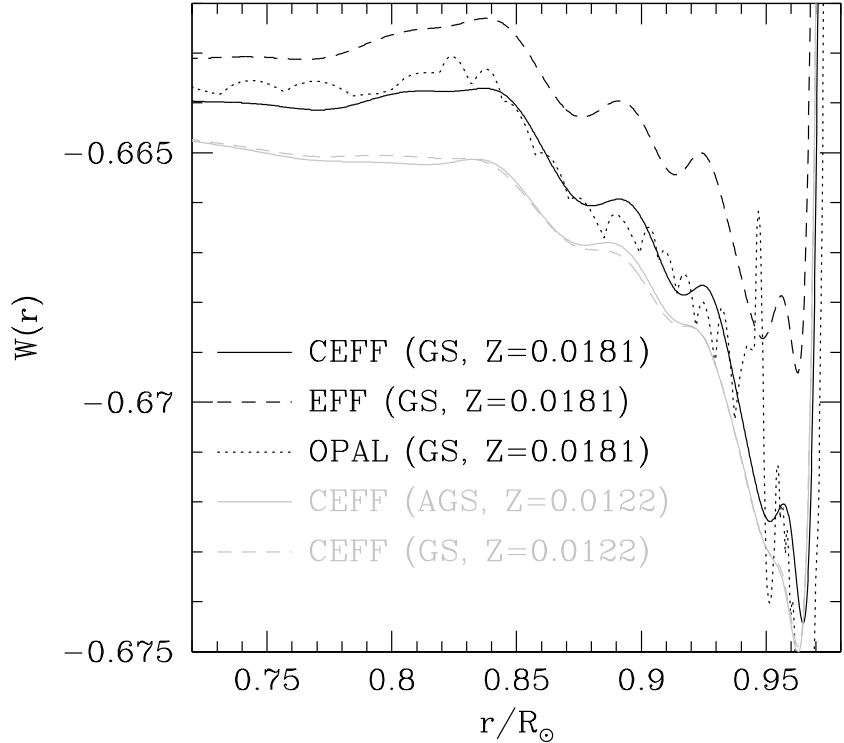}
\end{center}
\caption[]{The dimensionless sound speed-gradient, $W(r)$ in solar models
with different equations of state and heavy-element abundances.}
\label{fig:wreos}
\smallskip
\end{figure}

Figure~\ref{fig:wreos}
shows $W(r)$ for a few solar models using different equations of state and heavy element
abundances. The peak due to the He\,{\small II} ionization zone extends till $0.96R_\odot$,
and hence one  can  only look for
signatures from heavy elements below this region.
The broad peak in $W(r)$, which extends till
the convection-zone base, is caused by the combined effects of all heavy elements. Its height,
of the order of $10^{-3}$, is comparable to the total relative heavy element
abundance by numbers. 
The curve for the model constructed using the  CEFF equation of state shows small peaks 
that modulate the general peak, and these  are  caused by the  ionization zones of
the individual elements. The curve for the  model with
OPAL equation of state shows some additional fluctuations that are caused by
 interpolations needed to use  the OPAL tables.  These errors are
magnified when derivatives are calculated. The numerical errors arise 
from  the  interpolation in  the  grid of tabulated values and from  the finite
accuracy to which the  entries in tables are stored. The CEFF
equation of state and the derivatives are calculated analytically, and hence
does not suffer from this problem.
It is clear that  tables with higher accuracy are needed
to calculate  thermodynamic quantities using OPAL equation of state to sufficient accuracy.
However, for the same heavy element abundances, there is not much difference
between $W(r)$ using CEFF and OPAL equation of state in the radius range of interest in
this study. Thus CEFF is more useful than OPAL equation of state in determining $Z$. 
It is clear from Fig.~\ref{fig:wreos} that $W(r)$ depends on total heavy
element abundance $Z$, and that its dependence on the mixture of heavy
elements is more subtle. Although, the exact shape of the curve also
depends on the equation of state, the results for OPAL and CEFF models that
have the  same value of $Z$ are very similar.

\begin{figure}[t]
\begin{center}
\includegraphics[width=.8\textwidth]{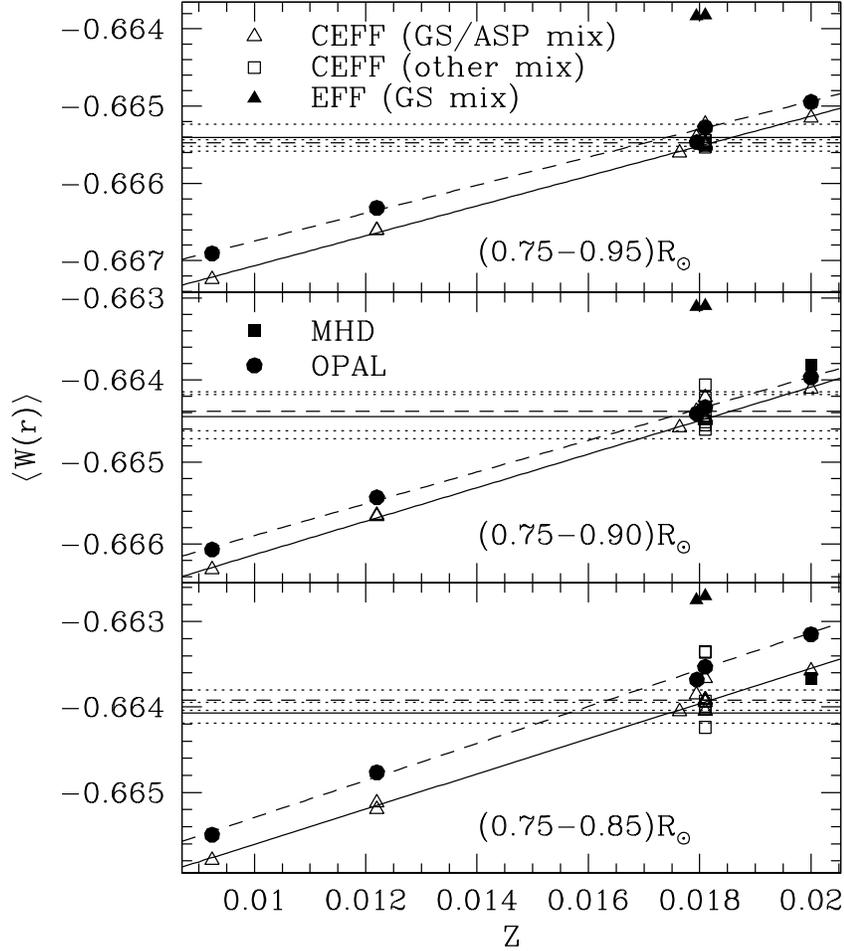}
\end{center}
\caption[]{The average value of $W(r)$ in different radius intervals is
shown as a function of $Z$ for different solar models. The horizontal
lines in each panel shows $\langle W(r)\rangle$ inferred through inversions
of observed solar frequencies. The dotted lines showing the $1\sigma$ error limits
on the helioseismic results.
The solid horizontal lines are for GONG data and dashed for MDI.}
\label{fig:wint}
\smallskip
\end{figure}

In order to get a quantitative estimate of $W(r)$, Antia \& Basu (2006)
proposed to use the average value of $W(r)$,  $\langle W(r)\rangle$, calculated over  different
radius ranges. They used three different radius intervals, 0.75--0.95$R_\odot$,
0.80--0.95$R_\odot$ and 0.85--0.95$R_\odot$.
They examined the
behavior of $\langle W(r)\rangle$ with $Z$ using solar envelope models with
fixed $X_s$ and $r_b$ but different  $Z$.  
The dependence was found to be almost linear  in the expected range of $Z$.
This can be seen in Fig.~\ref{fig:wint}. These calibration curves can be
used to estimate $Z$ from measured $\langle W(r)\rangle$.
The calibration curve, of course,
depends on the equation of state, but the difference between the OPAL and CEFF equation of state is
not very large. The calibration curve also depends on the mixture of heavy
elements used, but the difference between the mixtures obtained by
GS98 and AGS05 is very small hence the effect is not large. Antia \& Basu (2006) have estimated the
systematic errors expected from these and other   sources of uncertainties.  The largest source
of systematic errors are the equation of state and the depth of the convection zone. The
difference between OPAL and CEFF equations of state leads to an error of $0.0015$ in $Z$,
while a variation of $0.01R_\odot$ in the depth of the convection zone results in an error
of $0.0012$ in $Z$.

Using different  data sets of observed frequencies obtained from
GONG and MDI observations, and calibration models constructed with  CEFF and OPAL equations of state,
Antia \& Basu (2006) found $Z=0.0172\pm0.002$. The error bars include
systematic errors caused by uncertainties in the   equation of state, heavy element mixture 
and other uncertainties in the solar models. As can be seen from Fig.~\ref{fig:wrsun}, the function $W(r)$ in models
constructed with AGS05 abundances is inconsistent with that for the Sun as derived
from  observed frequencies. Models with GS98 abundances on the other hand,
fare much better. 
It should be noted that the discrepancy
between solar models and seismic data is essentially independent of opacities
and will not be improved  by increasing the opacity or diffusion coefficients
as described in the last section. Increasing Ne abundance will change
the equation of state  and this option too can be examined using $W(r)$.

\begin{figure}[t]
\begin{center}
\includegraphics[width=.8\textwidth]{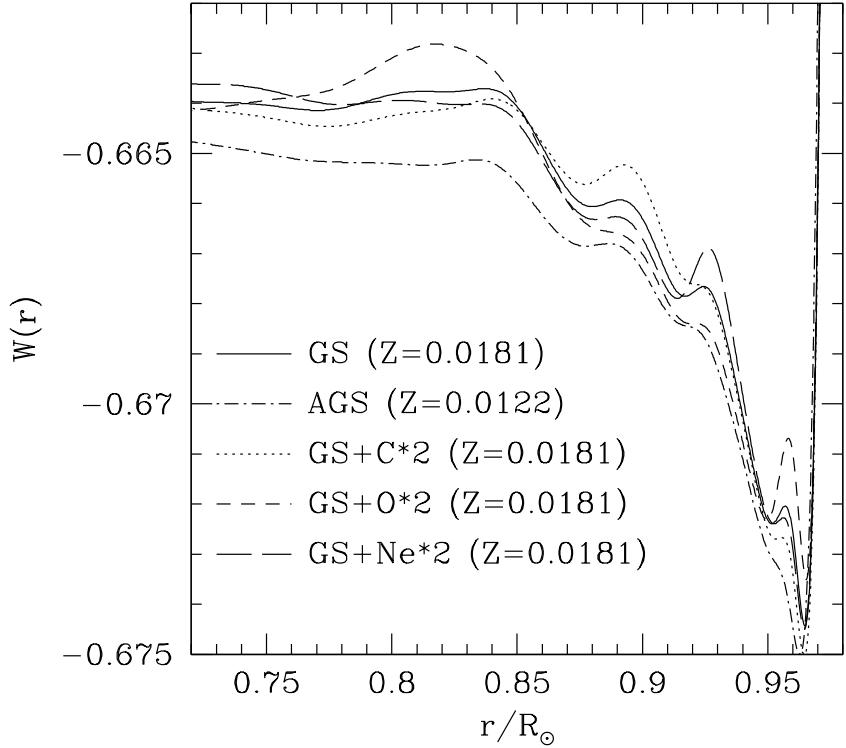}
\end{center}
\caption[]{The dimensionless sound-speed gradient, $W(r)$, of  solar models
with different heavy-element abundances and mixtures. Some of these
models were constructed by enhancing the abundance of an individual
element by a factor of 2. All models
were constructed with  the CEFF equation of state.}
\label{fig:wrz}
\smallskip
\end{figure}

The total heavy element abundance obtained by Antia \& Basu (2006) using $W(r)$ 
is consistent
with the GS98 abundances.
Basu  \& Antia (2006) examined whether or  not one could determine the abundances of 
individual
elements.
For this purpose they looked at
$W(r)$ of models constructed with different heavy-elements mixtures but with the
same value of $Z$.
Thus when the relative abundance of one element was increased,
the actual abundances of other elements were scaled down appropriately.
Fig.~\ref{fig:wrz} shows the results obtained for  models where
abundances of some
elements are  enhanced by a factor of two. It can be seen that increasing carbon
abundance leads to a small peak in $W(r)$ around $r=0.9R_\odot$ which is
probably due to the C\,{\small V} ionization zone. Similarly, increasing the oxygen abundance
gives a broad peak around $r=0.85R_\odot$ that is probably the ionization
zone for O\,{\small VII}, while increasing the neon abundance gives a peak around
$r=0.92R_\odot$. The main peak, caused by  Ne\,{\small IX}, occurs at the lower end of the
convection zone and is not in the radius-range where the  solar
$W(r)$ can be determined to the required precision. All  these peaks, however,
 are fairly small and it is difficult to
identify them in the solar $W(r)$ profile obtained from sound-speed inversions, and 
hence, cannot be used directly to determine the abundances of individual elements. 

Although, it may be difficult to isolated the peaks  due to individual elements from 
$W(r)$ obtained from inversions,  
it is clear from Fig.~\ref{fig:wrz} that the average
value of $W(r)$,  $\langle W(r)\rangle$, calculated over
different radius ranges changes by different amounts when
the mixture of elements is changed. If the calibration models are
constructed using the same mixture of heavy elements as that of  the Sun, and if the
equation of state of the models  also correctly represents the solar 
equation of state, then the estimated value of
$Z$ using  $\langle W(r)\rangle$ calculated over the three radius ranges will 
be identical. Thus, the differences in these estimates would imply a
difference in the mixture, differences in the  equation of state, 
or some other errors in the  solar models. Basu \& Antia (2006) found that 
$\langle W(r)\rangle$ calculated between  0.75--0.95$R_\odot$ is
fairly insensitive to differences in the mixture of heavy elements and is
essentially determined by the total value of $Z$. Thus if $Z_1$,$Z_2$ and $Z_3$ 
represent the values of $Z$ obtained using $\langle W(r)\rangle$ over the
radius ranges
0.75--0.85$R_\odot$, 0.75--0.90$R_\odot$, and  0.75--0.95$R_\odot$,
respectively, then the differences $Z_1-Z_3$ and $Z_2-Z_3$ will contain
some signature of the difference of the heavy-element mixture.
For example, when oxygen abundance is increased by a factor of 2,
$Z_1-Z_3=0.0025$, but $Z_2-Z_3=0.0010$. 
Basu \& Antia (2006) found using CEFF calibration
models constructed with  GS98 heavy element mixture 
that for solar data $Z_1-Z_3=0.00036\pm0.00088$
and $Z_2-Z_3=0.00013\pm0.00135$. Thus it again appears that the solar heavy element
abundances are consistent with GS98.

In principle, one can parameterize the heavy element mixture using three parameters
that can be determined from the three estimates $Z_1,Z_2,Z_3$. For example,
if we assume that these values are mainly determined by abundances of C, O and Ne,
then:
\bea
{\partial (Z_1-Z_3)\over \partial \ln Z_{\rm C}}{\delta Z_{\rm C}\over Z_{\rm C}}+
{\partial (Z_1-Z_3)\over \partial \ln Z_{\rm O}}{\delta Z_{\rm O}\over Z_{\rm O}}+
{\partial (Z_1-Z_3)\over \partial \ln Z_{\rm Ne}}{\delta Z_{\rm Ne}\over Z_{\rm Ne}}&=&
Z_1-Z_3\>,\\
\noalign{\smallskip}
{\partial (Z_2-Z_3)\over \partial \ln Z_{\rm C}}{\delta Z_{\rm C}\over Z_{\rm C}}+
{\partial (Z_2-Z_3)\over \partial \ln Z_{\rm O}}{\delta Z_{\rm O}\over Z_{\rm O}}+
{\partial (Z_2-Z_3)\over \partial \ln Z_{\rm Ne}}{\delta Z_{\rm Ne}\over Z_{\rm Ne}}&=&
Z_2-Z_3\>,\\
\noalign{\smallskip}
\delta Z_{\rm C}+\delta Z_{\rm O}+\delta Z_{\rm Ne}&=&\delta Z=0\>.
\eea
For the last equation we have assumed that the value of $Z$ is determined
from some average of $Z_1$, $Z_2$, and $Z_3$.
Using solar oscillations frequencies, Basu \& Antia (2006) found that
\be
{\delta Z_{\rm C}\over Z_{\rm C}}=0.37\pm0.86,\quad
{\delta Z_{\rm O}\over Z_{\rm O}}=0.04\pm0.31,\quad
{\delta Z_{\rm Ne}\over Z_{\rm Ne}}=-0.62\pm2.75.
\ee
It is clear that the estimated errors in the determination of individual elements
are rather large. In particular, the error in the estimated
 neon abundance is extremely large because of the relative insensitivity
of $W(r)$  to the neon abundance in the radius range
that is used.
The oxygen abundance has the least errors, which is to
be expected since it is the most abundant of the heavy elements and it also has
a  significant influence in the convection zone. The estimated oxygen abundance is
consistent with GS98 abundance. However, it has an error of about  
31\%, which is about a factor of 2
larger than the quoted uncertainties in the spectroscopic estimates listed in GS98
and AGS05, however the uncertainty is  about a factor of 1.5 smaller than the difference between
the GS98 and AGS05 abundances.

\begin{figure}[t]
\begin{center}
\includegraphics[width=.8\textwidth]{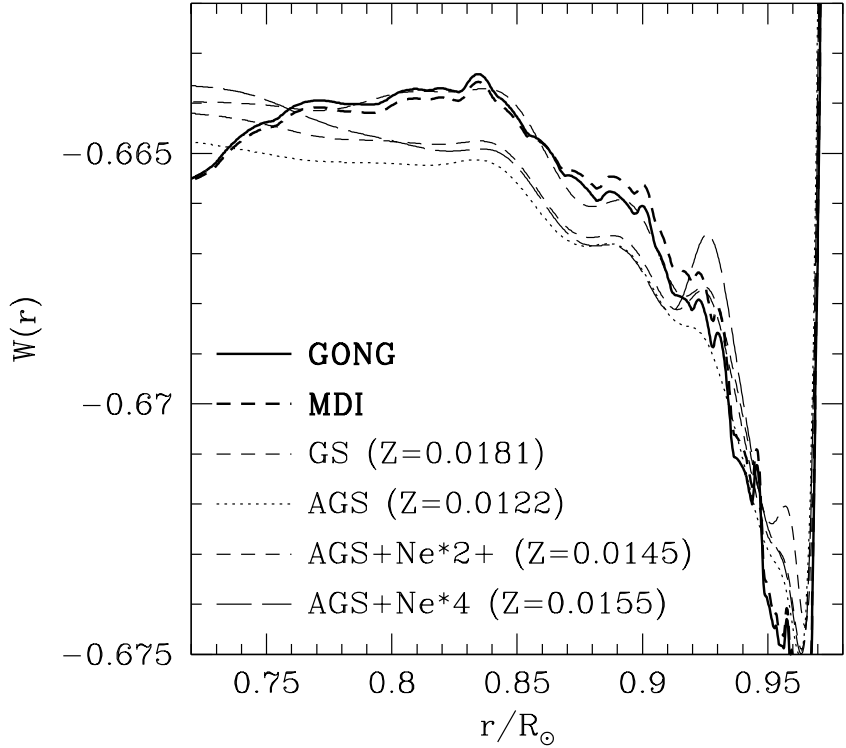}
\end{center}
\caption[]{The dimensionless sound-speed gradient, $W(r)$, of solar models
with different different heavy element abundances and mixtures  compared
with $W(r)$ obtained from inversions of GONG and MDI data. The model labeled
as ${\rm Ne}^*2+$ was constructed by enhancing the abundance of Ne by a
factor of 2 and increasing the C, N, O abundance by $1\sigma$ of AGS05
abundances.
All models, including the neon-enhanced ones,
were constructed with the CEFF equation of state.}
\label{fig:wrne}
\smallskip
\end{figure}

Basu \& Antia (2006) also examined the issue of the solar neon abundance.
From Fig.~\ref{fig:wrz} it is clear that the effect of increasing neon
abundance on $W(r)$ is quite different from that of increasing oxygen
abundance. Hence, the reduction in the oxygen abundance cannot be
be compensated by increasing the neon abundance, at least as far as $W(r)$
is concerned.
 Fig.~\ref{fig:wrne} compares $W(r)$ of models with increased
neon abundance with those of models with GS98 and AGS05 abundances. It is clear that
increasing the abundance of neon does not bring the  $W(r)$ of the
models   close to the solar $W(r)$, though it is better
than the model with normal AGS05 abundances.
Interestingly,  increasing the neon abundance in the GS98 mixture may 
still be consistent with observations. Thus, although, increasing neon 
abundance with the other abundances at the AGS05 level does not work,
 one cannot rule out an increased 
neon abundance beyond the GS98 value. Such an increase will also be consistent with other
seismic constraints.

In addition to $W(r)$, one can also use the adiabatic index, $\Gamma_1$, to constrain
solar heavy-element abundances, since, like $W(r)$, $\Gamma_1$ is also affected
in the ionization zones of heavy elements. The advantages of using $\Gamma_1$
are twofold. First is that unlike $W(r)$, which requires calculation of a derivative with $r$,
 $\Gamma_1$ can be directly inferred from seismic inversions.  The
second advantage is that using the equation of state, we can calculate the dependence of
$\Gamma_1$ on structure variables such as $P$ and $\rho$, and this can be 
subtracted to calculate the intrinsic difference in $\Gamma_1$
as explained in \S~\ref{subsec:helmethod}. This intrinsic difference,
$\delta\Gamma_{1,{\rm int}}/\Gamma_1$, depends only on the equation of state and 
heavy-element abundances. Thus the structure
dependence is explicitly filtered out, something that cannot be done for $W(r)$.
 While $W(r)$ shows a mild dependence
on opacities because of the effect of opacities on the density profile, 
$\delta\Gamma_{1,{\rm int}}/\Gamma_1$ does not show that.
Since $\delta\Gamma_{1,{\rm int}}/\Gamma_1$ 
is independent of differences in solar structure and depends only on
equation of state and the heavy-element abundances,  any discrepancy in $\Gamma_1$
between solar models and the Sun
will not be resolved by increasing opacity or diffusion coefficients.

As mentioned in \S~\ref{subsec:ion},
Lin et al.~(2007) showed that $\delta\Gamma_{1,{\rm int}}/\Gamma_1$
can be used to distinguish between models with different values of
$Z/X$ (see Fig.~\ref{fig:gam1}). They found that models constructed with
AGS05 heavy-element abundances have larger $\delta\Gamma_{1,{\rm int}}/\Gamma_1$ with
respect to the Sun than models with GS98 abundances.
 It is clear from Fig.~\ref{fig:gam1}
that in the region where reliable inversions are possible,
$\delta\Gamma_{1,{\rm int}}/\Gamma_1$ has essentially a uniform shift with change in 
$Z$, and, in principle, by measuring this shift one can estimate $Z$.
However, differences in the equation of state between the models and
the Sun  makes it difficult to get a perfect match with the Sun.

Lin et al.~(2007) also looked at the effect of different heavy-element
mixtures and  found peaks and
dips at different positions when various elements were enhanced separately.
But the peaks and dips for different elements often overlap and in some cases it is possible
to nearly cancel the effect of changes in the abundance of  one element by adjusting the
abundances of other elements suitably. Thus it may not be possible to
determine the abundances of individual elements using this technique,
though one may get a good estimate of total heavy element abundance, and it
appears that the heavy-element abundance is close to the GS98 value.

If the AGS05 abundances are applicable in the convection zone, then the discrepancy
in $W(r)$ and $\Gamma_1$ can only be attributed to errors in the equation of state. It is
difficult to estimate the possible errors in equations of state since these 
are theoretically
calculated and not obtained from laboratory measurements. If we assume that the differences 
between independently calculated
equations of state like OPAL and CEFF, or OPAL and MHD  are a measure of the error 
in the equation of state, we find that the differences are  not
sufficient to explain the discrepancy in $\Gamma_{1, int}$ between models with AGS05 abundances
and the seismic data.  Thus in addition to
 opacity, and diffusion coefficient etc., the equation
of state too will have  to be modified
to resolve the discrepancy between the  AGS05 models and  seismic data.

\section{Possible causes for the mismatch between seismic and spectroscopic abundances}
\label{sec:problems}

Since attempts to modify low-metallicity solar models to bring them in agreement
with seismic inferences have failed, and furthermore since seismic determinations of
heavy element abundances consistently yield higher values in agreement with
GS98, it is  worthwhile to examine the spectroscopic determinations that
lowered the solar abundances in some detail.
The problems could lie with the
solar-atmosphere model used by AGS05, or perhaps with the abundance of any one
of the elements, or perhaps in the physics inputs to stellar models and those
need to be re-examined. 
In this section, we discuss attempts to check the AGS05 abundances and
what needs to be done for that purpose.

It should be recalled that the lowered C, N and O abundances listed in
AGS05 were obtained using a 3D simulation of a small region of the lower solar
atmosphere instead of semi-empirical 1D atmospheric models
used for earlier abundance determinations. 
Asplund et al.~(2006)
claim that they are convinced that their results are correct for a number of different
reasons. These include the fact that they have used a full NLTE analysis. However, more
 important is the fact that they believe the 3D simulation they use is
more realistic than 1D empirical model atmospheres because the 3D model
includes the effects of turbulence. As a result of this, they could match the line profiles
without using any free-parameters for micro- and macro-turbulence and they could
match the line bisectors too. Another reason for their confidence in the results is that the 3D models fit the
line profiles better than 1D models. More importantly,  the abundances 
obtained from different lines are more consistent with each other than
those obtained with a similar 1D analysis. 
In addition to all these reasons,
Grevesse et al.~(2007) also claim that the AGS05 abundances  are more believable since the
lower abundances make solar abundances consistent with abundances of the surrounding
interstellar medium and nearby B stars.
While these are powerful arguments, the correctness of the abundance results
depends crucially on whether or not  the density and thermal structure of the simulations
used is indeed the  density and thermal structure of the Sun.
And although  3D models give similar results using different
lines for C and O,  the same is not true for N, where the 1D Holweger--Muller (1974) model actually
gives more consistent results using both the N\,{\small I} and NH lines
and these values agree with GS98. Furthermore, the AGS05 abundance
for Na is about 0.1 dex lower than the meteoritic abundance, while for Ne
the abundances obtained from Ne/O ratio in corona does not match that obtained
using Ne/Mg ratio. It is difficult to explain these discrepancies.
Thus it is clear that more work is needed to calculate
all heavy element abundances consistently.

Since the publication of the Allende Prieto et al.~(2001, 2002) and Asplund et al.~(2004)
papers and the  AGS05 table, there have been some attempts to independently
derive the solar oxygen abundance.
Mel\'endez (2004) determined the solar oxygen abundance using infrared OH lines. He used
a number of model atmospheres, such as a Kurucz model (Kurucz 1970; Sbordone et al.~2004;
Kurucz 2005a,b) with convective overshooting,
a temporally and spatially averaged version of the Asplund et al.~(2004) model he calls
$\langle 3D\rangle$, the MARCS model of
Asplund et al.~(1997),
and the Holweger-M\"uller (1974) model. He used non-LTE corrections from Asplund et al.~(2004).
The results obtained by Mel\'endez (2004) varied according to the model
atmosphere used. The $\langle 3D\rangle$ model yielded an abundance of 8.59 dex, the MARCS, Kurucz and
Holweger-M\"uller (1974) models give 8.61, 8.71 and 8.80 dex respectively.
However, since this work basically uses the 3D model of AGS05,
it cannot be considered to be an independent determination of the abundances.

Socas-Navarro \& Norton (2007) used spatially resolved spectropolarimetric observations
of the O\,{\footnotesize I} infrared triplet around 777.4 nm. They used Fe\,{\small I}
lines at 630.2 nm to reconstruct the three-dimensional thermal and magnetic structure
of the atmosphere. They  used a 1.5D approach which neglects
lateral radiative transfer.
They claim $\log \epsilon ({\rm O)}= 8.93$ for an LTE calculation and
$\log \epsilon ({\rm O)}=8.63$ for a non-LTE calculation using the
code of Socas-Navarro et al.~(2000).  Their error estimate is 0.1 dex.
But the most disconcerting feature of their results is that
they find that the oxygen abundance they obtain, whether LTE or non-LTE, is not
constant across the solar surface, with magnetic features showing higher
abundances, by as much as 0.3 dex. Their LTE results range for 
$\log \epsilon ({\rm O)}=8.7$ to 9.3, and non-LTE results vary from somewhat less than
$\log \epsilon ({\rm O)}=8.5$ to somewhat more than 8.9. Since there is no physical reason to expect
spatially varying abundances, the authors conclude that this variation must be an
artifact of the analysis, probably caused by imperfect modelling, especially
in the presence of magnetic fields.  

Ayres et al.~(2006) did a photospheric ``thermal profiling'' analysis using
rotational-vibrational bands of carbon monoxide as observed with the McMath-Pierce
Fourier Transform Spectrometer at Kitt peak, and from the space shuttle-borne
ATMOS experiment.
They find a high oxygen abundance, $\log\epsilon({\rm O})=8.85$,
which is consistent with GS98 value, but larger than AGS05.
They point out that this discrepancy may be
because the 3D model of Asplund et al.~(2000b) has too steep a temperature gradient in the
visible continuum forming layers. It is possible that the 3D atmospheric
models reproduce the velocity field in the solar atmosphere reasonably well
and hence they are able to predict correct line shapes, but the thermal
structure in these models may not match the solar atmosphere.
The  1D model perhaps  represent the thermal structure of the
solar atmosphere better because of
the adjustable parameters used in 1D model calculations. 
Ayres et al.~(2006) admit that there are considerable uncertainties in
deriving oxygen abundance using CO lines and more work is required to
get a reliable estimate of the oxygen abundance.
These results are now disputed by Scott et al.~(2006), who used the 
same lines as Ayres et al.~(2006) but the simulations of Asplund et al.~(2000b)
to get a lower abundance. Scott et al.~(2006) find low abundances, consistent
with AGS05, using
the weak lines, and they also find that the abundance is 
independent of the equivalent width when the 3D model is used. However, when
using the low-excitation lines, they find that the determined abundance does depend on the
equivalent width and the mean value is somewhat larger.
This discrepancy is likely to be due to deficiency in modelling higher  layers in the
solar atmosphere.
They also compared the results using different 1D models and found that in
all cases 1D models yield significantly larger abundances.
Thus all  recent investigations show that  there are large discrepancies in the solar 
oxygen abundance obtained from
spectroscopy.

The simulations used by  Allende Prieto et al.~(2001, 2002) and
Asplund et al.~(2004, 2005a) were obtained using the code of Nordlund and Stein,
which had been developed to study solar and stellar surface convection (see e.g.,
Nordlund \& Stein 1990; Stein \& Nordlund 1989, 1998), which has
 been described in \S~\ref{sec:solarz}.
It should be noted that the purpose of simulations done with this code was to study the
properties of solar convection, not abundances.
While there have been some tests of the simulations against the velocity spectrum obtained
at the solar surface, there are no direct tests to examine whether or not the
density and the vertical thermal structure obtained from the simulations matched that of
the Sun. 
The influence of magnetic
fields that permeate the solar photosphere has also not been properly estimated.
Borrero (2007) has attempted to estimate the effect of magnetic fields on
abundance estimates, to find that although Fe abundance can be affected
by up to 0.1 dex, the effect on oxygen abundance is small.

There are  two important tests available for validating  models
of the solar photosphere that can  be performed on the 3D model ---  one is
the center to limb variation and the other are the absolute fluxes (Allende Prieto 2007).
Preliminary results of Asplund et al.~(1999) and Allende-Prieto
et al.~(2004) on center to limb
variation were encouraging.
Koesterke et al.~(2007b), using a different spectral-synthesis code, found that for
many wavelengths,
the center-to-limb variation in abundances
obtained from these 3D models are comparable to that for theoretical 1D  models.
They
did not test the 3D models against semi-empirical  1D  models since the latter  show the
correct center-to-limb variation by design.
Another powerful diagnostic that can be applied to test the 3D models
is an analysis of spatially resolved solar spectra. The line profiles in
the granules would be different from that in the inter-granular regions
and if these are compared with model predictions, it can provide an
independent test of the models (Asplund 2005).

A hint that the thermal structure of the simulations do not exactly match that of the
Sun comes from the fact that hydrogen Balmer-line profiles and the cores
of strong lines do not match, and there is too little flux in the ultra violet
(Trampedach \& Asplund 2003).
While Koesterke et al.~(2007b) found that the center-to-limb variation of the
3D models are not worse than those of the theoretical 1D models,
they however, find that the models have 
temperature gradients that are too steep around $\tau=2/3$.
Ayres et al.~(2006) too have
argued that it is  important to calibrate all photospheric models against the
center-to-limb behavior since this quantity is a measure of the  temperature gradients 
of the regions from which emission occurs. They also claim that the 
temperature gradient of the averaged simulation used by Asplund  et al. is too steep in the layers
that form the visible continuum, however Koesterke et al.~(2007b) argue that the
arguments are not valid.
Scott et al.~(2006) examined  the line shapes of the strongest CO lines,
which form high in the solar photosphere to find some discrepancy which
probably implies that  temperature is overestimated  in the highest
layers of the simulation. 
Ayres et al.~(2006) point out that the discrepancy between their estimate
of the solar oxygen abundance and those obtained with the 3D model
could be a result of the fact that
the 3D model of Asplund et al.~(2000b) has too steep a temperature gradient in the
visible continuum forming layers.

One possible reason for an incorrect thermal structure of the 3D models is  the treatment
of radiative transfer in the simulations.
Because of the two extra dimensions in the horizontal direction, it is not possible to perform the radiative
transfer calculations with the same level of details as for the 1D stellar
atmosphere models (Trampedach 2006).
Trampedach \& Asplund (2003)
showed that the use of the multigroup method, i.e., opacity binning,
leads to systematic differences in radiative heating compared with mono\-chromatic
opacities. There are efforts underway to improve the radiative transfer in the
Nordlund-Stein simulation code (Trampedach 2006).  
In addition to assumptions used in the calculation of  radiative
transfer, all convection simulations have some ``hidden'' parameters. These 
include some type of numerical
viscosity to keep the code stable. Solar plasma on the other hand, has very small molecular viscosity. 
Also given that turbulence can occur
on scales much smaller than what the simulations usually resolve, a sub-grid
scale model of the quantities
must be specified.  Uncertainties in sub-grid scale physics can also give rise
to uncertainties in the thermal structure.
The effect of larger scales of convection, like supergranulation is also not
included in these simulations. 
All these can lead to differences in simulation results obtained by different codes. 
It is not clear yet how these factors affect abundance determinations using these simulations.
The results of
AGS05, therefore, need to be verified using simulations obtained from other codes that use different
models of viscosity and sub-grid scale physics, as well as different assumptions to
do the radiative transport. 
It would, therefore,  be very valuable if different groups would analyze the solar 
abundances using different simulations. The comparison 
between the results of different groups would permit a more robust estimation 
of the systematic uncertainties.

Although, the use of 3D atmospheric models is an improvement in stellar abundance calculations,
it should be noted that even the highest resolution simulations available today are
many orders of magnitude away from being able to reproduce
characteristic Reynolds numbers in the Sun. 
Scott et al.~(2006) found significant
changes in the line bisectors high in the atmosphere as they changed their resolution.
Asplund et al.~(2000a) found
sensitivity of up to 0.10 dex for abundances derived from intermediate
strong lines with significant non-thermal broadening.
Thus numerical tests with substantially increased resolution are required to test
abundance results obtained with 3D hydrodynamical models.
In addition to the use of 3D models, two other factors have played a role in 
the reduction of the oxygen abundance, atomic physics  
(including non-LTE effects), and problems with blended lines. These issues
also need to be examined critically.
In particular, a quantum mechanical treatment of cross-section for inelastic
electron and hydrogen collision is required to calculate the NLTE corrections.
Presently, NLTE calculations have not been performed for all lines and
a systematic study of these for all lines and elements is required.

In \S~\ref{sec:recon} we discussed attempts to modify solar interior models
with low abundances to bring them in agreement with seismic data. The same
exercise needs to be done with solar atmospheric models used in spectroscopic
determination of abundances to check how robust the results are.
Just like the standard solar models, the 3D atmospheric models do not have
any explicit free parameters, but the input physics, and the treatment of  radiative
transfer  can be modified to test the
sensitivity of these models. The atmospheric models depend on input physics
like the opacity, equation of state and other atomic data. Opacities
at these temperatures and densities are more uncertain than opacities
in the solar interior and their effect on the atmospheric models and the inferred
abundances needs to be studied. 
The equation of state may also play an important role in atmospheric model
since at these temperatures most of the free electrons are contributed by
heavy elements with low ionization potential and the electron pressure
plays an important role in line formation. The MHD equation of state used
in simulations of AGS05 probably does not use the correct mixture of heavy
elements. Thus it is necessary to use
equation of state with correct mixture of heavy elements to get the
correct electron pressure.
These models also depend on how radiative transfer is handled.
Because of the extremely time-consuming nature of radiative transfer calculations,
simplifying assumptions are used to solve the relevant equations. 

In the absence of extensive tests it is
difficult to estimate systematic errors in these abundance calculations.
We think that it is necessary to do a Monte-Carlo simulation for atmospheric models,
of the kind done by Bahcall et al.~(2006) for solar interior models, to get a
realistic estimate of errors in abundance calculations.  To a
first approximation, the difference between results obtained by 1D and 3D
atmospheric models may be considered to be an estimate of the errors
caused by   uncertainties
in the atmospheric models. The difference in abundances between the GS98 and
AGS05 tables  are generally of order of $2\sigma$ and as
such we can conclude that there is no serious disagreement. However,
the precision of helioseismic data are much higher, and hence 
the relatively small changes in the  heavy element-abundances (compared to their error estimates) 
result in changes that are much larger than the errors in seismic inferences.

It is heartening to note that there
are at least some efforts to determine solar abundances using the same technique as
Asplund et al.~(2004) etc., but using simulations obtained by different codes.
Ludwig \& Steffen (2007) report a preliminary estimate of
$\log\epsilon({\rm O})= 8.72\pm 0.06$ which is consistent with $8.66\pm 0.05$ determined by AGS05, 
and somewhat lower than
the GS98 value of $8.83\pm 0.06$. However, Ludwig \& Steffen (2007) also
used opacity binning in their simulations and hence the results probably have uncertainties
related to the thermal structure.
Caffau et al.~(2007) have used 3D atmospheric models to find phosphorus
abundance using the P\,{\small I} multiplet at 1051--1068 nm to find
$\log\epsilon({\rm P})=5.46\pm0.04$, which is consistent with the GS98 value
but higher than the AGS05 estimate. They used 3D models computed with
CO${}^5$BOLD code (Freytag et al.~2002; Wedemeyer et al.~2004).
They find that the corrections due to
use of 3D models is small for these lines. It is not clear why their
results are about 0.1 dex higher than AGS05. Thus much more work needs to be done to understand
the results.

Assuming that the 3D solar atmosphere models have the correct thermal structure, and that the
NLTE corrections have been applied correctly and hence the new C, N, and O abundances
are correct, the possibility still remains that the abundance of one key
element is incorrect. As discussed in \S~\ref{subsec:neon}, judiciously increasing the
abundance of neon  can bring models with low C, N and O abundances in better agreement
with seismic data.  There still is  considerable uncertainty in the abundance of neon, which
is an important source of opacity in solar interior. Photospheric
models are not relevant for determining neon abundance since there are
no photospheric lines of neon,  good coronal
models are needed for this purpose. More work is
needed to understand the abundance variations in the solar corona,
solar energetic particles and
solar wind to identify the various processes giving rise to fractionation
in the upper solar atmosphere.
Similar problems were encountered in determining the solar helium abundance,
which is now measured seismically. Determining neon abundance seismically
is more difficult because of its low abundance, though the possibility
should be explored.

There is, of course, also  a need for  more
detailed (and independent) studies of systematic errors involved in 
seismic determination of the solar
heavy element abundance. 
Opacity effects have been studied by different
investigators independently  using independent solar models, and there 
is good agreement between
the results. However, the more subtle effects caused by the equation of state have not been studied
independently. This is particularly important since  these effects are the
ones that 
can rule out most of the input-physics modifications that have been
suggested to resolve the discrepancy between AGS05 models and the Sun.
Current seismic studies in this direction are
based on the  CEFF equation of state, which is not the best that is available,
but is the best that can be used for these studies.
Much better results could be obtained if tables for more 
sophisticated equations of state like MHD
or OPAL are available with different mixtures of heavy elements so that these
studies can be repeated with these equations of state. This, coupled with improved understanding
of systematic errors should lead to improved seismic estimates of heavy
element abundances.

\section{Concluding thoughts}
\label{sec:concl}

The discussions presented in the  previous sections clearly indicate that there is a large discrepancy between the
heavy-element abundances needed to make solar models consistent with
helioseismic data and the lowered solar heavy element abundances
as compiled by AGS05. The disagreement
between the models and the Sun, as determined from helioseismology,
covers almost the whole of the solar interior. 
As discussed in
\S~\ref{sec:abund}, the most
obvious discrepancies between these models and the Sun are the
position of the base of the convection zone and the helium abundance
in the convection zone. Careful investigations, however, have shown that the
models are discrepant in the ionization zones in the upper part of the
convection zone, and in the core as well.
Various suggestions have been put forward to bring the models using reduced abundances back in
concordance with helioseismic data, none of the suggestions however, seem to work very well.

Since the publication of the AGS05 abundances, there have been attempts to determine
solar abundances from helioseismic data, as has been discussed in \S~\ref{sec:seisz}.
All the seismic estimates point towards a high solar $Z$ compared with AGS05.
Since the $Z$-dependence of
solar  oscillations frequencies is not direct, but occurs through the  influence of $Z$ on 
input microphysics such as opacities, equation of state, nuclear reaction rates
etc., it is  possible, in principle,  that the errors in the input 
physics results in discrepant values of $Z$ obtained from helioseismology on one hand and
the analysis technique used by Allende Prieto et al.~(2001, 2002) and
Asplund et al.~(2004, 2005a) on the other. However, the seismic $Z$ determinations
have been done using techniques that depend on different inputs, and
despite the differences in techniques and dependences on different inputs,
 all seismic estimates of $Z/X$ are consistent
with the higher GS98 abundances, and they agree with each other as well.
Thus the question arises as to  what could be causing the discrepancy between
the seismic abundances on one hand and the abundances determined from spectral lines
listed in AGS05. 

If the convection-zone abundances of the Sun are indeed consistent with
the low abundances compiled by AGS05, then almost all the input physics that goes into construction
of stellar models are much more uncertain than they are normally assumed to be.
On the other hand, if the GS98 abundances are correct
then the currently known input physics is consistent with seismic data.
Thus  it is either a  sheer coincidence that the errors in the 
input physics cancel each other when GS98 abundances are used, or the
recently revised abundances due to AGS05 need to be revised upwards.
The only other option is that some fundamental process is missing in the
theory of stellar structure and evolution,
but it is difficult to speculate what that could be. It is easier to 
find reasons  that could cause the thermal stratification of the new three-dimensional 
model atmospheres to be modified and give results different from AGS05.

The serious disagreements between helioseismic estimates and recent spectroscopic
estimates of the solar heavy-element abundance requires a careful examination of
 solar  atmospheric models, and also  models of the solar interior.
This is important not only to resolve the discrepancy, but also
 because solar abundances
are used as the scale for abundances of other astronomical objects. 
A lot of improvements in solar models during the last three decades of the
previous century were fueled by the solar neutrino problem. The discrepancy
between the observed neutrino fluxes and those predicted by the solar models
led to critical examination of all input physics that goes in construction
of solar models. These developments, including the seismic studies,
finally led to confirmation of new physics beyond the standard model of
particle physics.
We can expect that the discrepancy caused by revision of solar heavy element
abundances will lead to further improvements in  models of the solar
atmosphere and perhaps of the solar interior as well.

\section*{Acknowledgments}

The authors thank the referee for comments on the first version of this review.
They would like to thanks the OPAL project for making it possible to calculate 
opacities for different relative abundances.
The authors would also like to thank Profs. Pierre Demarque and Sabatino Sofia for their
comments on this article.
Most of the results quoted in this work and the figures shown were obtained using data from
the GONG and MDI projects. MDI is an instrument on board  the Solar
and Heliospheric Observatory (SOHO).  SOHO is a project of
international cooperation between ESA and NASA.  MDI is supported by NASA grant NAG5-8878
to Stanford University. The GONG Program is managed by the National Solar 
Observatory, which is operated by AURA, Inc. under a cooperative agreement with 
the National Science Foundation. The data were acquired by instruments operated by 
the Big Bear Solar Observatory, High Altitude Observatory, 
Learmonth Solar Observatory, Udaipur Solar Observatory, Instituto de 
Astrofisica de Canarias, and Cerro Tololo Interamerican Observatory.
Some of the figures use data from the Birmingham Solar Oscillations Network (BiSON) which is
funded by the UK Science and Technology Facilities Council (STFC).
SB acknowledges NSF grant ATM 0348837 for partial support.

%Journal names
\newcommand{\apj}{Astrophys.\ J.\ }
\newcommand{\apjl}{Astrophys.\ J.\ }
\newcommand{\apjs}{Astrophys.\ J. Suppl.\ Ser.\ }
\newcommand{\apss}{Astrophys.\ Space Sci.\ }
\newcommand{\aap}{Astron.\ Astrophys.\ }
\newcommand{\aaps}{Astron.\ Astrophys.\ Suppl.\ Ser.\ }
\newcommand{\aj}{Astron.\ J.\ }
\newcommand{\an}{Astron.\ Nacht.\ }
\newcommand{\mnras}{Mon.\ Not.\ Roy.\ Astron.\ Soc.\ }
\newcommand{\pasj}{Publ.\ Astron.\ Soc.\ Japan \ }
\newcommand{\physrep}{Phys.\ Rep.\ }
\newcommand{\prd}{Phys.\ Rev.\ D\ }
\newcommand{\prl}{Phys.\ Rev.\ Lett.\ }
\newcommand{\nat}{Nature\ }
\newcommand{\araa}{Ann.\ Rev.\ Astron.\ Astrophys.\ }
\newcommand{\ssr}{Space Sci.\ Rev.\ }
\newcommand{\rmp}{Rev.\ Mod.\ Phys.\ }
\newcommand{\basi}{Bull.\ Astron.\ Soc.\ India\ }
\newcommand{\jgr}{J. Geophys.\ Res.\ }

\end{document}